\documentclass[prd,preprintnumbers,amsmath,amssymb]{revtex4}

\usepackage{array,mathtools,amssymb,booktabs,multirow,diagbox,hhline}
\usepackage{graphicx}
\usepackage{dcolumn}
\usepackage{bm}
\usepackage{bbold}
\newcommand{\Od}{{\cal O}}
\newcommand{\pint}{\int \frac{d^3 \vec{p}}{(2\pi)^3}}
\newcommand{\tr}{\mbox{tr}}
\newcommand{\re}{\mbox{Re}\,}
\newcommand{\intT}{\int_0^\beta d\tau \int d^3 \vec{x}}

\newcommand{\mean}[1]{\left\langle{#1}\right\rangle}
\newcommand{\parent}[1]{\left({#1}\right)}

\newcommand{\condl}{\mean{\bar q q}_l}
\newcommand{\conds}{\langle \bar s s \rangle}

\newcommand{\quarkcorlmod}{\langle \bar \psi_l \psi_l (x)  \bar \psi_l \psi_l (0)\rangle}
\newcommand{\im}{\mbox{Im}}
\newcommand{\lsim}{\raise.3ex\hbox{$<$\kern-.75em\lower1ex\hbox{$\sim$}}}
\usepackage{environ}
\NewEnviron{myequation}{%
\begin{equation}
\scalebox{1}{$\BODY$}
\end{equation}}

\begin{document} 

\title{Aspects on Effective Theories and the  QCD transition}
\author{A. G\'omez Nicola}
\email{gomez@fis.ucm.es}
\affiliation{Departamento de F\'{\i}sica Te\'orica and IPARCOS. Univ. Complutense. 28040 Madrid. Spain}

\begin{abstract}
We~review recent advances in the understanding of the Quantum Chromodynamics (QCD) transition and its nature, paying special attention to the analysis of chiral symmetry restoration within different approaches based on effective theories. After presenting some of the main aspects of the current knowledge of the phase diagram from the theoretical, experimental and lattice sides, we discuss some recent problems where  approaches relying on effective theories have  been particularly useful. In~particular, the~combination of ideas such as Chiral Perturbation Theory, unitarity and Ward Identities allows us to describe successfully several observables of interest. This is particularly relevant for  quantities expected to be dominated by the light meson components of the hadron gas such as the scalar and topological susceptibilities. In~addition, ward identities and   effective Lagrangians  provide systematic results regarding chiral and $U(1)_A$ partner degeneration properties which are of great importance for the  interplay between those two transitions and the nature of chiral symmetry restoration. Special attention is paid to the connection of this theoretical framework with   lattice simulations.
\end{abstract}

\maketitle

\section{Introduction and Motivation: Remarks on the QCD Phase Diagram}
\label{sec:intr}

One of the most striking features of strongly interacting  matter is its behaviour under extreme external conditions. A prominent example is the physical system created in an ultrarelativistic heavy ion collision and its subsequent expansion, which undergoes several phases and regimes, starting from an initial nonequilibrium high-density state where the relevant degrees of freedom are  deconfined quarks and gluons, the~Quark-Gluon Plasma (QGP),  followed by a transition to a hadron gas through a locally thermalized expansion \cite{Adams:2005dq}. During such expansion, the~produced yields of particles and their spectral properties carry useful information about the thermal system and the phases involved~\cite{Adcox:2004mh}. The~main experimental heavy-ion programs are nowadays developed at the Relativistic Heavy Ion Collider (RHIC) at Brookhaven National Laboratory and the Large Hadron Collider (LHC) at CERN. 

\begin{figure}[h]
\centering
\includegraphics[width=10 cm]{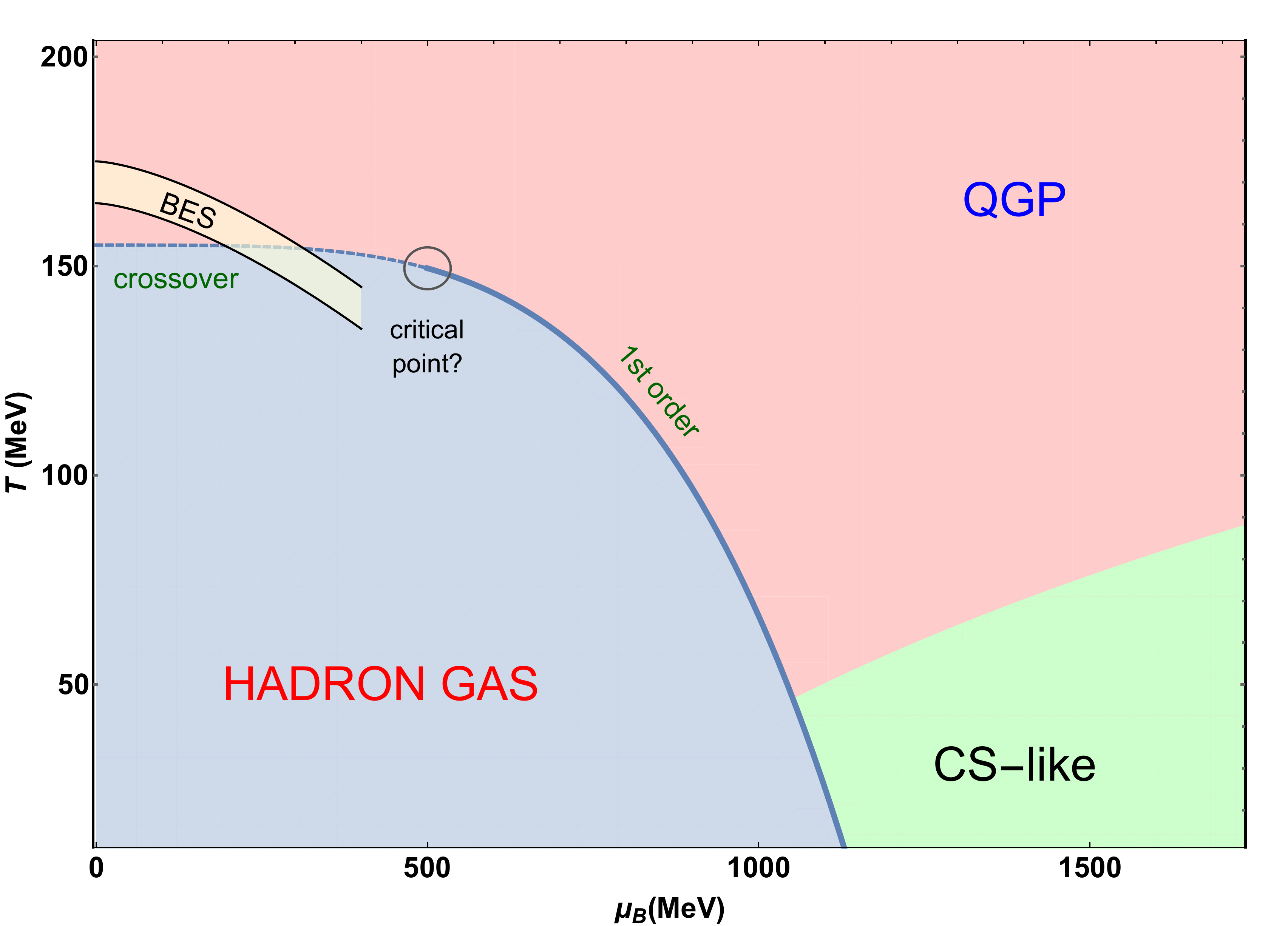}
\caption{Schematic representation of the Quantum Chromodynamics (QCD) phase diagram in the plane of temperature and baryon chemical potential.}
\label{fig:phasediag} 
\end{figure}

The~statistical variables of the system, namely temperature and chemical potentials, label the phase diagram  of Quantum Chromodynamics (QCD),  depicted schematically in Figure \ref{fig:phasediag} 
 in the $(T,\mu_B)$ plane, with $\mu_B$ the chemical potential associated to the conservation of baryon number.

As shown in the figure, a transition curve $(T_c,\mu_c)$ is expected, signaling deconfinement and chiral symmetry restoration, two phenomena of very different nature but connected in  a way still not totally understood. 
Chiral symmetry restoration reflects the transition from an ordered phase of low $T$ and low $\mu_B$ (and hence low baryon density)  where the chiral symmetry $SU_L(N_f)\times SU_R(N_f)$ is broken down to the vector one $SU_{L+R}(N_f)$ (isospin symmetry for $N_f=2$) with $N_f$  the number of light flavors, to a disordered one where such symmetry is restored. The~critical line corresponds very likely to a first-order phase transition for large values of $\mu_B$, presumably ending in the so called QCD critical point where the transition would become a smooth crossover for low $\mu_B$. The~existence and properties of the QCD critical point is actually one of the major goals in the current research activity in this field~\cite{Ratti:2018ksb,Bazavov:2019lgz}

The~region of  high baryon density and  small temperature is not so well understood and is  accessible nowadays only  within the realms of astrophysical objects such as neutron stars. In~that region, novel phases of QCD such as Color Superconductivity (CS) appear, with different possibilities for Cooper pairing quark states and for quark matter types \cite{Alford:2007xm}.

The~best explored region of the phase diagram is that of vanishing or low baryon density. Our~current knowledge in that region comes essentially from two complementary approaches to hot and dense QCD matter: lattice simulations and direct experimental information extracted from Heavy-Ion~data.

Thus, on the one hand, lattice simulations have been over many years the only source of information on the QCD phase diagram to compare with from theoretical models and analysis. At~$\mu_B=0$, the~analysis of observables directly related to confinement and chiral symmetry restoration has provided a very useful insight. That is the case of the Polyakov loop, thermodynamic pressure, the~entropy and the trace anomaly, all showing deconfinement features  \cite{Ratti:2018ksb,Bazavov:2016uvm}  and  derived from the free energy density at finite temperature $T$,
\begin{equation}
z(T)=-\lim_{V\rightarrow\infty}(\beta V)^{-1}\log Z,
\label{z}
\end{equation} 
with $\beta=1/T$ $Z$ the QCD partition function.  The~thermodynamic pressure and the trace anomaly are then given by
\begin{eqnarray*}
P(T)&=&z(T=0)-z(T) \nonumber \\
\langle \theta(T) - \theta (0) \rangle&=&T^5\frac{d}{dT}\left(\frac{P(T)}{T^4}\right)
\end{eqnarray*}
where $\theta=T_\mu^\mu$ with $T_{\mu\nu}$ the energy-momentum tensor. The~trace anomaly, also denoted as interaction measure, is a very interesting quantity from the point of view of the deconfinement transition. It~parameterizes, through the QCD conformal anomaly, the~breaking of conformal invariance which is meant to be enhanced when quarks and gluon degrees of freedom are liberated. In~fact, the~peak in the trace anomaly observed in lattice simulations is a reflection of that mechanism \cite{Ratti:2018ksb}.

 As for chiral symmetry restoration, the~main observables to be considered are the light quark condensate $\condl=\langle \bar u u + \bar d d \rangle$ and the scalar susceptibility $\chi_S$:
\begin{eqnarray}
\condl(T)&=&\frac{\partial}{\partial \hat m}z(T),
\label{condef}\\
\chi_S (T)&=&-\frac{\partial}{\partial \hat m} \condl(T)=\int_T {dx \left[\quarkcorlmod-\condl^2(T)\right]},
\label{susdef}
\end{eqnarray}
where $\displaystyle \int_T dx\equiv \intT$ at finite temperature $T=1/\beta$, $\mean{\cdot}$ denote Euclidean finite-$T$ correlators, $\psi_l^T=(u,d)$ is the fermion flavor doublet field and $\hat m=m_u=m_d$ is the light quark mass in the isospin limit. Note that $\chi_S$ is the total susceptibility that can be separated into connected and disconnected parts in terms of quark diagrams as $\chi_S=2\chi_{con}+4\chi_{dis}$ for $N_f=2$, where $\chi_{dis}$ is expected to diverge at chiral restoration  \cite{Smilga:1995qf}.

The~light quark condensate is the order parameter in the light chiral limit $m_u=m_d=0$, where the transition is most likely a second-order one for $\mu_B=0$~\cite{Pisarski:1983ms}. In~that limit, at the critical point  the quark condensate vanishes and the  scalar susceptibility diverges \cite{Smilga:1995qf}. In~the physical quark mass case, simulations show a crossover transition at $T_c=154\pm 9$ MeV from the study of various observables   \cite{Ratti:2018ksb,Aoki:2009sc,Borsanyi:2010bp,Bazavov:2011nk,Bazavov:2018mes} corresponding to an inflection point of the quark condensate and a peak of the scalar susceptibility. Recent simulations in the light chiral limit \cite{Ding:2019prx} show a reduction of the transition temperature, as expected from the absence of explicit chiral symmetry breaking, down to $T_c^0\simeq$ 132~MeV. 

\begin{figure}[h]
\centering
\includegraphics[width=7.5 cm]{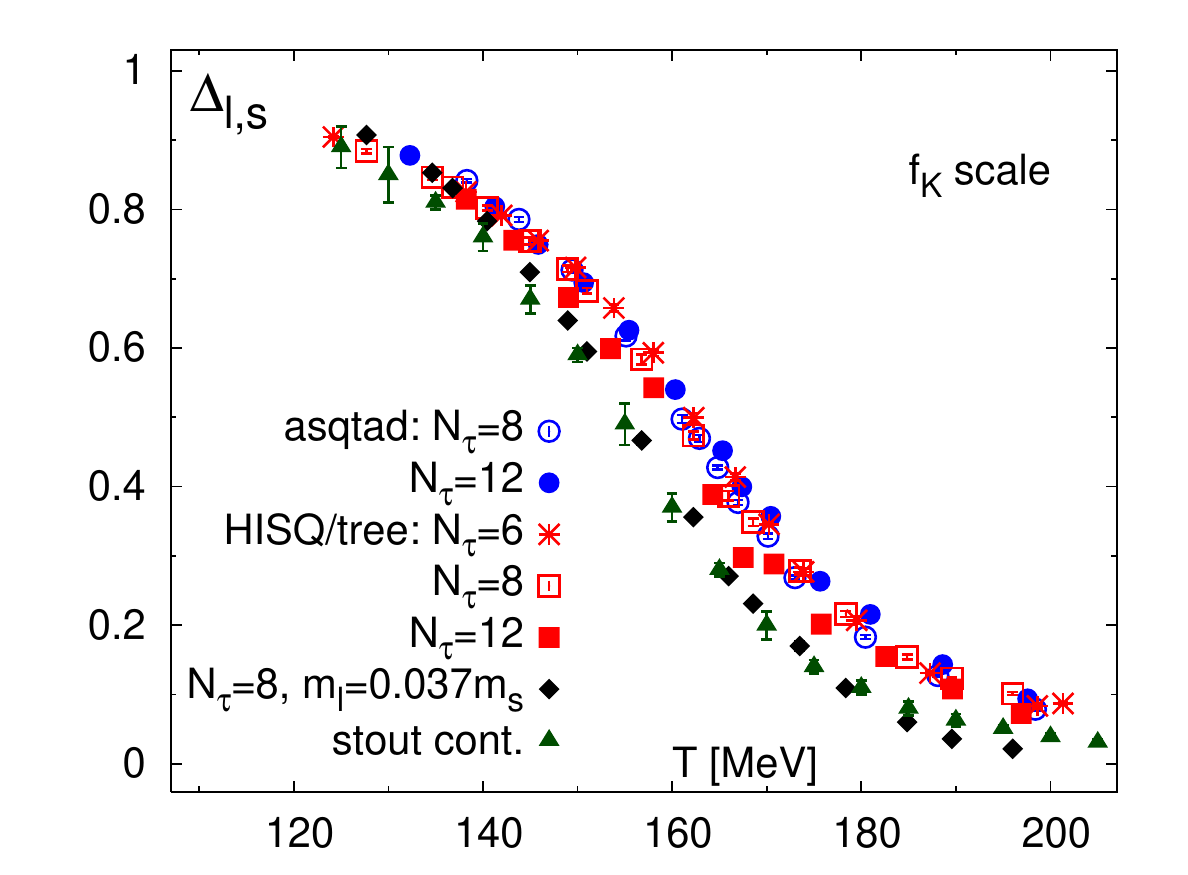}
\includegraphics[width=7.1 cm]{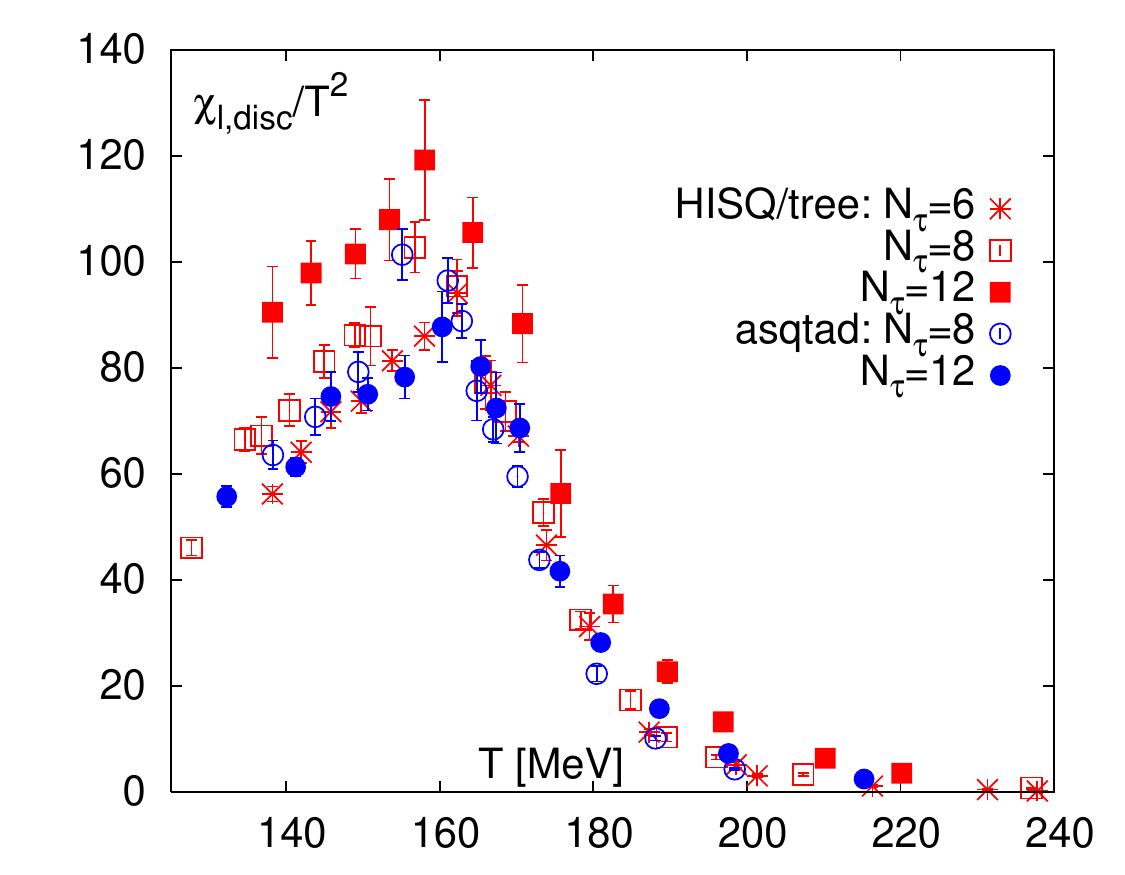}
\caption{Subtracted quark condensate and disconnected scalar susceptibility from the lattice work~\cite{Bazavov:2011nk} plotted for different actions and lattice resolutions.}
\label{fig:condandsuslatt} 
\end{figure}

In~Figure \ref{fig:condandsuslatt} we show the results for the lattice simulations in  \cite{Bazavov:2011nk} regarding the subtracted quark condensate and  the disconnected part of the scalar susceptibility (which carries the critical behaviour) for different choices of the lattice action and the number $N_\tau$ of points in the euclidean time direction. We~refer to that work for the definitions of those observables and their normalizations used. In~particular, $\Delta_{l,s}$ is obtained by subtracting  $\condl-(2\hat m/m_s)\conds$ with $\conds$ the strange quark condensate, in order to avoid finite-size lattice divergences. Such subtracted condensates  carry out the expected transition behaviour and will play an important role for our discussion on chiral partners and screening masses in Section \ref{sec:patt}. 

\vspace{-6pt}

Simulations at $\mu_B\neq 0$ present a  strong technical limitation known as the sign problem, related to the imaginary part added to the euclidean action which renders statistical weights complex. Approaches to circumvent this problem include analytical continuation from imaginary $\mu_B$ \cite{DElia:2002tig,deForcrand:2006pv}, reweighting methods \cite{Fodor:2004nz}, stochastic quantization \cite{Aarts:2008wh} and Taylor expansions around $\mu_B=0$ \cite{Bazavov:2018mes}. The~possible location and nature of the critical point is still under debate and is definitely one of the major open problems regarding the QCD transition \cite{Bazavov:2019lgz}.

On the other hand, very useful information can be extracted from the experimental analysis of the evolution of particle yields and their ratios as the energy of the collision is varied, the~so called Beam Energy Scan (BES) program \cite{Adamczyk:2017iwn}. A crucial step in this direction has been the observation that the physical conditions for the chemical freeze-out of net baryon number $B$, electric charge $Q$ and strangeness $S$ conservation  (i.e., when the system is dilute enough so that interactions conserve those quantities and the corresponding chemical potentials are built out) overlap significatively with the boundary of the phase diagram \cite{Bazavov:2018mes}.   This is schematically depicted in the ``BES'' region in \mbox{Figure \ref{fig:phasediag}}.   In~this context, hadron statistical models fit hadron yields from the STAR (at RHIC) and ALICE (at LHC) 
 experiments very well \cite{Andronic:2017pug}.  The~result of such fits is consistent with the critical line obtained in the recent lattice analysis within the Taylor expansion method \cite{Bazavov:2018mes}. Therefore, we are currently exploring the phase diagram directly from experimental information, which represents a major advance in this~field.

In~addition, the~study of fluctuations of conserved charges has opened up interesting possibilities~\cite{Ratti:2018ksb}.  In~particular, certain combinations of crossed  $BS$ and $QS$ susceptibilities provide a relation between chemical potentials $\mu_{B,S,Q}$ which can be directly tested from experimental hadron yields at chemical freeze-out   \cite{Bazavov:2014xya}. Experimental data on fluctuations also offer an additional way to explore the critical point \cite{Luo:2017faz}.

With the above motivation in mind, the~purpose of the present work is to review recent theoretical approaches based on effective theories to describe the physics below the transition. We~will be particularly interested in observables related to chiral symmetry restoration for which  the lightest degrees of freedom provide the dominant effect. Thus, as we will show here, the~light meson sector offers already a very useful picture to understand crucial aspects of relevance in this context such as the interplay between chiral and $U_A(1)$ symmetries and its relation with the nature of the transition, chiral partners, screening masses, the~scalar susceptibility peak, closely related to the role of thermal resonances, and the topological susceptibility.  The~connection of these theoretical analyses and lattice results will be specifically emphasized throughout this work. 

The~paper is structured as follows: In~Section \ref{sec:effthe} we will review the main existing approaches to the description of the physics below the transition in terms of effective theories. The~rest of the paper deals with specific topics where recent advance has been achieved in the understanding of the phase diagram and their properties in this context. Thus,  one the main effects needed to explain several important physical results are the thermal bath interactions leading to corrections in the spectral properties of hadrons, specifically thermal resonances, which will be analyzed in Section \ref{sec:reso}. In~particular, the~role of the thermal $f_0(500)$ state for chiral symmetry restoration is analyzed in that section where thermal unitarity will be a key ingredient. Section \ref{sec:patt} is dedicated to another important topic in this context: the nature or pattern of the transition and its connection with the degeneration of different partners (meson states) under chiral and $U(1)_A$ transformations. The~use of Ward Identities will be crucial in that part of the discussion and the connection of  those results with the description of screening masses and the topological susceptibility will be analyzed in detail. Our main conclusions are presented \mbox{in Section \ref{sec:conc}}.

\section{Effective Theories below the Transition} 
\label{sec:effthe}

Before going through the details of  different approaches describing the physics below the QCD transition, it is important to clarify our meaning of effective theories in this work. By that general concept we will be referring to  approaches where only a certain number of hadronic degrees of freedom is considered under particular theoretical assumptions. A more restrictive case, included in the above denomination, is that of Effective Field Theories (EFT) whose main ingredients are, on the one hand, a well-defined power counting rendering the theory perturbative for energies below a given scale, and, on the other hand, an effective Lagrangian constructed order by order in the most general way compatible with the underlying symmetries, with  low-energy parameters which ensure renormalizability and can be fixed with experimental or phenomenological information. The~most prominent example of EFT is Chiral Perturbation Theory (ChPT) which will play an essential role within  the present review. EFT constructed in this way are model-independent albeit limited to the energy domain around which the perturbative scheme is well defined. On the contrary, although effective theories are generally based on EFT, the~ additional assumptions incorporated will introduce some model dependency. Examples of the latter will be also discussed throughout this work.

\begin{figure}[h]
\centering
\includegraphics[width=10 cm]{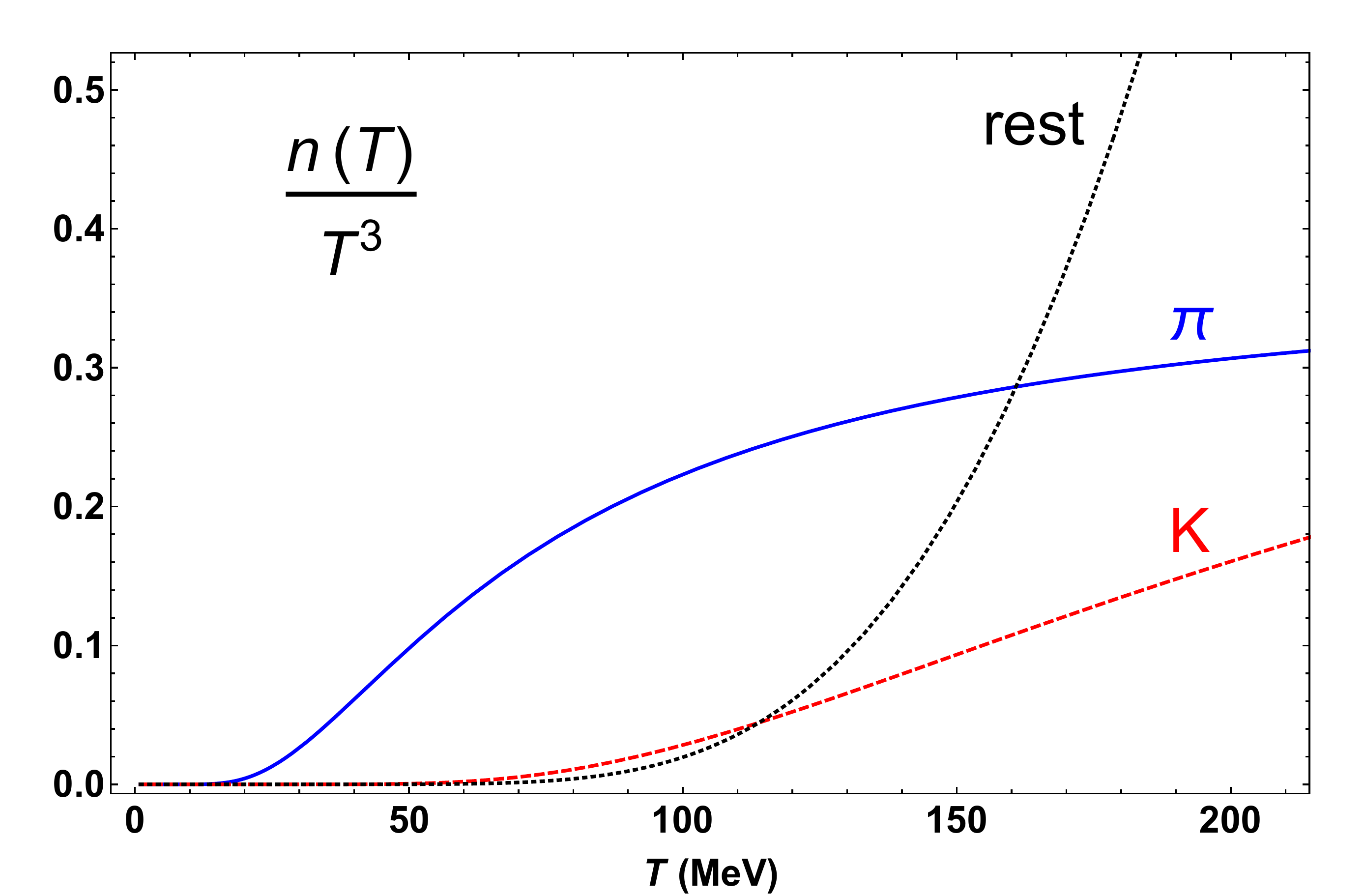}
\caption{Particle density for pions, kaons and the rest of states in the Particle Data Group (PDG) with masses below 2 GeV.}
\label{fig:number} 
\end{figure}

The~theoretical effort to describe the region of the phase diagram around $\mu_B=0$ has  been intense. In~that context, it is important to identify the relevant degrees of freedom for the description of particular observables.   In~principle, below the transition, hadron states of increasing masses and their interactions become more important as the temperature is increased, according to their characteristic Boltzmann weight. In~Figure \ref{fig:number} we plot the particle density of the pion and kaon component of the hadron gas, compared with the rest of hadron states quoted by the Particle Data Group (PDG) with masses below 2 GeV \cite{Tanabashi:2018oca} where we use for simplicity the free particle density with vanishing chemical~potentials. 

\vspace{-9pt}

It is clear that, generically, one should include as many hadrons as possible to describe correctly the thermodynamics below the transition, at least for thermodynamic quantities such as the pressure growing with the number of degrees of freedom. This is actually the spirit 
of the so called Hadron Resonance Gas (HRG) approximation  \cite{Hagedorn:1968zz,Karsch:2003zq,Karsch:2003vd,Tawfik:2005qh,Leupold:2006ih,Huovinen:2009yb,Megias:2012kb,Jankowski:2012ms}  where the free energy density and the pressure are described as a free gas of all the relevant states and resonances  up to a given mass, assuming that hadron interactions can be encoded through the corresponding resonant channels, i.e., 
\begin{eqnarray}
z^{HRG}(T)&=&z_M(T)+z_B(T),\nonumber\\
z_{M,B}(T)&=&\pm T \sum_{M,B} d_i \pint \log\left[1\mp e^{-\beta E_i(p)}\right],
\label{zHRG}
\end{eqnarray}
where $E_i=\sqrt{\vert\vec{k}\vert^2+M_i^2}$, $M,B$ stand for the meson and baryon contributions, the~upper sign is for mesons and the lower one for baryons. The~sum extends to  hadron states with degeneracy $d_i$ and mass $M_i$. The~HRG is  not a proper EFT, since it lacks a Lagrangian description and relies on assumptions like the dominance of resonant channels and the low-density regime  (see below).

This simple HRG description already accounts for the description of many thermodynamical observables below the transition, like the quark condensate,  with proper parametrizations of the dependence of hadron masses upon quark masses \cite{Tawfik:2005qh,Jankowski:2012ms} the pressure and trace anomaly, with HRG hadron masses  adapted suitably to the lattice ones to reduce discretization effects \cite{Huovinen:2009yb}, the~Polyakov loop \cite{Megias:2012kb}  and $\mu_B\neq 0$ Taylor expansion coefficients \cite{Karsch:2003zq,Ratti:2018ksb}.

There are, however, various examples of physically relevant issues in this context for which a  HRG-like description may not be enough or it is simply not applicable.  This is the case of some of the problems that we will consider here, such as chiral partners or thermal resonances. Actually, for certain observables dominated by the lightest states, such as the scalar and the topological susceptibilities, the~essential physical behaviour is captured from an appropriate thermal description of those states rather than by including higher mass ones. As a general rule, the~HRG approach gives rise to monotonically decreasing or increasing functions of temperature, not reproducing some of the characteristic transition features, such as the inflection point of the quark condensate or the peaks of the scalar susceptibility and the trace anomaly. This is intrinsic to the HRG approach, since it is expected to work only below $T_c$ by construction.

One particularly important aspect in that context is the effect of interactions, which may be crucial for certain observables. As mentioned, the~HRG only considers the  interactions between stable hadrons through their resonant channels, e.g., the~$\rho(770)$ $I=J=1$ channel in $\pi\pi$ scattering. However, this misses, on the one hand, non-resonant channels and, on the other hand, does not include the width of the resonances, which is not negligible in some significant cases such as the $f_0(500)$ (formerly known as $\sigma$) or the $I=1/2$, $J=0$ state $K^* (700)$ (or $\kappa$) showing up in $\pi K$ scattering \cite{Tanabashi:2018oca}. These~two states will actually play a significant role in chiral symmetry restoration, as we will discuss in \mbox{Sections  \ref{sec:reso} and \ref{sec:patt}}.   Actually, the~effect of the width of the resonances has proven to be important  within thermal statistical models, for which  the decay channels of those resonances feed the hadron yields at chemical freeze-out \cite{Andronic:2005yp}. Corrections to the HRG with repulsive interactions provided by excluded hadron volume corrections also allows us to achieve a better description of lattice data~\cite{Andronic:2012ut,Huovinen:2017ogf}.

A full consistent treatment of the thermodynamics of the hadron gas including  interactions would require the use of effective chiral Lagrangians \cite{Weinberg:1978kz} combined with the techniques of quantum field theory at finite temperature and density, often denoted  as thermal field theory \cite{galekapustabook}. Thus, chiral symmetry dictates in principle the dynamics of hadron states under the general formalism of ChPT, starting from the lightest octet of pseudoscalars ($\pi,K,\eta$)  which corresponds to the would-be Nambu-Goldstone bosons of the chiral symmetry $SU_L(N_f)\times SU_R(N_f)\rightarrow SU_{L+R} (N_f)$ where $N_f$ is the number of light flavors ($N_f=2,3$) and $SU_{L+R} (2)$ is the isospin symmetry \cite{Gasser:1983yg,Gasser:1984gg}, which can be suitably extended to higher mass states such as vector mesons or nucleons \cite{Ecker:1988te,Ecker:1989yg,Meissner:1993ah,Bernard:1995dp,Ecker:1994gg,Pich:1995bw}.   A very useful extension for our present discussion is the $U(3)$ ChPT framework, where the  $\eta_0$ (singlet) state  is included formally as ninth Goldstone Boson, combining  the chiral low-energy expansion with the large-$N_c$ framework  \cite{Witten:1979vv,DiVecchia:1980ve,HerreraSiklody:1996pm,Kaiser:2000gs}.

Thus, ChPT is a genuine EFT in the sense explained at the beginning of this section. The~ChPT chiral expansion consists generically in a combined expansion of, on the one hand,  effective Lagrangians up to a given order in masses and derivatives, and on the other hand higher order loop contributions, which provide a systematic and model-independent framework where the loop divergences can be consistently absorbed in the Low Energy Constants (LECs) multiplying the different allowed operators in the chiral Lagrangian at a given order \cite{Weinberg:1978kz,Gasser:1983yg,Gasser:1984gg}.   An updated review on the determination and numerical values of those LECs, as well as many other low-energy parameters involved in this framework, can be found in \cite{Aoki:2019cca}. 

The~chiral expansion is usually parametrized by powers $p^n$ where $p$ is a generic energy scale denoting particle masses, momenta or temperature. The~lowest order Lagrangian is $\Od(p^2)$ and is given by the familiar Non-linear Sigma Model, which for $SU(3)$ in the absence of external sources~reads
\begin{equation}
{\cal L}_2=\frac{F^2}{4}\tr \left[ \partial_\mu U^\dagger \partial^\mu U + 2B_0 {\cal M}(U+U^\dagger)  \right]
\label{l2}
\end{equation}
where $U=\exp(i \lambda^a \phi_a/F)$ with $\phi_a$ the Nambu-Goldstone Boson (NGB) fields, ${\cal M}=\mbox{diag} (m_u,m_d,m_s)$ is the quark mass matrix,  $F$ is the pion decay constant in the chiral limit, i.e., to leading order (LO) in the chiral expansion, $B_0=M_{0\pi}^2/(2\hat m)=-\condl^0/(2F^2)$ where $M_{0\pi}$ is the LO pion mass and $\condl^0$ is the LO light quark condensate. $F,B_0$ are the only low-energy parameters at this order, while at the next order ${\cal L}_4$ contains different LECs named customarily as $l_i$ for $SU(2)$, where the NGB are just the pion fields,  and $L_i$ for $SU(3)$.

Thus, the~main advantage of  ChPT is that it provides a model-independent and systematic framework to calculate hadron observables. Its main limitation is of course that it is built as a low-energy scheme, its range of validity being then restricted to energies not far from the corresponding particle thresholds or temperatures below the excitation of new states. Considering higher orders in the ChPT chiral expansion compensates partially for this limitation, although in principle at the expense of introducing more new LECs and hence losing predictability.

\begin{figure}[h]
\centering
\includegraphics[width=7 cm]{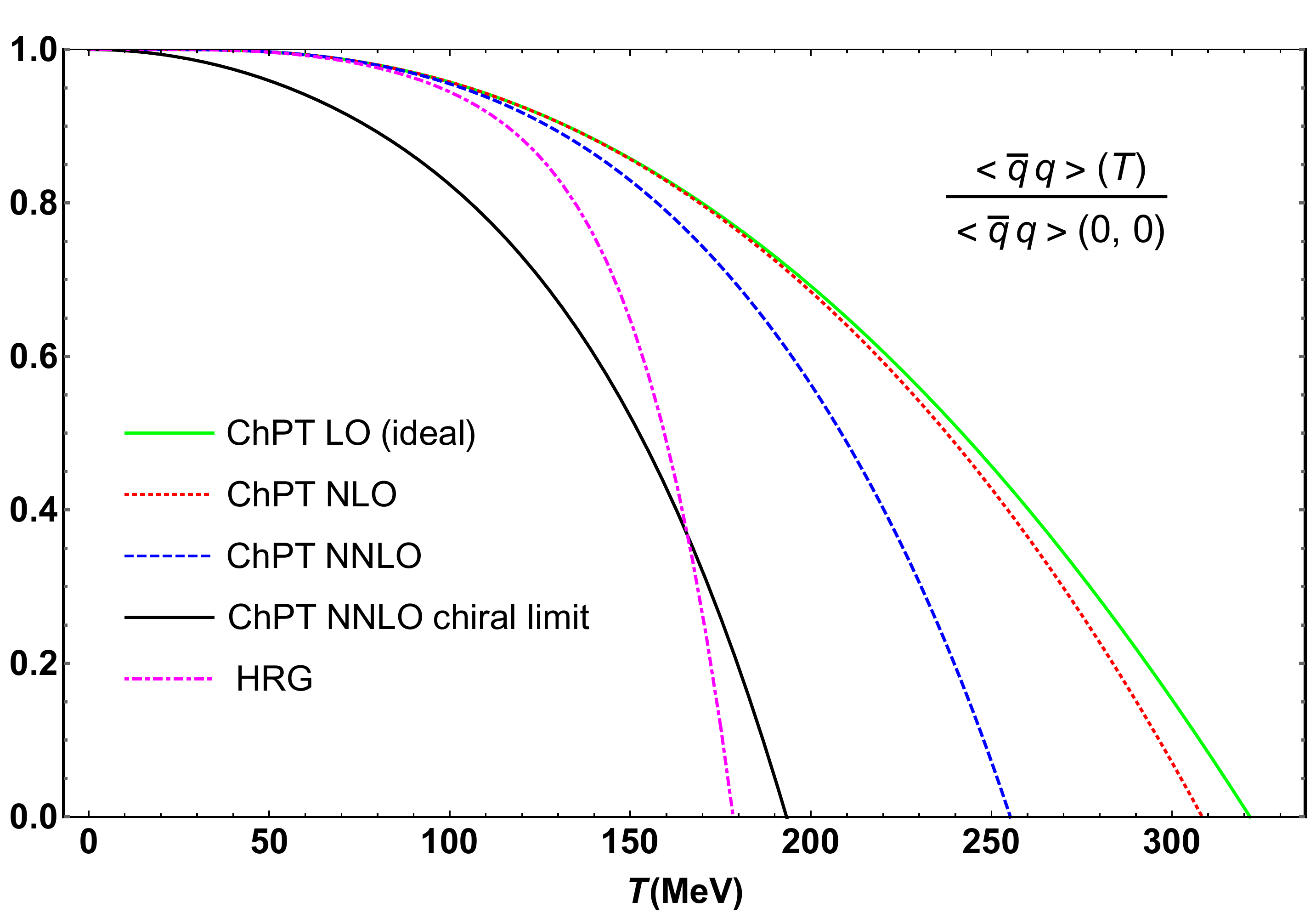}
\includegraphics[width=7.5 cm]{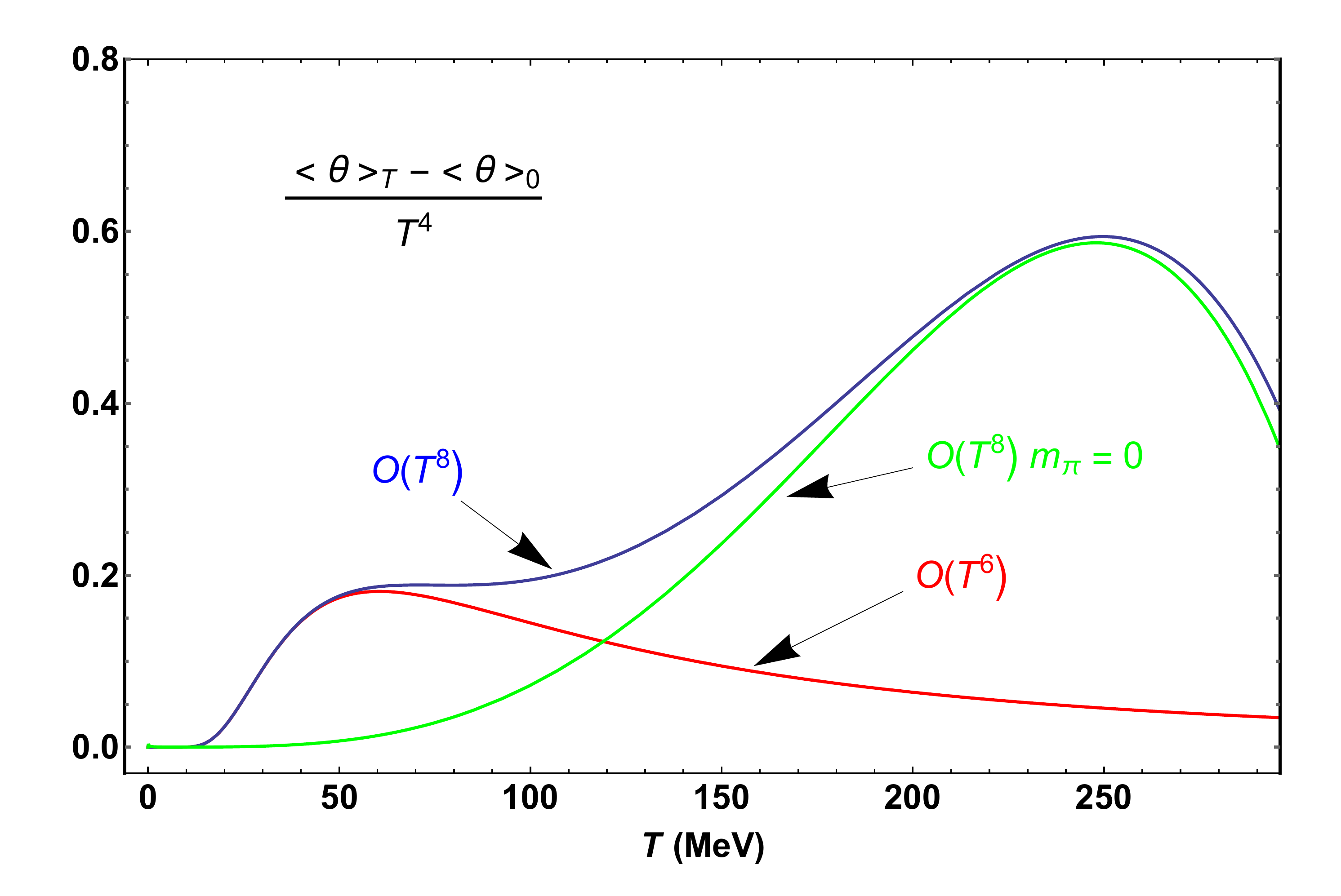}
\caption{(\textbf{Left}): quark condensate in SU(2) Chiral Perturbation Theory (ChPT) (pion gas) at different orders from \cite{Gerber:1988tt} compared to the Hadron Resonance Gas approach in  \cite{Jankowski:2012ms}. (\textbf{Right}): trace anomaly for the pion gas at different orders in ChPT from the results in \cite{Gerber:1988tt}.}
\label{fig:condensateanomaly} 
\end{figure}

Within that framework, it is worth mentioning that the thermodynamics of the pion gas has been studied up to order $\Od(T^8)$, which implies a three-loop calculation for the free energy density~\cite{Gerber:1988tt}.  The~pion gas reproduces,  at least qualitatively, the~main physical properties that would be expected for the more general hadron gas. Thus, as the temperature approaches the transition region, the~thermodynamic pressure  increases  with respect to the free gas and the quark condensate decreases also sharply.  For nonzero quark masses,  ChPT does not capture the inflection point of the condensate, which  tends to vanish at a given temperature.  In~the light chiral limit those effects are enhanced, as expected, and the transition temperature, defined where the condensate vanishes, decreases with respect to the massive case. In~Figure \ref{fig:condensateanomaly} we show the different orders from the analysis in \cite{Gerber:1988tt}. The~LO (ideal gas) and the next-to-leading order (NLO) one curves are close because the NLO 
 interactions can  be absorbed in the renormalization of the physical parameters in the  free gas expressions \cite{Gerber:1988tt}. At~the next-to-next-to-leading order (NNLO), additional interactions set in, implying stronger restoration effects, as seen clearly in the figure. We~also compare the the ChPT results with a HRG analysis where the quark mass dependence of hadrons is taken from \cite{Leupold:2006ih,Jankowski:2012ms}, where the dependence of pseudo Nambu-Goldstone Bosons, i.e., pion, kaon and eta masses, is extracted directly from the one-loop ChPT calculation \cite{Gasser:1984gg} while the masses of the rest of hadrons are taken to scale within a constituent quark picture within a Nambu-Jona-Lasinio (NJL) approach. We~also follow \cite{Jankowski:2012ms} for the assignments of the hadron strangeness content for open and hidden strange mesons, as well as for singlet and octet members.  It is clear that including higher mass states is crucial in the case of the quark condensate, their effect being to decrease the transition temperature towards values closer to the lattice results. Actually, although one may first think that introducing massive states should increase the condensate due to explicit chiral symmetry breaking, that sort of ferromagnetic-like effect is known to be overshadowed by the paramagnetic-like disorder introduced by adding degrees of freedom to the system. 
 
 Another interesting result of the ChPT analysis is that to $\Od(T^8)$ the trace anomaly does actually generate a peak around the transition region, which is neatly seen in the chiral limit \cite{Leutwyler:1987th}. That peak is not seen at the $\Od(T^6)$ order, which reveals again, as commented above, the~importance of considering the relevant order of interactions to describe some of the transition features. These results are also showed in Figure \ref{fig:condensateanomaly}. Interestingly, the~curve of the trace anomaly presents an additional, lower temperature peak in the massive case at $\Od(T^8)$ which disappears in the massless case and turns out to be directly related to the quark contribution of the conformal anomaly, that two-peak structure being also directly correlated to the behaviour of the bulk viscosity at finite temperature in the pion gas \cite{FernandezFraile:2008vu,FernandezFraile:2009mi}. The~numerical values of the LECs used in Figure  \ref{fig:condensateanomaly} are the same as in \cite{FernandezFraile:2008vu,FernandezFraile:2009mi}. In~turn, we remark that the analysis of transport coefficients of the hadron gas through different formalisms within effective theories is another important field of research or which an accurate description of the interactions is crucial  to obtain results consistent with  heavy-ion phenomenology \cite{FernandezFraile:2008vu,FernandezFraile:2009mi,FernandezFraile:2005ka,Dobado:2008vt,Dobado:2011qu,Dobado:2011wc,Das:2011vba,Tolos:2016slr,Abhishek:2017pkp}. 
  
 \vspace{-6pt}
 
 An interesting alternative to the effective Lagrangian framework has been the use of the virial expansion, since the hadronic gas meets the conditions for the dilute gas regime to be applicable. Thus, an expansion in fugacities $\xi=\exp\left[\beta(\mu_i-M_i)\right]$ with $\mu_i$ and $M_i$ generic  chemical potentials and masses, allows us to write the thermodynamic pressure to second order in the virial expansion,  \mbox{as \cite{Dashen:1969ep,Gerber:1988tt,Venugopalan:1992hy,Dobado:1998tv,Pelaez:2002xf,GarciaMartin:2006jj}}
  \begin{equation}
   \beta P=\sum_{i}{\parent{B_i^{(1)}\xi_i+B_{i}^{(2)}\xi_i^2+\sum_{j\geq i}{B_{ij}^{\mathrm{int}}\xi_i\xi_j}}} + \Od(\xi^3),
\label{virialpressure}
\end{equation}
 where the virial coefficients above are given, on the one hand,  by (for simplicity we take $\mu_i=0$ but the corresponding extension is straightforward) 
  \begin{equation} 
B_i^{(n)}=\frac{d_i}{2\pi^2 n}\int_0^\infty{dp\,p^2 e^{-n\beta(\sqrt{p^2+M_i^2}-M_i)}},
\label{virialfreecoeff} 
\end{equation}
which are nothing but the coefficients of the expansion in fugacities of the free pressure given by the HRG expression \eqref{zHRG}.

On the other hand,  the coefficients $B_{ij}$ account for the binary interactions occurring in the gas. The~interesting advantage of this method is that those interaction coefficients can be cast into the $T=0$ scattering phase shifts of the particles considered, which could even be extracted from experiment avoiding model dependency, without having to go through the thermal field theory calculation. For meson interactions, 
\begin{equation}
B^{\mathrm{int}}_{ij}=\frac{\xi_i^{-1}\xi_j^{-1}}{2\pi^3}\int_{M_i+M_j}^\infty{dE\,E^2K_1\left(E/T\right)\Delta_{ij}(E)}
\label{virialintcoeff}
\end{equation}
where $K_1$ is the first modified Bessel function and
 \begin{equation}
\Delta_{ij}(E)=\sum_{I,J}(2I+1)(2J+1)\,\delta_{IJ}^{ij}(E),
\label{Deltaphases}
\end{equation}
where the $\delta^{ij}_{IJ}$ are the $ij\rightarrow ij$ elastic scattering phase shifts (chosen so that $\delta=0$ at threshold $E_{th}=M_i+M_j$) of a state $ij$  with quantum number $I,J,$ (isospin, and angular momentum). The~extension to other hadron species such as baryons  follows the same guidelines. 

The~virial expansion allows us then to obtain a fairly decent description of the thermodynamics of the hadron gas including interactions and then correcting the HRG description. In~fact, it was already shown in 
\cite{Gerber:1988tt} that the ChPT calculation up to $\Od(T^8)$ is equivalent to considering the virial expansion with the $T=0$ scattering amplitudes calculated up to $\Od(p^4)$ (i.e., up to one loop) for the dominant channels in ChPT, namely $IJ=00,20,11$.  One interesting feature of the pion gas is that there is an almost exact cancellation in the sum in \eqref{Deltaphases} between the scalar $00$ and $02$ contributions (attractive and repulsive respectively) which prevails even for unitarized interactions (see below) in the relevant temperature range and holds both for the quark condensate and for the scalar \mbox{susceptibility \cite{Gerber:1988tt,Venugopalan:1992hy,GomezNicola:2012uc,Broniowski:2015oha}}. A similar cancellation holds between the scalar attractive and repulsive scalar channels in $\pi K$ scattering, i.e., the~channels $I=1/2, J=0$ and $I=3/2, J=0$ \cite{Venugopalan:1992hy,Broniowski:2015oha}. The~$00$ and $1/2,0$ channels correspond to the resonances $f_0(500)$ ($\sigma$) and $K^*(700)$ ($\kappa$) respectively and therefore, this cancellation arguments supports that those resonances are not included in the usual HRG description. In~fact, due to this argument, the~pion gas component within the virial framework turns out to be almost completely dominated by the $\rho(770)$. However, as we will see in detail below, the~inclusion of thermal effects in the scattering amplitudes and then in the resonance parameters, changes this picture close to the transition, at least for some particular observables like the scalar susceptibility, dominated by the $00$ channel. Actually, neither the HRG nor the ChPT or virial approaches reproduce a peak for the scalar susceptibility, all those approaches giving a monotonically increasing function for that quantity \cite{GomezNicola:2012uc,Ferreres-Sole:2018djq}.

As it has become clear from our previous discussion, a proper treatment of interactions  is crucial for a correct description of the hadron gas thermodynamics. Based on  chiral symmetry, resonances can be consistently included at the Lagrangian level, as well as their interactions with other hadrons.  This is for instance the spirit of the Vector Meson Dominance (VMD) and resonance saturation \mbox{approaches \cite{Ecker:1988te,Ecker:1989yg}} which provide an accurate description mostly for vector mesons. In~that framework one can estimate for instance the contribution to the LECs coming from integrating out those heavy degrees of freedom, which allows us for useful determinations of those constants in terms of the masses and widths of resonances. However, that picture is  not adequate for scalar mesons. A clear example is  the lowest lying scalar resonance, the~$f_0(500)$ (known simply as $\sigma$ for many years) which over recent years has been a subject of many analysis and certain dose of controversy about its nature and main properties \cite{Albaladejo:2012te,Pelaez:2015qba}.

Attempts of modeling this state go back to the Linear Sigma Model (LSM) or vector $O(4)$ model in terms of $(\sigma,\pi^a$) states \cite{GellMann:1960np} which indeed exhibits chiral restoration features, since the $\sigma$ state tends to degenerate with the pions in the limit of exact chiral restoration \cite{Hatsuda:1985eb,Bernard:1987im} and $\mean{\sigma}$ scales like the quark condensate \cite{Bochkarev:1995gi,Ayala:2000px}. The~scalar susceptibility calculated within the LSM shows also chiral restoration features, since it increases with temperature as $T$ approaches the transition \cite{Ferreres-Sole:2018djq}. Although the peak of the transition is not reproduced either by the LSM, this model can be used as a suitable testbed, as we will discuss in detail in Section \ref{sec:reso}. 
  
The~main drawback of the Lagrangian-based analysis of broad resonances such as the $f_0(500)$ or the $K^*(700)$ is that their strongly coupled nature reflects somehow in the parameters and couplings of the theory, preventing from a consistent perturbative quantum field theory analysis. In~fact, the~$f_0(500)$ is nowadays understood as a broad resonance showing up in the $\pi\pi$ scattering amplitude second Riemann sheet (2RS), the~main determinations of its position in the complex plane being quoted by the PDG \cite{Tanabashi:2018oca}. Thus, although within the LSM one can actually generate a pole  in the complex plane from the $\sigma$ self-energy, it is not possible to accommodate the coupling constant $\lambda$ of the theory to get properly the real and imaginary parts of that state pole quoted in the PDG, the~closest to those values being obtained for $\lambda\sim$ 10--20 which as we have just commented, reflects the strong-interacting regime~\mbox{\cite{Masjuan:2008cp,Pelaez:2015qba,Ferreres-Sole:2018djq}}.

A modern and phenomenologically reliable way to generate dynamically resonances and at the same time enlarging the applicability range of ChPT is the so called Unitarized Chiral Perturbation Theory (UChPT) framework 
  \cite{Truong:1988zp,Dobado:1989qm,Dobado:1996ps,Oller:1997ti,Oller:1998hw,GomezNicola:2001as,Pelaez:2003xd,GomezNicola:2007qj,Albaladejo:2010tj}.  This scheme relies on ChPT as the low-energy EFT, demanding exact unitarity as an additional requirement, which introduces some model dependency, as we are about to see.  Thus, the~general principle within this framework is to construct scattering amplitudes which satisfy unitarity exactly while keeping its main analytical properties and being consistent at low energies with the ChPT expansion. Such requirements allow us in most cases to generate resonances in the 2RS of the unitarized amplitudes, whose spectral properties can be rendered in agreement with the PDG values with suitable fits of the LECs involved. Since the resonant contributions are well captured by this method, it naturally allows us to enlarge considerably the range of energies to which experimental data on phase shifts and inelasticities can be phenomenologically fitted. 

\begin{figure}[h]
\centering
\includegraphics[width=9cm]{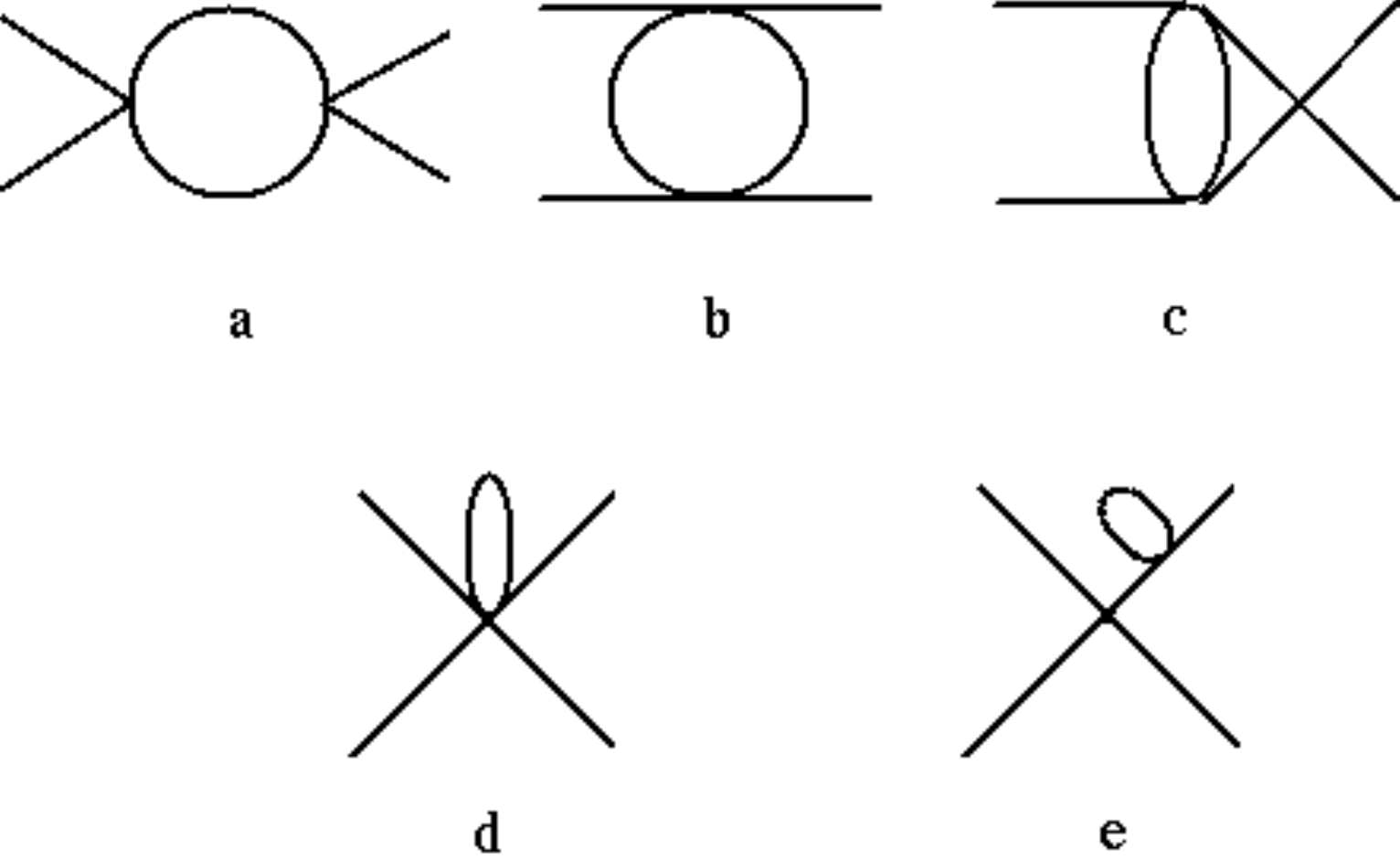}
\caption{One-loop Feynman diagrams contributing to two-particle scattering to order $\Od(p^4)$ in Chiral Perturbation Theory. The different diagrams correspond to (\textbf{a}): $s$-channel, (\textbf{b}): $t$-channel, (\textbf{c}): $u$-channel, (\textbf{d}): tadpoles from six-point vertices, (\textbf{e}): tadpoles from external legs renormalization. All vertices come from the ${\cal L}_2$ lagrangian.}
\label{fig:diag} 
\end{figure}

A remarkable example, which we will be using extensively here, is the Inverse Amplitude Method (IAM). Consider the simple case of elastic scattering of identical particles, say $\pi\pi$ pion scattering. The~condition of unitarity for the $S$ matrix, i.e., $S^\dagger S= {\bf 1}$, translates into the following condition for partial waves $t^{IJ}$ of well-defined isospin $I$ and angular momentum $J$:
\begin{equation}
\mbox{Im} t^{IJ}(s+i0^+)=\sigma_{\pi} (s) \vert  t^{IJ}(s) \vert^2 \quad (s\geq 4M_\pi^2)
\label{unitpw} 
\end{equation}
where $\sigma_{i}(s)=\sqrt{1-4M_i^2/s}$ is the phase space for two identical particles of mass $M_i$. Note that the ChPT  expansion $t=t_2+t_4+\dots$ only satisfies the unitarity condition perturbatively, i.e., 
\begin{equation}
\im t_2(s)=0, \quad \im t_4(s+i0^+)=\sigma_{\pi}(s) \left[t_2(s)\right]^2 \quad (s\geq 4M_\pi^2)
\label{unitpertzero}
\end{equation}
and so on, where the diagrams contributing to $t_4=\Od(p^4)$ to one loop are those depicted in~Figure~\ref{fig:diag}. The~imaginary part contribution in \eqref{unitpertzero} corresponds to the unitarity cut $s\geq 4M_\pi^2$ arising from the $s$-channel diagram (a) in Figure \ref{fig:diag}, whereas $t$ and $u$ channel diagrams (b) and (c) give rise to a left \mbox{cut $s\leq 0$}.

The~unitarity condition \eqref{unitpw} can be written more conveniently in terms of the inverse amplitude as $\mbox{Im} t^{-1}=-\sigma_{\pi}$ (we drop the $IJ$ indices for simplicity). Therefore, by construction, any amplitude of the form $t_U=\left[\mbox{Re} \left(t_U^{-1}\right)-i\sigma_{\pi}\right]^{-1}$ is automatically unitary. The~different unitarization methods amount essentially to different approximations for $\mbox{Re} \left(t^{-1}\right)$, which introduces certain model dependency. When performing those approximations it is important to ensure that the full unitarized amplitude remains analytic so that, in particular, it is meaningful to consider the amplitude in the  different Riemann sheets in terms of the unitarity cuts at different thresholds, in a general multichannel case. The~latter is particularly important for the determination of resonances and their parameters. For~instance, the~2RS amplitude $t^{II}(s)$ for the single channel case is constructed as customary so that it connects continuously with the first sheet $t(s)$ across the unitarity cut, i.e., $\im t^{II}(s-i\epsilon)=\im t(s+i\epsilon)$ for $s\geq 4M_\pi^2$.  The~IAM is constructed by requiring that the unitarized amplitude matches the ChPT expansion up to a given order. The~most common choice is that it does up to fourth order,  resulting in the following unitarized~amplitude:
\begin{equation}
t_U^{IAM}(s)=\frac{t_2(s)^2}{t_2(s)-t_4(s)}
\label{IAMT0}
\end{equation}
which incidentally corresponds to the $[1,1]$ Pad\'e approximant. The~IAM amplitude can also be justified from the use of dispersion relations with proper subtractions, incorporating additional analytic features of the amplitude such as Adler zeros \cite{GomezNicola:2007qj}. This method can also be extended to multiple scattering channels  including the corresponding inelasticities, by considering a matrix formulation of \mbox{Equations \eqref{unitpw} and \eqref{IAMT0}} in the internal space of scattering states \cite{GomezNicola:2001as}.  A complementary approach in this context is the  large NGB number expansion, which provides a resummation for NGB scattering giving rise to a exact unitary amplitude in the $IJ=00$ channel in the chiral limit, consistent with the phase shift and $f_0(500)$ pole experimental data \cite{Dobado:1992jg,Dobado:1994fd,Cortes:2015emo}. This approach has also be used recently to analyze the pion gas thermodynamics \cite{Cortes:2016ecy}.

The~key ingredients described in this section, namely, effective theories for the hadron gas and  unitarization, will be exploited in the rest of the paper in order to gain more insight into specific problems pertaining chiral symmetry restoration such as the importance of thermal properties of resonances and the interplay between chiral and $U_A(1)$ restoration.

\section{The~Role of Thermal Resonances}
\label{sec:reso}
\vspace{-9pt}
\subsection{Spectral Properties of Hadrons in the Thermal Bath} 

One important consequence of considering the interactions among the thermal bath components is that the spectral properties of both stable states  and resonances are modified. Generally speaking, self-energies receive thermal field theory corrections through loop contributions \cite{galekapustabook}, which  give rise to important observable consequences, as we will discuss here.

Thus, within the pion gas, modifications to the pion dispersion relation can be calculated in the ChPT framework. To one loop, the~only modification is a tadpole-like thermal shift in the pion mass, which gives a slowly growing function with $T$ \cite{Gasser:1986vb}.  A two-loop analysis reveals a more complicated dispersion relation \cite{Schenk:1993ru} which includes an absorptive imaginary part defining a thermal collision rate and hence a mean free path for pions in the thermal bath \cite{Goity:1989gs}. Actually, relating  the pion pole parameters and the forward pion scattering amplitude  \cite{Schenk:1993ru}  allows us to include unitarized amplitudes in the analysis, which introduce significant corrections near the transition, mostly for the damping rate. Those spectral modifications have indeed physical consequences. For instance, the~mean free path can be used to estimate thermal freeze-out conditions and is crucial to describe  transport coefficients within the diagrammatic or Kubo formalism \cite{FernandezFraile:2009mi}. In~addition, the~absorptive part is actually related to the imaginary part of the pion decay constant $F_\pi$ defined in the thermal environment, where Lorentz covariance is lost so that the space and time components of $F_\pi$ generally differ  \cite{Pisarski:1996mt}. 
 
A very important physical example where thermal corrections to self-energies are crucial comes from  the description of the dilepton and photon spectra in heavy-ion collisions.  Resonances decaying within the thermal environment in the hadron phase undergo a substantial modification of their spectral properties, which contribute to explain in particular the low-energy excess  observed in the dilepton spectrum mostly  around  the  $\rho(700)$ resonance peak, as seen in recent data from various experimental collaborations at different collision energies \cite{Adamczyk:2015lme,Adare:2015ila,Acharya:2018nxm,Adamczewski-Musch:2019byl}. In~fact, both the dilepton and photon yields are entirely encoded in the vector electromagnetic spectral function, whose thermal modifications during the hadron phase give rise to observable consequences. Therefore, the~calculation of in-medium spectral functions is crucial for a realistic analysis of thermal dilepton rates at finite temperature and density \cite{Rapp:1999ej,Jung:2016yxl}. A significant in-medium broadening of the $\rho$ meson is achieved from all the theoretical analysis, which is a key ingredient to describe correctly the data on the dilepton spectrum, which  are also strongly correlated to the thermalization and time-evolution of the expanding plasma  \cite{Rapp:2014hha}.

  The~description of thermal photons, i.e., the~part of the direct photon spectrum coming from hadron decays inside the expanding plasmas, can also be rather accurately described through the use of effective chiral theories for the determination of the relevant hadronic in-medium decays, combined with suitable models for the hydrodynamic evolution  \cite{Turbide:2003si,Paquet:2015lta}. The~agreement with experimental data is quite good  both at RHIC and LHC energies.

   One interesting consequence of  the analysis of the spectral properties of vector and axial-vector states involved in dilepton and photon spectra is the connection with chiral symmetry restoration,  directly related to our present discussion. In~particular, the~thermal spectral functions of the $\rho (770)$ and the $a_1(1260)$ mesons shows that those states become degenerate at the chiral transition \cite{Rapp:1999ej,Jung:2016yxl} which is expected since the bilinear quark operators with their same quantum numbers are chiral partners (see our discussion in Section \ref{sec:patt}).

  The~above discussion about thermal modifications confirms that for vector and axial-vector  resonances, it is fine to consider an effective Lagrangian treatment where those states are explicitly included and their widths can be treated perturbatively within a Breit-Wigner (BW) approach, being narrow resonances. However, as we have explained in Section \ref{sec:effthe}, for the case of light scalar resonances, such treatment is  definitely not adequate. That is one of the main reasons motivating the recent development of a program to analyze those thermal resonances and their properties as dynamically generated from UChPT scattering at finite temperature and density \cite{GomezNicola:2002tn,Dobado:2002xf,FernandezFraile:2007fv,Cabrera:2008tja,Nicola:2013vma,Cortes:2015emo,Ferreres-Sole:2018djq}.  

Let us explain the main ideas and results of that approach. At finite temperature, one can consider the same one-loop diagrams as for  $T=0$ scattering and include the temperature corrections carrying out the standard Matsubara sums replacing the time-momentum integrals in the imaginary-time formalism of thermal field theory \cite{galekapustabook}. The~result is an analytic function which defines the thermal amplitude by analytic continuation in the external line energies and application of the standard LSZ reduction formula  for the external asymptotic states \cite{GomezNicola:2002tn}.   Such thermal amplitude can be projected as usual into partial waves $t^{IJ}(s;T)$ in the reference frame $\vec{p_1}=-\vec{p_2}$ where the incoming particles 1,2 are at rest with the thermal bath. The~corresponding ChPT series $t^{IJ}_2(s)+t^{IJ}_4(s;T) + \dots$ ($t^{IJ}_2$ is just a tree-level amplitude and hence it gets no loop $T$-dependent corrections) satisfies a perturbative unitarity relation, which in the case of one-channel elastic scattering  is similar to \eqref{unitpertzero}  extended to the finite-$T$ case as
\begin{equation}
\im t_2(s)=0, \quad \im t_4(s+i0^+)=\sigma_{\pi}(s;T) \left[t_2(s)\right]^2 \quad (s\geq 4M_\pi^2)
\label{unitperttemp}
\end{equation}
where
\begin{equation}
\sigma_{\pi}(s;T)=\sigma_{\pi}(s)\left[1+2n_B(\sqrt{s}/2;T)\right] 
\label{therphasespace}
\end{equation}
is the so-called thermal phase space and $n_B(x;T)=\left[\exp(x/T)-1\right]^{-1}$ is the Bose-Einstein distribution function. The~above contribution stems from loop integrals of the form
\begin{equation}
J\left(M_{i};k_0,\vec{k},T\right)=T\sum_{n=-\infty}^{\infty}\int\dfrac{d^3\vec{p}}{\left(2\pi\right)^3}\dfrac{1}{p^2-M_i^2}\dfrac{1}{\left(p-k\right)^2-M_{i}^2},
\label{Jter}
\end{equation}
coming from the loop integrals (a,b,c) in Figure \ref{fig:diag}. It satisfies, once analytically continued  to continuous $k_0$,  $\im J(M_i,k_0,\vec{k}=\vec{0},T)=\mbox{sgn}(k_0)\theta(k_0-2M_i) \sigma_i(k_0^2;T)/(16\pi)$ and so on for similar relations involving loop integrals with different four-momenta combinations in the numerator of the integrand \cite{GomezNicola:2002tn}. The~equivalent loop integrals for different masses of the incoming particles receive additional imaginary parts which are purely thermal, i.e., they vanish at $T=0$, coming from the so called Landau \mbox{cuts \cite{Weldon:1983jn,Ghosh:2009bt}}. The~effect of those cuts has to be taken into account in addition to the standard unitarity cuts, like that in \eqref{unitperttemp}, and the left cut $s\leq 0$.

The~thermal enhancement of the phase space in \eqref{therphasespace} can be interpreted as  $(1+n_B)^2-n_B^2$ which corresponds to the difference between  processes of `` emission" (enhanced probability of the final two-pion state by collisions of  thermal pions) and ``absorption" (of the initial two-pion state through scattering with thermal pions)  inside the thermal bath, pretty much in the same way as a particle decaying in-medium \cite{Weldon:1983jn}. 

The~requirement of exact thermal unitarity, i.e., 
\begin{equation}
\mbox{Im} t^{IJ}(s+i0^+;T)=\sigma_{\pi} (s;T) \vert  t^{IJ}(s;T) \vert^2 \quad (s\geq 4M_\pi^2)
\label{unitpwther} 
\end{equation}
plus the matching condition with the ChPT series yields then the IAM unitarized amplitude at finite temperature, extending the $T=0$ result in \eqref{IAMT0} as
\begin{equation}
t_U^{IAM}(s;T)=\frac{t_2(s)^2}{t_2(s)-t_4(s;T)}
\label{IAMT}
\end{equation}

\begin{figure}[h]
\centering
\includegraphics[width=7.5 cm]{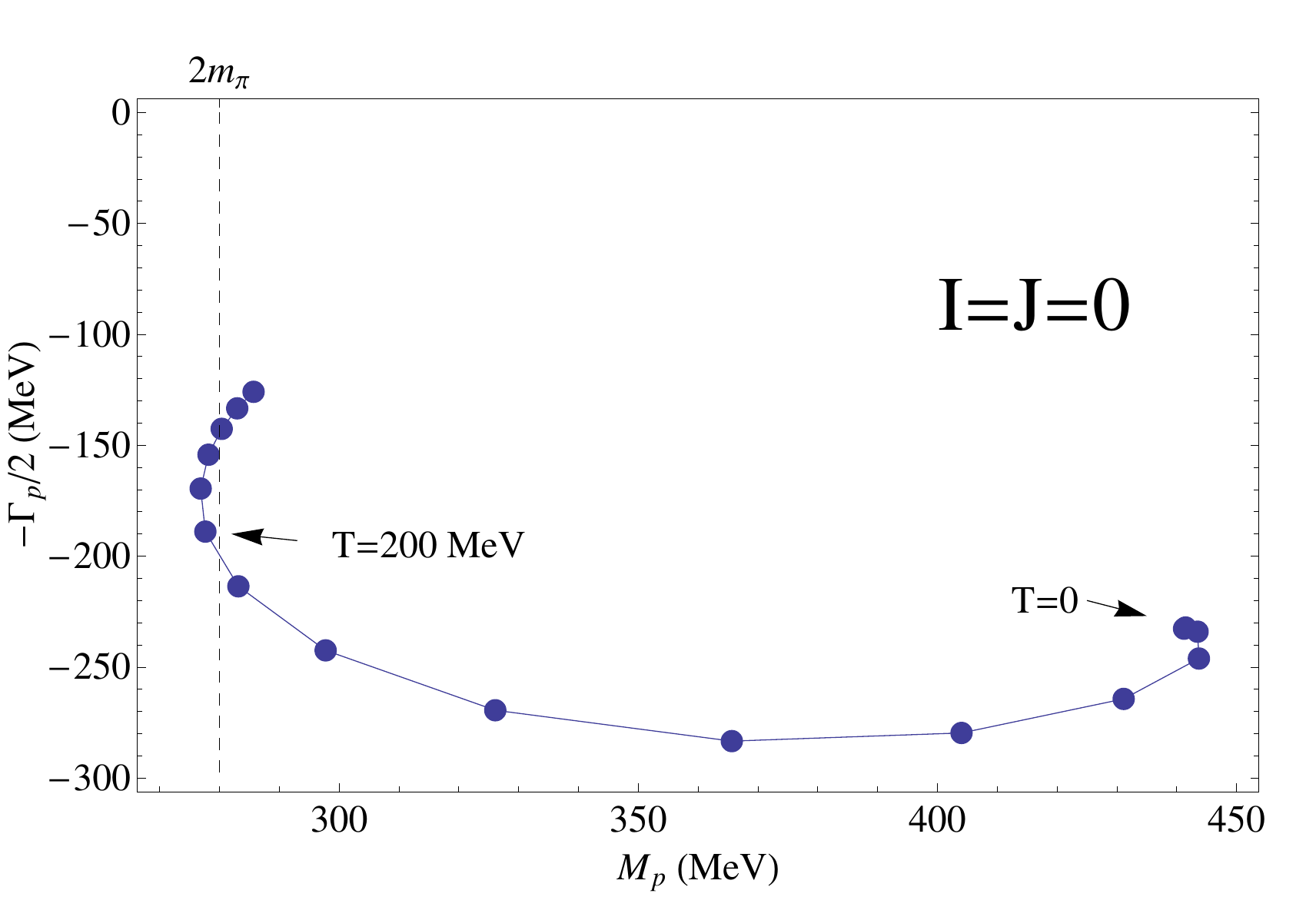}
\includegraphics[width=7.5  cm]{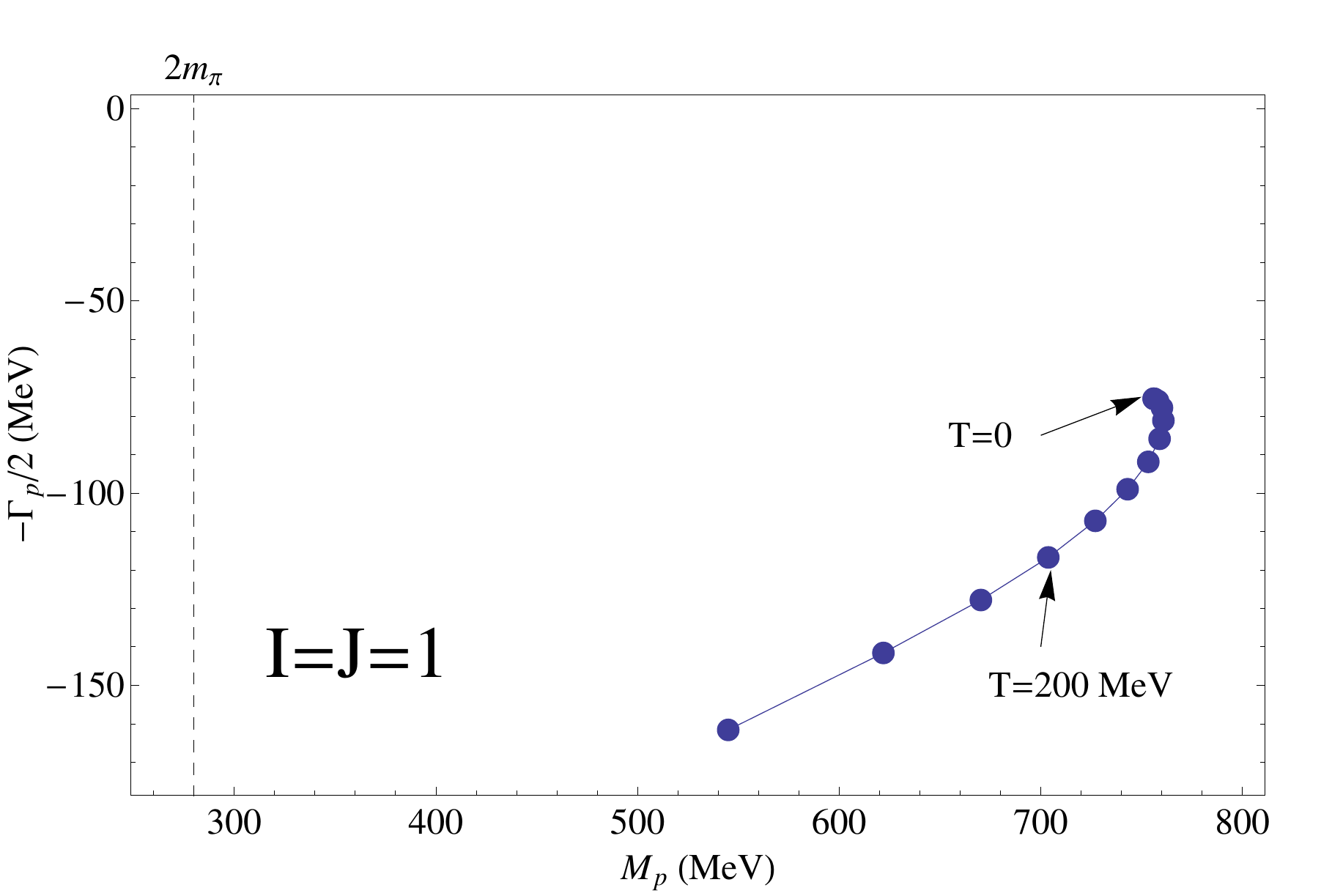}
\caption{Thermal evolution of the thermal poles $s_p=(M_p-i\Gamma_p/2)^2$ for the $IJ=00$ (\textbf{left}) and $IJ=11$ (\textbf{right}) channels in the complex $s$ plane, from \cite{Dobado:2002xf,FernandezFraile:2009mi}.}
\label{fig:thermalpoles} 
\end{figure}

\begin{figure}[h]
\centering
\includegraphics[width=7.5 cm]{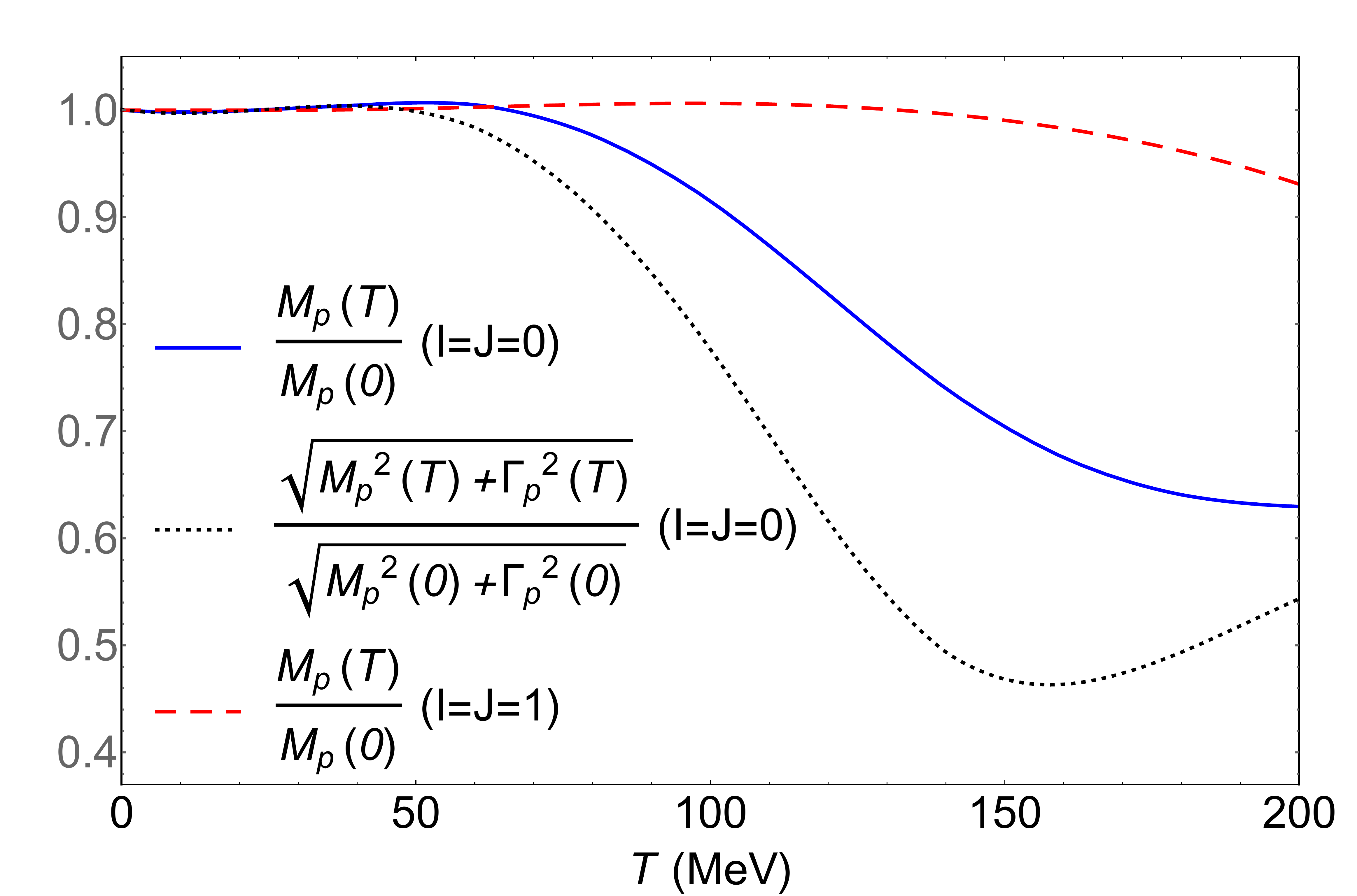}
\includegraphics[width=7.5 cm]{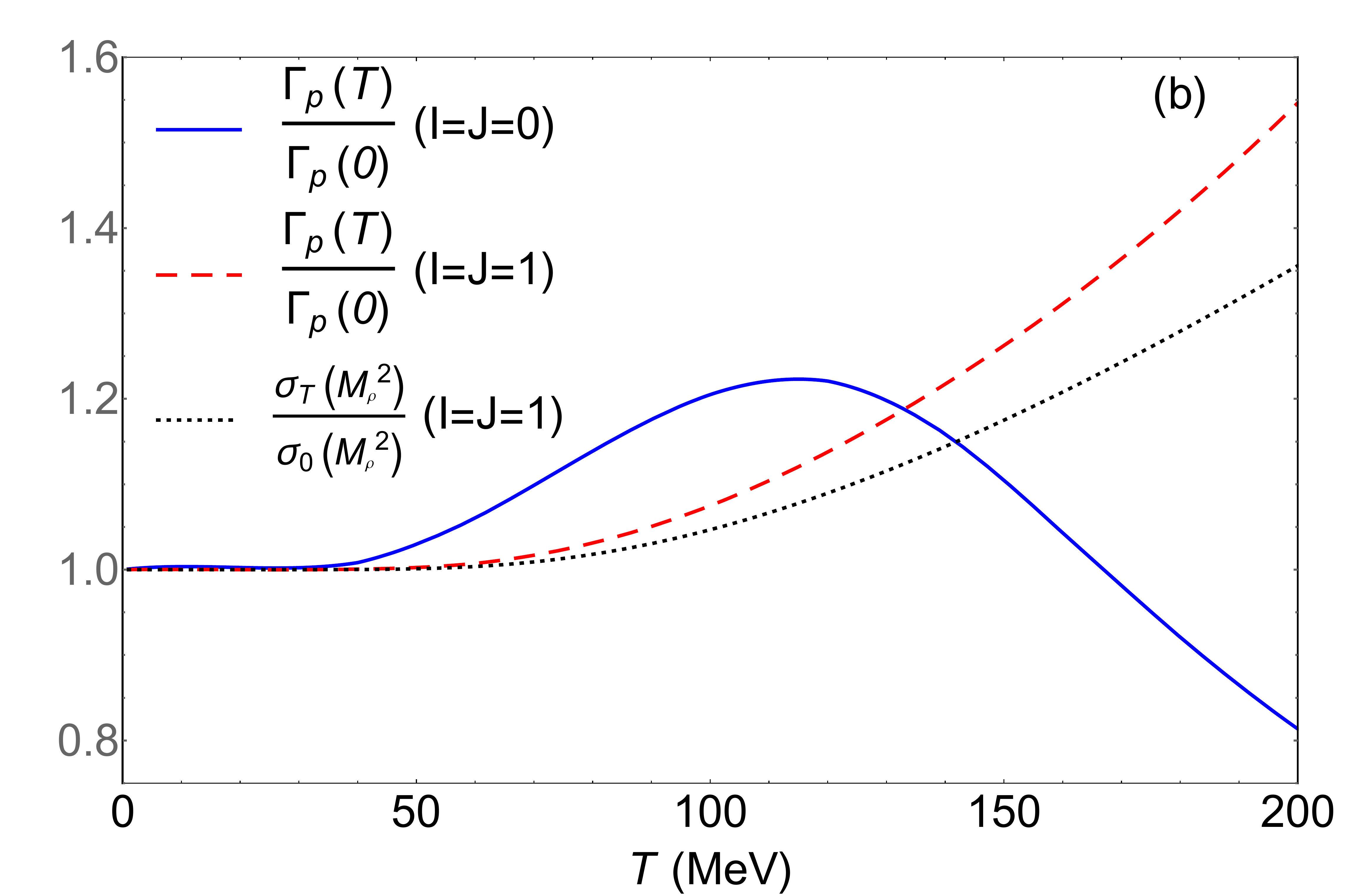}
\caption{Thermal evolution of the real and imaginary parts of the thermal pole $s_p=(M_p-i\Gamma_p/2)^2$ for the $IJ=00$ and $IJ=11$ channels, from \cite{Dobado:2002xf,Nicola:2013vma}. (\textbf{Left}): real part  $M_p$ and the scalar mass corresponding to the real part of the self-energy as explained in the text. (\textbf{Right}): imaginary part $\Gamma_p$, compared to the thermal phase space in the vector channel.}
\label{fig:thermalpolepar} 
\end{figure}

The~poles of this unitarized amplitude in the second Riemann sheet correspond to the thermal resonances \cite{Dobado:2002xf}, whose parameters (corresponding to states at rest) are the real and imaginary part of the pole position parametrized as customary as  $s_p (T)=\left[M_p(T)-i\Gamma_p(T)/2\right]^2$ so that $M_p$ and $\Gamma_p$ correspond to the mass and width in the narrow resonance (BW) case. In~Figures \ref{fig:thermalpoles} and \ref{fig:thermalpolepar} we show the results for the thermal poles in the $I=J=0$ and $I=J=1$   channels, obtained first \mbox{in \cite{Dobado:2002xf}}. The~corresponding resonances are the $f_0(500)$ in the scalar channel and the $\rho(770)$ in the vector channel and the LECs entering the IAM formula have been chosen so as to yield $T=0$ values for those resonances compatible with the PDG. 

\vspace{-9pt}
\unskip

 In~the vector channel  $\Gamma_p<<M_p$ for the relevant temperature range  and therefore the $\rho$ can be considered a narrow BW resonance with $M_p$ and $\Gamma_p$ its mass and width respectively. Actually, $M_p^2(T)$ decreases very slightly with $T$. However, the~thermal effect in $\Gamma_p(T)/\Gamma_p(0)$ is much more sizable, increasing with $T$ as it can be clearly seen in the figures. Such broadening is dominated at low and moderate temperatures  by the increase of thermal phase space from the $n_B$ term in \eqref{therphasespace} evaluated close to $s=M_\rho^2$ while for higher temperatures a significant increase with $T$ of the effective $\rho\pi\pi$ coupling  explains the additional increase seen in Figure \ref{fig:thermalpolepar} \cite{Dobado:2002xf}. The~thermal broadening of the $\rho$ obtained through this approach is consistent with our discussion above regarding the dilepton spectrum. That~connection is more clearly established through the analysis of the temperature dependence of the electromagnetic form factor, whose squared modulus is directly related to the imaginary part of the electromagnetic current-current correlator if only pion degrees of freedom are considered. As~mentioned before, such correlator encodes the physical information about the photon and the dilepton thermal yield. The~calculation of the form factor at finite temperature within ChPT and UChPT has been performed in 
 \cite{GomezNicola:2004gg}. Once unitarized, the~squared modulus of the thermal form factor exhibits a significant broadening dictated again by the $\rho$ thermal pole, which as stated, is consistent with the expectations from the dilepton spectrum. That analysis is consistent with previous ones \cite{Song:1996dg} based on the VMD approach. 
 
 The~scalar channel has a completely different behaviour. The~$f_0(500)$ pole remains a broad state also at finite temperature, since $M_p$ and $\Gamma_p$ remain comparable, so its BW interpretation does not apply, as it also happens at $T=0$. One immediate consequence of the finite-$T$ analysis in this channel is actually that the thermal contributions of the different diagrams involved becomes much stronger than the $\rho$ case because the energy scales involved, to be compared with the temperature, apart from  the pion mass,  are $M_p,\Gamma_p$ which are lower than the vector case. This is clearly seen in Figure \ref{fig:thermalpolepar}. The~behaviour of $M_p(T)$ is clearly decreasing, which might be thought of a signal of chiral restoration if one naively identifies $M_p$ with the $\sigma$ mass, which as described above, would tend to become degenerate with the pion and decrease as the quark condensate, for instance within a LSM approach. However, as we are going to discuss in detail below, life is not that simple and the interpretation of the $f_0(500)$ thermal pole is much more subtle, as we will see in Section \ref{sec:f0500}. As for $\Gamma_p(T)$, one can see in Figures \ref{fig:thermalpoles} and \ref{fig:thermalpolepar} that it develops non-monotonic behaviour, with a maximum around $T\sim$ 120 MeV. A qualitative explanation for such behaviour is that there are two effects competing here. On the one hand, the~increase of thermal phase space provided by the $n_B$ contribution in \eqref{therphasespace} tends to increase $\Gamma_p (T)$ (although as said before, this is just qualitative because the BW approximation is not valid here). On the other hand, as the would-be ``mass" $M_p(T)$ approaches the threshold $2m_\pi$ one expect a reduction of such phase space, which is the dominant contribution for higher temperatures. The~delicate balance between those two effects turns out to be crucial for the discussion of the role of the thermal $f_0(500)$ state in chiral symmetry restoration as will be discussed in detail in Section \ref{sec:f0500}. 
 
 The~calculation of the thermal scattering poles in the $I=J=0$ channel within the large-NGB approach in the chiral limit confirms the previous findings \cite{Cortes:2015emo}. The~resumation of diagrams provided by that approach gives rise to an exactly unitary amplitude satisfying the unitarity condition \eqref{unitpwther}, namely, with a suitable choice of the renormalization scale $\mu$, the~partial wave in this channel can be written as
 \begin{equation}
t_{00}(s;T)=\frac{s f[ I_\beta]}{32\pi F^{2}}\frac{1}{1-\frac{s f[I_\beta]}{32\pi^2F^2}\left[\ln\left(\frac{\mu^{2}}{-s}\right)+16\pi^2\delta J(s;T)\right]} + \Od(1/N)
\label{t00TlargeN}
\end{equation}
where $N$ is the number of NGB, $I_\beta=T^2/12$ is the NGB tadpole at finite temperature in the chiral limit, $\delta J(s;T)=J(0,\sqrt{s},\vec{0},T)-J(0,\sqrt{s},\vec{0},0)$ with the $J$ integral defined in \eqref{Jter} and
\begin{equation}
f(I_{\beta})=\frac{1}{1-I_\beta/F^2}.
\end{equation}

The~pole position within this approach follows a similar trajectory as the IAM one discussed above, in the chiral limit, for different choices of the LECs involved \cite{Cortes:2015emo}.

 In~this context, it is worth mentioning  also a recent analysis of other scattering processes at finite temperature and the corresponding evolution of thermal poles, performed recently in \cite{Gao:2019idb}. In~particular,  the $f_0(500)$, $K^*(700)$, $f_0(980)$ and $a_0(980)$ states are analyzed, which require the finite-$T$ amplitudes  for binary processes involving $\pi, K, \eta$ states. Those amplitudes are not calculated fully  in ChPT at finite-$T$, only a subset of contributions is considered, essentially the loops giving rise to thermal unitarity, i.e., diagrams of type (a) in Figure \ref{fig:diag} as well as tadpole contributions of the type (d) and (e) in that figure.  Once the amplitudes are unitarized, the~results show a behaviour of the $K^*(700)$ pole quite similar to the $f_0(500)$, while the   $f_0(980)$ and $a_0(980)$ are little affected by the thermal corrections, as one could anticipate from their higher mass and small width.

 \subsection{The~Thermal $f_0(500)$ and Chiral Symmetry Restoration} 
 \label{sec:f0500} 
 
 In~this section we will review some recent work on the connection of the thermal $f_0(500)$ state with chiral symmetry restoration. We~will show that the  properties of the thermal pole described in the previous section have direct phenomenological implications, in particular, they allow for a description of the scalar susceptibility $\chi_S(T)$ in accordance with lattice measurements, improving over other effective theory descriptions around the transition.  This connection with the scalar susceptibility is somehow naturally expected since $\chi_S(T)$ in \eqref{susdef} appears as the correlator 
of the operator $\sigma_l\sim \bar \psi_l \psi_l$, which has the same quantum numbers as the $f_0(500)$, and indeed of the vacuum, and hence some relation would be expected between $\chi_S$ and the $f_0(500)$ self-energy. Actually, in the discussion about degeneration of chiral partners that we will present in Section \ref{sec:patt}, $\chi_S$ will play also a crucial role as it becomes degenerated with $\chi_\pi$, the~susceptibility corresponding to the pion-like operator  $\pi^a\sim \bar \psi_l \tau^a \psi$.

The~above mentioned connection can be better understood using as a testbed model the LSM. Although, as mentioned in Section \ref{sec:effthe}, this model does not provide a fully satisfactory low-energy  phenomenology, one can speak more properly about the self-energy of the $\sigma$ pole and its temperature dependence. Consider then the LSM Lagrangian,
\begin{equation}
{\cal L}_{LSM}= \frac{1}{2}\partial_\mu\Phi^T\partial^\mu \Phi-\frac{\lambda}{4}\left[\Phi^T\Phi-v_0^2\right]^2+h\sigma,
\label{lsm1}
\end{equation}
with $\Phi^T=(\sigma,\vec{\pi})$. The~$\sigma$ direction is the symmetry-broken one from $O(4)\rightarrow O(3)$, where $O(4)\approx SU_L(2)\times SU_L(2)$ and $O(3)\approx SU_{L+R}(2)$. The~$h$ term breaks explicitly the chiral symmetry, with $h$  proportional to the pion mass squared and the potential minima at $\Phi^2=v^2\neq 0$  implement spontaneous chiral symmetry breaking.  Following the standard procedure of shifting the field  as $\tilde\sigma=\sigma-v$, one has $\mean{\tilde\sigma}(T)=v(T)-v\neq 0$, which in particular implies that one-particle reducible (1PR) diagrams enter in the calculation of correlators \cite{Ayala:2000px,Ferreres-Sole:2018djq}. From the shifted Lagrangian, one can on the one hand calculate the scalar susceptibility by taking derivatives with respect to the light quark mass $\hat m$, related to the tree-level pion mass as $M_{0\pi}^2=2B_0 \hat m$ where $\condl (T=0)=-2B_0 F^2(1+\Od(M_\pi^2/M_\sigma^2))$. Thus one gets,
\begin{eqnarray}
\chi_S (T)&=&4B_0^2 \left[ -\frac{1}{2B_0 M_{0\sigma}^2}\frac{2M_{0\sigma}^2-3M_{0\pi}^2}{M_{0\sigma}^2-M_{0\pi}^2} \condl(T)+v^2\left(\dfrac{M_{0\sigma}^2}{M_{0\sigma}^2-M_{0\pi}^2}\right)^2 \Delta_\sigma (k=0;T)\right],
\label{susmod}
\end{eqnarray}
where 
\begin{eqnarray}
\Delta_\sigma (k;T)
=\frac{1}{k^2+M_{0\sigma}^2+\Sigma(k_0,\vec{k};T)}
\end{eqnarray}
is the Euclidean propagator of the $\tilde\sigma$ field and  $\Sigma(k_0,\vec{k};T)$ is the self-energy, which  depends separately on the space and time components of the four-momentum $k$ at finite $T$ \cite{galekapustabook}.

The~first term in \eqref{susmod} is actually negligible near the transition with respect to the second one. First, it vanishes proportionally to  $\condl(T)$. Second, near the transition $\chi_S\sim \chi_\pi$, which can also be related to the condensate using a Ward Identity that will be discussed in Section \ref{sec:patt}, namely \eqref{wipi}. Thus, replacing in the first term $\condl(T)\rightarrow - M_{0\pi}^2 \chi_S(T)/(2B_0)$  yields a numerical suppression of order $\Od\left(M_{0\pi}^2/M_{0\sigma}^2\right)$ for that term with respect to the second one. Either way, we arrive to the conclusion that near the transition,
  \begin{equation}
 \frac{\chi_S (T)}{\chi_S (0)}\simeq \dfrac{M_{0\sigma}^2+\Sigma\left(k=0;T=0\right)}{M_{0\sigma}^2+\Sigma\left(k=0;T\right)}
\label{LSMsaturated} 
\end{equation}

This relation shows, within the LSM, the~connection between the scalar susceptibility and the $\sigma$ self-energy around the transition, which one can interpret as a as saturation approach, where the self-energy of the  lightest thermal state, the~$\sigma$ in the LSM, saturates the $k=0$ correlator defining the scalar susceptibility,  providing a first hint towards the corresponding saturated definition in the case of the $f_0(500)$ described within UChPT. Note that the self-energy is real at $k=0$. 

On the other hand, in order to compare the LSM saturated scalar susceptibility \eqref{LSMsaturated} with a purely perturbative calculation within that model, we need an explicit calculation of the self-energy, which can be carried out within the $\lambda$ expansion. Although, as mentioned before, the~values of $\lambda$ are  large to comply with phenomenology, the~one-loop corrections to the self-energy lie around  15\% at $T=0$ \cite{Pelaez:2015qba}, which is still reasonably under control. Besides, this analysis serves to reach interesting theoretical conclusions parametrically in $\lambda$.  A detailed perturbative calculation of the $\sigma$ self-energy at finite temperature up to one-loop can be found in \cite{Ayala:2000px,Ferreres-Sole:2018djq}.  
 The~renormalization of the pion and sigma masses allows us to get a finite and scale-independent result for the self-energy within dimensional regularization, namely,
  \begin{eqnarray}
\Delta_\sigma^{-1}&=&M_\sigma^2+\Sigma(k_0,\vec{k};T),\nonumber\\
\Sigma(s,T=0)&=&\dfrac{3\lambda}{16\pi^2}(M_{\sigma}^2-M_{\pi}^2)\left[\sigma_{\pi}(s)\log \left(\dfrac{\sigma_{\pi}(s)+1}{\sigma_{\pi}(s)-1}\right)+3\hspace{0.1cm}\sigma_{\sigma}(s)\log \left(\dfrac{\sigma_{\sigma}(s)+1}{\sigma_{\sigma}(s)-1}\right)\right. \nonumber \\
&+&\left. \log\left(\dfrac{M_{\pi}^2}{M_{\sigma}^2}\right)-\dfrac{13}{3}\right]+\Od(\lambda^2),
\nonumber\\
\Sigma(k_0,\vec{k};T)&=&\Sigma(s,T=0)+ 3\lambda\left\{\frac{3M_\pi^2-2M_\sigma^2}{M_\sigma^2}\left[g_1(M_\pi,T)+g_1(M_\sigma,T)\right]\right.
\nonumber\\
&-&\left.\left(M_\sigma^2-M_\pi^2\right)\left[\delta J(M_\pi;k_0,\vec{k},T)+3\delta J(M_\sigma;k_0,\vec{k},T)\right]\right\}+\Od(\lambda^2),
\label{selfenergylsm}
\end{eqnarray}
where 
\begin{equation}
g_1(M,T)=\frac{T^2}{2\pi^2}\int_{M/T}^\infty dx  \frac{\left[x^2-(M/T)^2\right]^{1/2}}{e^{x}-1}
\label{g1}
\end{equation}
is the finite-$T$ part of the pion tadpole and $\delta J$ stands for$J(T)-J(T=0)$ with $J$ the thermal integral defined in \eqref{Jter}.

 From the self-energy, we can readily calculate the pole of the propagator and its temperature dependence, namely $s_p=M_\sigma^2+\Sigma(k^2=M_\sigma^2)=(M_p-i\Gamma_p/2)^2$. For our purposes here we only need the pole at $\vec{k}=\vec{0}$. Determining the pole position allows us to calibrate whether the model parameters can be chosen so that such pole complies with the PDG values. As commented, this is a clear drawback of the model. Fixing the pion  decay constant and mass to their physical values, the~coupling $\lambda$ needs to be kept in a range $\lambda\sim 10-20$ in order that $M_p$ and $\Gamma_p$  lie not far from the PDG. Outside that range at least one of them deviates more than 200 MeV from their PDG value.  
 
\begin{figure}[h]
\centering
\includegraphics[width=10 cm]{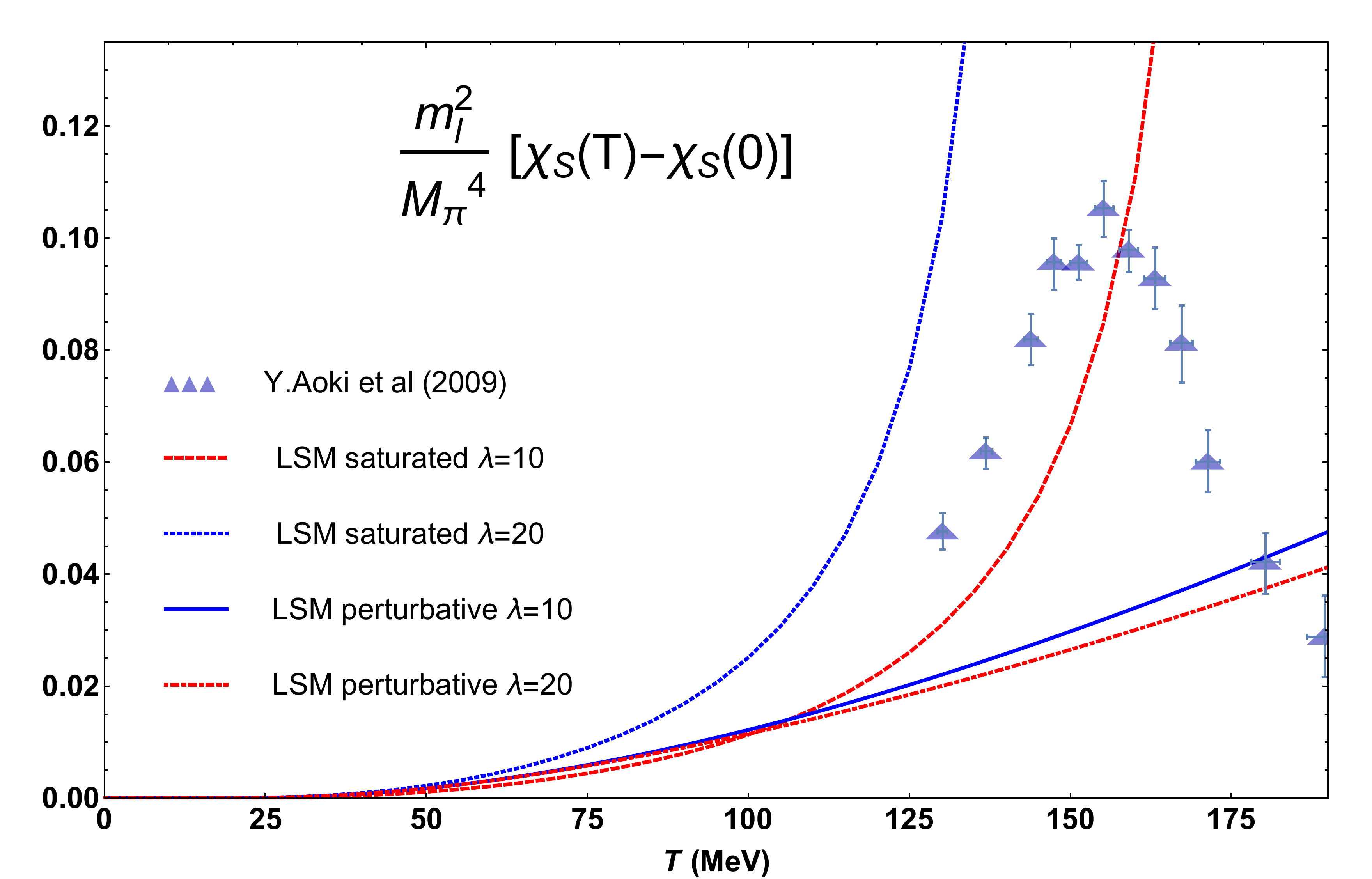}
\caption{Comparison between the saturated and perturbative scalar susceptibility in the Linear Sigma Model (LSM), from \cite{Ferreres-Sole:2018djq}. Lattice points and errors are taken from \cite{Aoki:2009sc}.}
\label{fig:LSMsus} 
\end{figure}

 In~Figure \ref{fig:LSMsus} we plot the result of the scalar susceptibility, properly normalized for an easier comparison with lattice data, within the LSM in the saturated approach given by \eqref{LSMsaturated} and the strictly perturbative one, which corresponds to retaining only the $\Od(\lambda)$ in that expression. We~have also represented the lattice points obtained in \cite{Aoki:2009sc}. The~main conclusion, holding also in the UChPT  approach below, is that the saturated result grows must faster than the perturbative one near the transition, actually it eventually diverges even in the massive case. The~curve does not reproduce then the transition peak, although the range of $\lambda$ mentioned before lies reasonably close to lattice data. This LSM description already provides support for the idea of a saturated thermal pole analysis, which might account for the lattice values including only the lightest state, i.e., the~$f_0(500)$.  As we are about to see, the~UChPT framework based on the same idea shares this feature and actually improves considerably the description around the transition.  It is important to mention also that evaluating the self energy at the pole position $s_p$ instead of  $k=0$ gives rise to the same qualitative behaviour for the saturated susceptibility. This comment is relevant for the UChPT case, as we will see below.  Note that  $M_\sigma^2+\re\Sigma$  corresponds to the $T$-dependent scalar mass, which drops below the transition approaching the pion mass \cite{Bochkarev:1995gi}. 
 
 \vspace{-6pt}
 
 The~previous ideas can be applied  within the UChPT framework \cite{Nicola:2013vma,Ferreres-Sole:2018djq} as follows.  As discussed above, we expect the scalar susceptibility $\chi_S (T)$ to be proportional to the inverse of $\Sigma (k=0)$, the~self-energy of the $f_0(500)$ state at vanishing momentum. Although the thermal $f_0(500)$ state generated  within UChPT shows up as a pole in the 2RS of the corresponding partial wave rather than from a thermal correlator, one can still think of the pole parameters in terms of an effective resonance exchange. Thus, around the pole,
  \begin{equation}
t^{II}=\frac{1}{16\pi}\frac{g_{\sigma\pi\pi}^2}{s-s_p}+\dots,
\label{unitpoleexp}
\end{equation}
where the superscript ``II" denotes the amplitude in the 2RS, the~dots denote subdominant terms around $s\sim s_p$ and the effective $\sigma\pi\pi$ vertex is $g_{\sigma\pi\pi}$ defined as the residue at the pole. We~can then interpret \eqref{unitpoleexp} as the exchange of a scalar $f_0$ state with self-energy satisfying $\Sigma_{f_0}(s_p)=s_p$ (including in $\Sigma$ the free mass). Since $\im\Sigma_{f_0} (k=0)=0$  due to the absence of decay channels at vanishing momentum, there is some uncertainty in this approach  on the sensitivity of $\re\Sigma_{f_0}$ from $s=s_p$ to $s=0$. Such uncertainty, as we will see below, can be reasonably kept under control, lying  within the typical uncertainty range of  this method, which includes also the sensitivity to the unitarization method or the uncertainties in the LECs, all of them analyzed in detail in \cite{Ferreres-Sole:2018djq}. With those ideas in mind, we define 
 the saturated (unitarized) scalar susceptibility as
  \begin{equation}
\chi_S^U(T)=A\frac{M_\pi^4}{4m_l^2}\frac{M_S^2(0)}{M_S^2(T)},
\label{susunit}
\end{equation}
normalized for a better comparison with lattice results, where now the scalar thermal pole mass is defined as
\begin{equation}
M_S^2(T)=\re s_p (T)=M_p^2(T)-\frac{1}{4}\Gamma_p^2(T),
\label{scalarmass}
\end{equation}

\begin{figure}[h]
\centering
\includegraphics[width=10 cm]{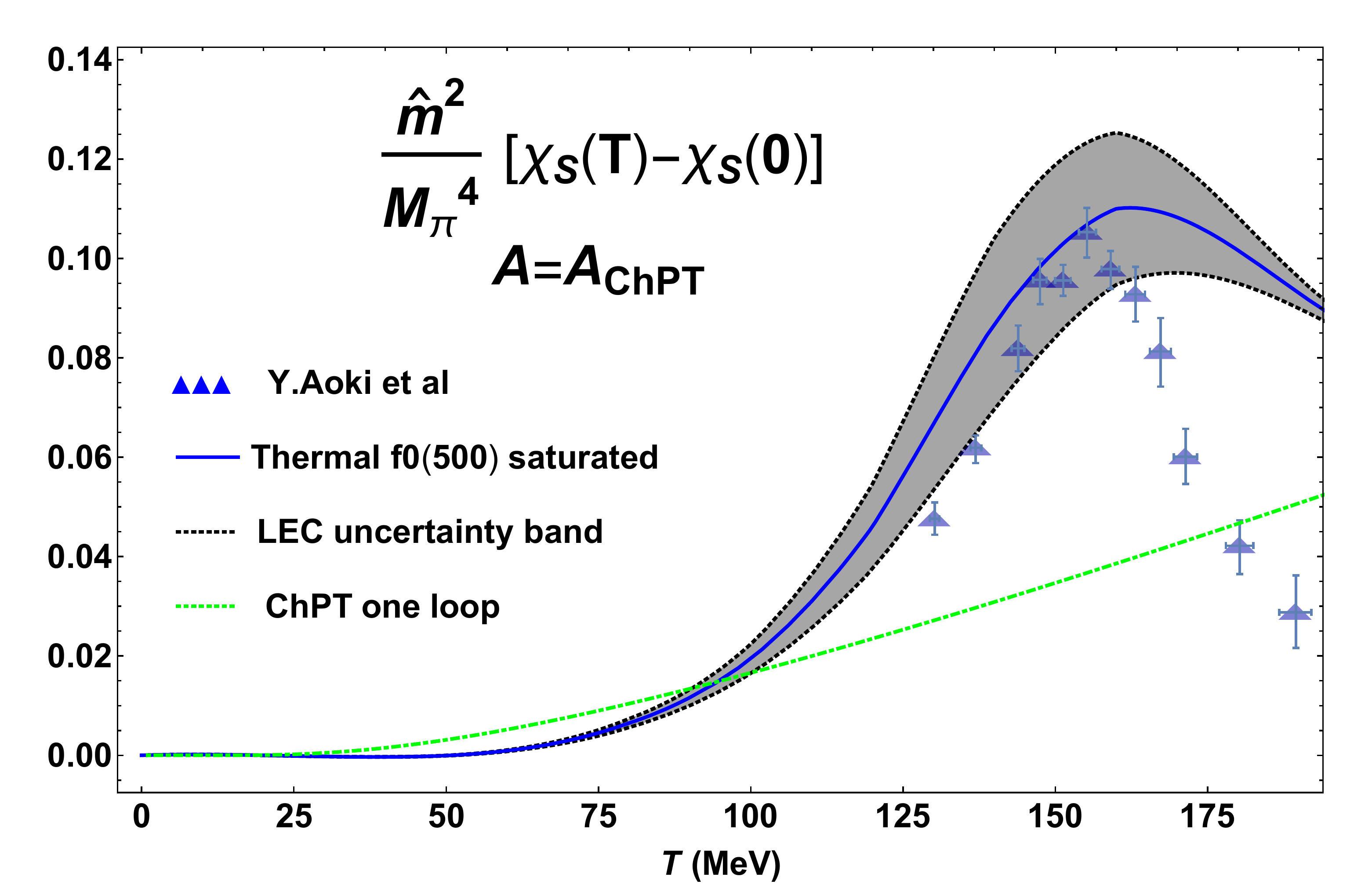}
\caption{Scalar susceptibility within the Unitarized Chiral Perturbation Theory (UChPT) saturated approach \cite{Ferreres-Sole:2018djq}. The~Low Energy Constant (LEC) band uncertainty corresponds to the $l_1^r$ and $l_2^r$ values \mbox{in \cite{Hanhart:2008mx}}. Lattice points and errors are from \cite{Aoki:2009sc}. }
\label{fig:unitsus} 
\end{figure}

Such thermal mass drops even faster than $M_p(T)$, as seen in Figure \ref{fig:thermalpolepar}, but more importantly, it develops a minimum around the transition, which stems from the relative behaviour of $M_p(T)$ and $\Gamma_p(T)$ explained in Section \ref{sec:reso}. The~normalization constant $A$ can be fixed to the $T=0$ value of the scalar susceptibility calculated within ChPT,  giving $A_{ChPT} \simeq 0.15$ \cite{Nicola:2013vma}, which already provides an excellent description around the transition region, as showed in Figure \ref{fig:unitsus}. 

 \vspace{-6pt}

\begin{figure}[h]
\centering
\includegraphics[width=7.5 cm]{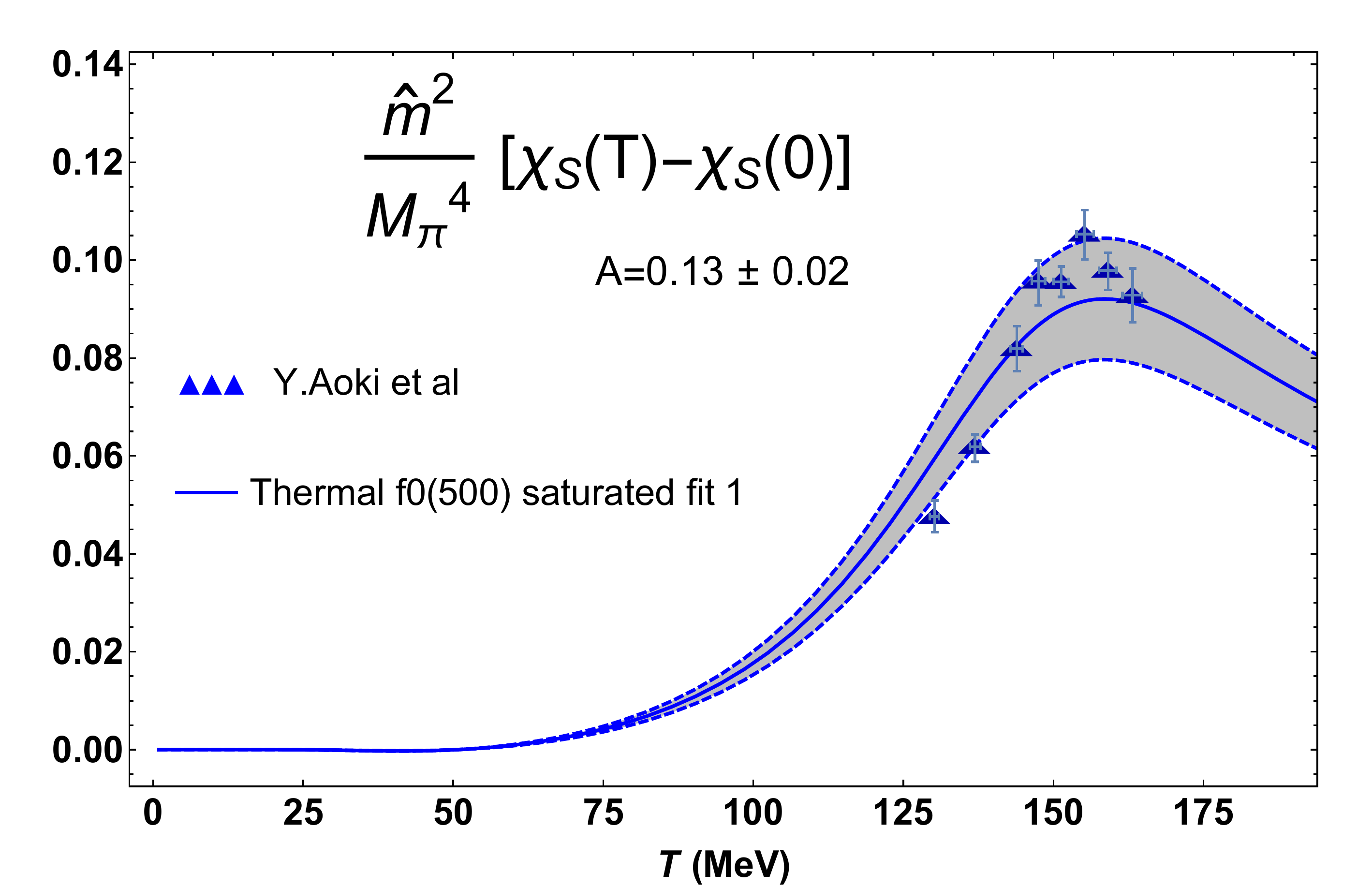}
\includegraphics[width=7.5 cm]{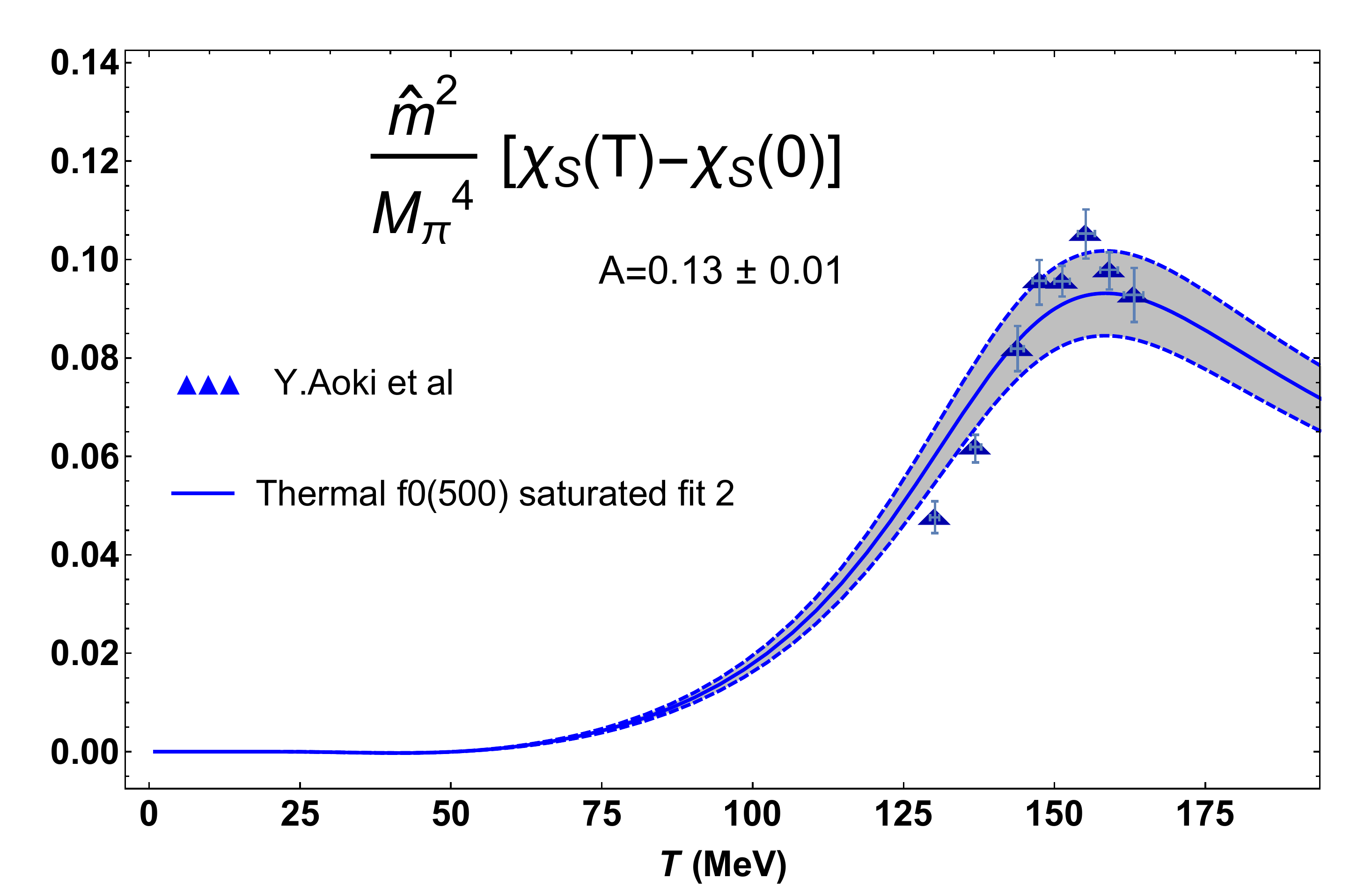}
\caption{Scalar susceptibility within the UChPT saturated approach fitting the $A$ parameter to lattice data \cite{Ferreres-Sole:2018djq}  with the central values of the LEC given in \cite{Hanhart:2008mx}. Fit 1 corresponds to fitting data up to $T\leq T_c=155$ MeV while in fit 2 we include two more lattice points, up to $T=163$ MeV. The~uncertainties in  $A$  and the bands correspond to the 95\% confidence level of the fit. The~lattice data and errors are from \cite{Aoki:2009sc}. The~$\chi^2$/dof values are 6.2 and 4.9 for fits 1 and 2 respectively.}
\label{fig:unitsusfit} 
\end{figure}

An even more accurate picture can be obtained by fitting the $A$ parameter in \eqref{susunit}. The~result of such fits is showed in Figure \ref{fig:unitsusfit}. Once more, we see that the saturated description accounts rather satisfactorily for lattice data, including the position of the transition peak, just with the thermal $f_0(500)$ state. Note that the value of $A$ obtained in the fits is compatible with the ChPT one. 

 \vspace{-6pt}

\begin{figure}[h]
\centering
\includegraphics[width=7.5 cm]{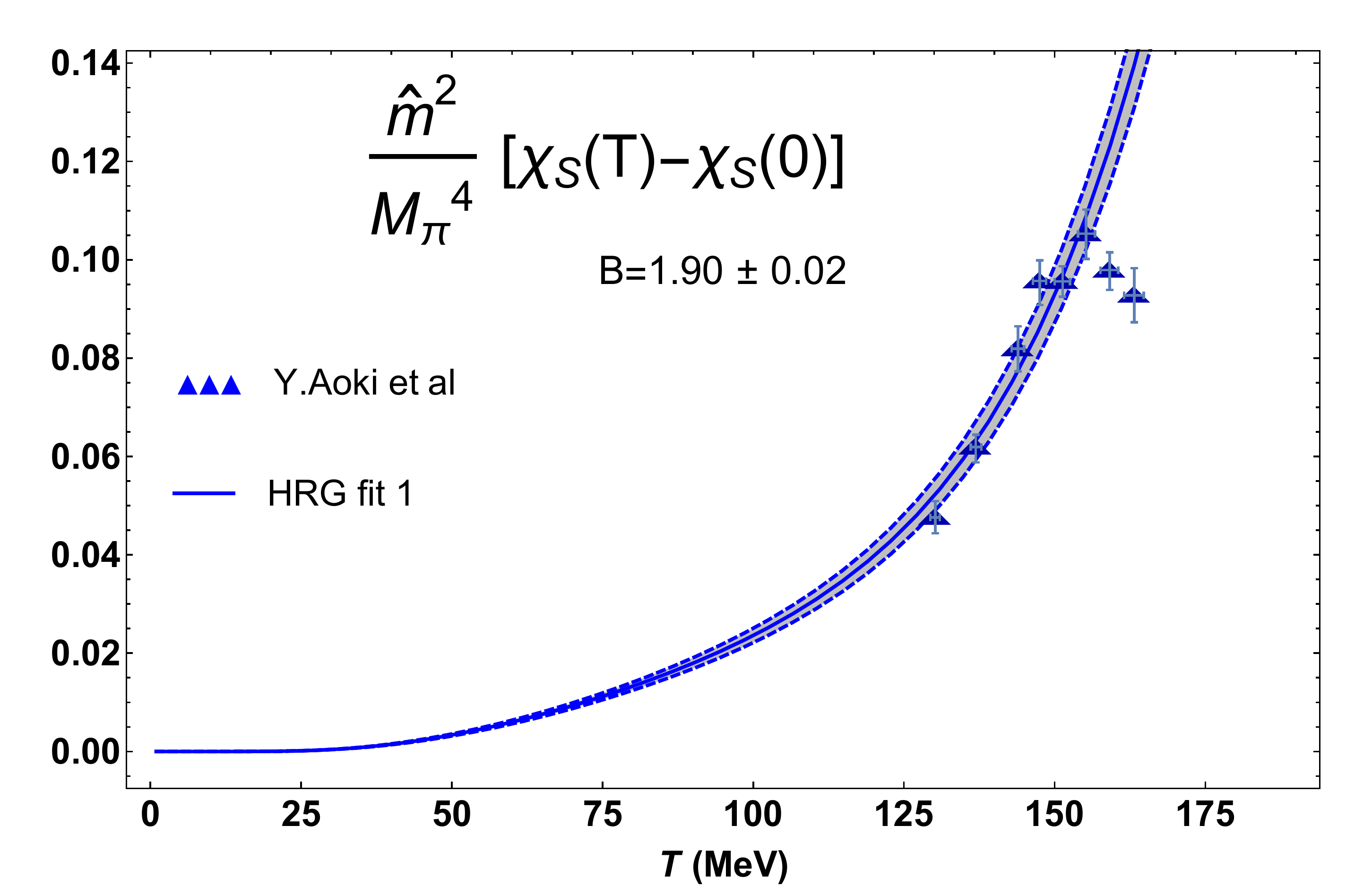}
\includegraphics[width=7.5 cm]{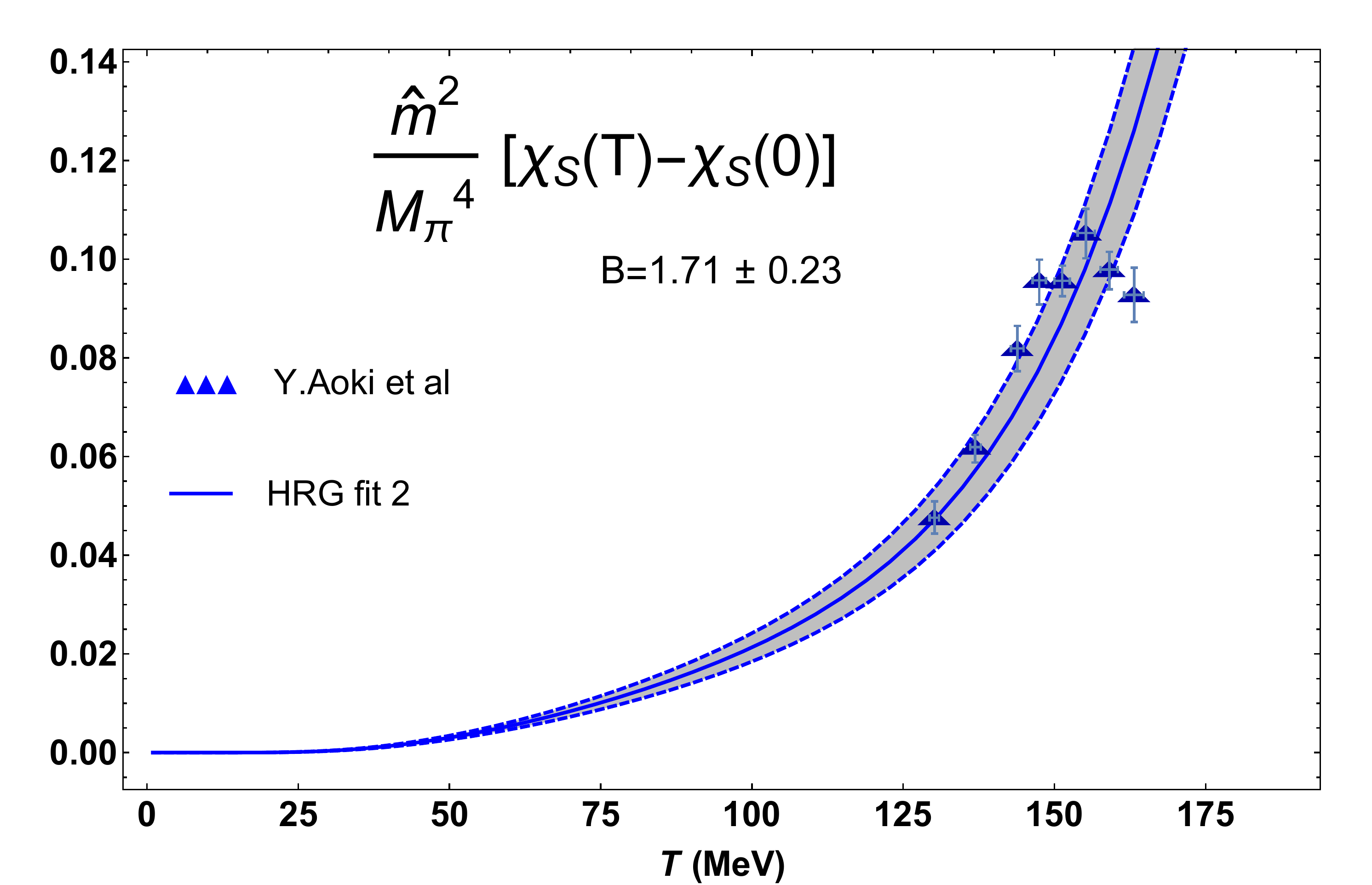}
\caption{HRG fits for the scalar susceptibilty  \cite{Ferreres-Sole:2018djq}.  Fit 1 corresponds to fitting data up to $T\leq T_c=155$ MeV while in fit 2 we include two more lattice points, up to $T=163$ MeV. The~uncertainties in  $B$  and the bands correspond to the 95\% confidence level of the fit. The~lattice data and errors are from \cite{Aoki:2009sc}. The~$\chi^2$/dof values are 1.3 and 10.3 for fits 1 and 2 respectively.}
\label{fig:hrgfit} 
\end{figure}

It is instructive to compare the thermal $f_0(500)$ description with one based on the HRG, i.e., including many more resonant states, as explained in Section \ref{sec:effthe}, but with their $T=0$ masses, not including the $T=0$ $f_0(500)$ for the reasons explained there. For that purpose, we consider the HRG   free energy given in \eqref{zHRG} and the quark mass dependence considered 
in \cite{Leupold:2006ih,Jankowski:2012ms}  and explained in Section \ref{sec:effthe}. We~include also a normalization fit parameter $z\rightarrow Bz$ in that case.  The~results are showed \mbox{in Figure \ref{fig:hrgfit}}. 

Although the HRG fit improves over the thermal $f_0(500)$ one for fit 1, i.e., including lattice points only up to the transition, the~situation changes drastically when adding two more lattice points around $T_c$ (fit 2) where the saturated $f_0(500)$ becomes clearly more competitive. Besides, as showed in~\cite{Ferreres-Sole:2018djq}, fitting the quark condensate lattice data with the HRG gives $B = 1.06\pm 0.12$ with $\chi^2$/dof = 3.8, that value of $B$ being incompatible with those in Figure  \ref{fig:hrgfit}. Thus, the~HRG description of the quark condensate and the scalar susceptibility are difficult to reconcile with each other. 

 \vspace{-6pt}

The~above analysis, both in the LSM and UChPT approaches, emphasizes the importance of the thermal  $f_0(500)$  state as far as chiral symmetry restoration is concerned.  In~the case of the scalar susceptibility,   the saturated approach allows for a description of lattice data competitive with respect to the HRG, accounting therefore for the most relevant part of the hadronic spectrum with the lightest thermal state.  We~would like to remark here that $\chi_S$ is quite appropriate for that description, because of its expected $M_S^{-2}$ dependence, so that clearly a similar description for other thermodynamical observables such as the quark condensate would not be adequate. Another important comment is that the inclusion here of the thermal corrections to the $\pi\pi$ scattering amplitude is ultimately responsible for the importance of the $f_0(500)$ state, which is perfectly compatible with our discussion in Section~\ref{sec:effthe} regarding the virial expansion and the cancellation of the $I=J=0$ contribution at $T=0$ which justifies not to include the $T=0$ $f_0(500)$ state within the HRG framework.

\section{The~Nature of the Transition:  Partners and Patterns} 
\label{sec:patt}

We~address in this section an interesting open problem within the context of chiral symmetry restoration, which involves its very nature, namely to determine what is the universality class (or~pattern) of the transition. This is not merely a theoretical question, since, as we will comment below, it~has phenomenological consequences which can be observed in lattice analysis. 

The~transition pattern depends strongly on whether the $U(1)_A$ anomalous symmetry is sufficiently restored at $T_c$, which may even affect the order of the transition itself \cite{Pisarski:1983ms,Pelissetto:2013hqa,Eser:2015pka,Fejos:2015xca}. Thus, in the regime where $U(1)_A$ and chiral restoration are effective, a  $SU_L(N_f)\times SU_R(N_f)\times U(1)_A$ symmetry breaking pattern would be expected, which corresponds to $O(4)\times U(1)_A$ for two light flavors. Thus, for $N_f=2$ a second-order  transition in the $O(4)$ universality class would be preferred in a scenario with  $U(1)_A$ breaking at $T_c$, while a second-order one in the $U(2)\times U(2)$ universality class would correspond to a $U(1)_A$ restored situation. The~latter  may even degenerate into a first order transition for strong enough $U(1)_A$ restoration. 

The~$U(1)_A$ symmetry restoration is theoretically possible at finite temperature, at least asymptotically for high enough $T$. Early proposals in this directions suggested a mechanism for such asymptotic restoration  driven by the vanishing of the instanton density~\cite{Gross:1980br}.  Some additional consequences of a sizable $U(1)_A$ restoration at the critical point include the position and properties of the $\mu_B\neq 0$ critical point \cite{Mitter:2013fxa} and effects related to the expected reduction of the $\eta'$ mass, which would then become a ninth Goldstone Boson in that regime  with  implications  in particular on the dilepton and diphoton spectra \cite{Kapusta:1995ww}. Such reduction of  the $\eta'$ mass in the hot medium has been actually been observed experimentally~\cite{Csorgo:2009pa} and in the lattice \cite{Kotov:2019dby}. 

A useful way to analyze the interplay between chiral and $U(1)_A$ restoration is through the study of the degeneration of the corresponding partners under those symmetries. We~have seen already a significant example in Section \ref{sec:reso} regarding the $\rho-a_1$ degeneracy. 

A sector which has been analyzed in detail and will be of great importance for our present analysis is the pseudoscalar/scalar nonet one, namely, 
\begin{eqnarray}
P^a &\rightarrow&I=0: \eta_l=i\bar\psi_l\gamma_5 \psi_l, \ \eta_s=i\bar s \gamma_5 s, \quad  I=1: \pi^a=i\bar\psi_l\gamma_5\tau^a\psi_l \ (a=1,2,3),
 \nonumber \\
&&I=1/2: K^a=i\bar\psi  \gamma_5 \lambda^a \psi \ (a=4,5,6,7),
 \label{Pop}\\
S^a &\rightarrow& I=0: \sigma_l=\bar\psi_l \psi_l, \ \sigma_s=\bar s  s, \quad I=1: \delta^a=\bar\psi_l \tau^a \psi_l \ (a=1,2,3),
 \nonumber \\
 && I=1/2: \kappa^a=i\bar\psi  \lambda^a  \psi \ (a=4,5,6,7).
 \label{Sop}
\end{eqnarray}
with $\psi_l^T=(u,d)$, $\psi^T=(u,d,s)$. The~above states correspond  to  the quantum numbers of the pion ($\pi^a$), $a_0(980) (\delta^a)$, light ($\eta_l$) and strange ($\eta_s$) component of the $\eta/\eta'(958)$, light ($\sigma_l$)and strange ($\sigma_s$) components of the $f_0(500)/f_0(980)$, kaon ($K^a$) and $K(800)$ or $\kappa$ ($\kappa^a$). For the isospin $I=0,1$ sectors, chiral and $U(1)_A$ transformations connect the bilinears as
\begin{eqnarray}
\pi^a\,&\xleftrightarrow{SU(2)_A}&\sigma, \quad \delta^a\xleftrightarrow{SU(2)_A}\eta_l, \label{chideg} \\ \pi^a&\xleftrightarrow{U(1)_A}& \delta^a,  \quad \sigma\xleftrightarrow{U(1)_A}\eta_l. 
\label{ua1deg}
\end{eqnarray}

The~$U(1)_A$ symmetry restoration at finite temperature including implications for partner degeneration has been actually studied in different theoretical works. In~\cite{Shuryak:1993ee,Cohen:1996ng}  it was already proposed, through analysis of the spectral properties of the quark propagator and related ideas, that $U(1)_A$ partners can degenerate in an ideal chiral restoring scenario. In~the massive case,  gauge configurations  responsible for a residual 
 $U(1)_A$ breaking are analyzed in \cite{Lee:1996zy}. Further analyses have been carried out within the  LSM famework \cite{Meggiolaro:2013swa}, the~renormalization group \cite{Heller:2015box} and the NJL model~\cite{Ishii:2015ira}.  At high enough temperatures, the~degeneration of  additional partners connected not only by chiral and $U(1)_A$ transformations has been proposed as an indication of restoration of the so-called chiralspin symmetry of QCD \cite{Glozman:2018jkb}.

 A complementary recent  line of research, which we will discuss in more detail in Section  \ref{sec:ward} is the use of Ward Identities (WIs) \cite{Nicola:2016jlj,Azcoiti:2016zbi,GomezNicola:2017bhm} to establish useful conclusions regarding partner degeneration in terms of the susceptibilities associated to the relevant quark bilinears, namely,  
\begin{eqnarray}\label{eq:chip-def}
  \chi_P^{ab}(T)&=&\int_T dx  \langle  P^a (x) P^b (0)\rangle,\\
  \tilde\chi_S^{ab}(T)&=&\int_T dx  \left[\langle S^a (x) S^b (0)\rangle-\mean{S^a}\mean{S^b}\right],\label{eq:chis-def}
\end{eqnarray}
where $P$ ($S$) denotes a pseudoscalar (scalar) operator, similarly to the definition \eqref{susdef}, where $\chi_S(T)=\tilde \chi^{ll} (T)$. A full analysis within the ChPT framework, which requires the use of the $U(3)$ extension to incorporate properly the $\eta'$ state and its properties,  has been developed in \cite{Nicola:2016jlj,Nicola:2018vug} including the ChPT realization of  those WIs. We~will present the results of that analysis in Section \ref{sec:u3chpt}. 

Most of the above theoretical analyses, although not conclusive, are compatible with $U(1)_A$ restoration around $T_c$ in the light chiral limit $m_{ud}\rightarrow 0$ while the situation is not so clear in the physical massive case with respect to whether the $U(1)_A$ breaking strength around the transition is enough to generate observable effects. 

Similar conclusions are reached in lattice analysis:  for $N_f=2+1$ flavors and physical quark masses, the~analysis of~\cite{Buchoff:2013nra} shows degeneracy of $U(1)_A$ partners well above the $O(4)$ ones. On~the other hand, $N_f=2$ works~\cite{Aoki:2012yj,Cossu:2013uua,Tomiya:2016jwr,Brandt:2016daq} point to $U(1)_A$ restoration at $T_c$ in the chiral limit, while for massive quarks in those works the strength of $U(1)_A$ breaking increases with the volume  \cite{Brandt:2019ksy}. In~\cite{Lee:2017uvl}   $U(1)_A$ restoration above $T_c$ is obtained  for two flavors and two colors in terms of screening masses. The~effect of $U(1)_A$ restoration including baryon matter has been also investigated in the lattice\mbox{ in \cite{Chandrasekharan:2010ik}} where the phase diagram and the critical point are studied. Degeneration of chiral partners in the lattice for the nucleon sector has been considered  in \cite{Aarts:2015mma,Aarts:2017rrl}.

\subsection{Ward Identities, $O(4)$ vs. $U(1)_A$ Restoration and Screening Masses}
\label{sec:ward}

The~use of WIs to establish symmetry constraints on the theory is a powerful and model-independent tool, which was first suggested to play a role regarding chiral restoration\mbox{ in \cite{Bochicchio:1985xa,Boucaud:2009kv}} within the lattice context. Recently, a full program for the study of  relevant WIs for pseudoscalar and scalar operators, including the $U(1)_A$ anomaly operator, and their properties has been carried out in \cite{Nicola:2016jlj,GomezNicola:2017bhm,Nicola:2018vug}. Those identities are obtained generically by taking suitable axial and vector transformations over pseudoscalar and scalar QCD bilinears. Thus, one gets the following generic WI for expectation values involving pseudoscalar and scalar operators $O_P$ and $O_S$:
\begin{myequation}
\left\langle\frac{\delta\mathcal{O_P}(x_1,\cdots,x_n)}{\delta \alpha_A^a(x)}\right\rangle=
-\left\langle\mathcal{O_P}(x_1,\cdots,x_n)\bar\psi(x)\left\{\frac{\lambda^a}{2},\mathcal{M}\right\}\gamma_5\psi(x)\right\rangle 
+i\frac{\delta_{a0}}{\sqrt{6}}\left\langle \mathcal{O_P}(x_1,\cdots,x_n) A(x)\right\rangle,
\label{wigenP}
\end{myequation}
\begin{align}
&\left\langle\frac{\delta\mathcal{O_S}(x_1,\cdots,x_n)}{\delta \alpha_V^a(x)}\right\rangle=\left\langle\mathcal{O_S}(x_1,\cdots,x_n)\bar\psi(x)\left[\frac{\lambda^a}{2},\mathcal{M}\right]\psi(x)\right\rangle.
\label{wigenS}
\end{align}
where 
\begin{equation}
  A(x)=\frac{3g^2}{32\pi^2}G_{\mu\nu}^a\tilde G^{\mu\nu}_a,
  \label{anomaly}
\end{equation}
is the anomaly (pseudoscalar) operator entering the anomalous divergence of the $U(1)_A$ current, with the gluon fields $G_\mu^a$, $G_{\mu\nu}^a=\partial_\mu G_\nu^a-\partial_\nu G_\mu^a-gf_{abc}G_\mu^b G_\nu^c$ and $\tilde G^a_{\mu\nu}=\epsilon_{\mu\nu\alpha\beta}G^{\alpha\beta, a}$ the dual gluon tensor. 

Thus, setting in the above equations $n$-point functions for the $O_{P,S}$ operators, the~WI relates $n$ and $n+1$ operators. In~particular, replacing for  $O_P\rightarrow P^a$ in \eqref{Pop}  and the $A$ operator in \eqref{anomaly}, as well as   $O_S\rightarrow S^a$ in \eqref{Sop}, we get the more relevant set of identities for our present discussion:
\begin{eqnarray}
\chi_P^\pi(T)&=&-\frac{\condl(T)}{\hat m},
\label{wipi}\\ 
\chi_P^K(T)&=&-\frac{\condl (T)+2\conds (T)}{\hat m + m_s},
\label{wiK}\\
\chi_P^{ ss}(T)&=&-\frac{\conds(T)}{m_s}-\frac{1}{m_s^2}\chi_{top}(T),
\label{wiss}\\
\chi_S^\kappa (T)&=&\frac{\condl (T)-2\conds (T)}{m_s-\hat m},
  \label{wikappa} \\
\chi_P^{ls}(T)&=&-2\frac{\hat m}{m_s} \chi_{5,disc}(T)=-\frac{2}{\hat m m_s}\chi_{top}(T),
\label{wils5}
\end{eqnarray}
where 
$\chi_{5,disc}(T)=\frac{1}{4}\left[\chi_P^\pi(T)-\chi_P^{ll}(T)\right]$ is one of the possible parameters measuring $O(4)\times U(1)_A$ restoration, since it should vanish if both symmetries are realized,  according to \eqref{chideg} and \eqref{ua1deg}, and $\chi_{top}$ is the topological susceptibility
\begin{equation}
\chi_{top}(T)\equiv -\frac{1}{36}\chi_P^{AA}(T)=-\frac{1}{36}\int_T dx  \langle A(x) A(0) \rangle.
\label{chitopdef}
\end{equation} 
which will be dealt with in detail in Section \ref{sec:topsus}. 

The~identity \eqref{wipi} is the one obtained first in \cite{Bochicchio:1985xa,Boucaud:2009kv} and it was used in \cite{Nicola:2013vma} to check the $\pi$-$\sigma$ degeneration at chiral restoration straight from lattice data on the light quark condensate and the scalar susceptibility. The~validity of that identity, as well as that of \eqref{wiss},  have been recently checked in the lattice \cite{Buchoff:2013nra}  correcting for finite-size effects. We~will discuss below an important application of the identities \eqref{wipi}--\eqref{wikappa} in terms of lattice screening masses. Before that, let us examine the consequences of those identities for our previous discussion about $U(1)_A$ and chiral restoration. In~particular, consider the identity \eqref{wils5} and the following  $SU(2)_A$ transformation acting on the $\eta_l$ bilinear:
\begin{equation}
\eta_l(x)\rightarrow i\bar\psi_l(x)\gamma_5 e^{i\gamma_5 \alpha_a\tau^a}\psi_l(x)=i\bar\psi_l(x)\gamma_5 \cos (\alpha_a\tau^a)\psi_l(x)-\bar\psi_l(x) \sin (\alpha_a\tau^a)\psi_l(x),
\label{chilsvanishing1}
\end{equation}
with
\begin{equation}
\alpha_b=\pi/2 \quad{\rm and}\quad \alpha_{a\neq b}=0 \qquad 
\label{trans}
\end{equation}
so that 
\begin{equation}
\eta_l(x)\rightarrow -\bar\psi_l(x) \tau^b \psi_l(x)=-\delta^b (x)\Rightarrow \chi_P^{ls} \rightarrow -\int_T dx \mean {\delta^b (x) \eta_s(0)} =0,
\label{chilsvanishing2}
\end{equation}
 where we have used that $\eta_s$  is invariant under $SU(2)_A$ transformations and the last correlator vanishes by parity invariance of the vacuum, being a product of $P$-even and $P$-odd operators.

Therefore, from \eqref{wils5}, the~conclusion is that for exact chiral restoration, i.e., when the axial $SU(2)_A$ symmetry is realized in the particle spectrum,   $\chi_{ls}$ would vanish according to \eqref{chilsvanishing2}. Then, the~WI~\eqref{wils5} implies that  $\chi_{5,disc}$ should vanish as well so that the $O(4)\times U(1)_A$ pattern is realized. Of course, the~regime of ideal $SU(2)_A$ symmetry is never realized for physical quark masses, but it should be the case in the chiral limit for two massless flavors at $T_c$, consistently with the lattice results in ~\cite{Aoki:2012yj,Cossu:2013uua,Tomiya:2016jwr,Brandt:2016daq,Brandt:2019ksy}.  For $N_f=2+1$ flavors and physical masses, the~strangeness contribution and the large uncertainties for $\delta-\eta_l$ degeneration~\cite{Buchoff:2013nra} might explain a stronger $U(1)_A$ breaking, consistently also with the chiral limit analysis of that collaboration \cite{Bazavov:2018mes}.

Very interesting consequences for chiral and $U(1)_A$ restoration can also be extracted from the previous WI in the $I=1/2$ sector, which has been much less explored by lattice collaborations. Actually,  the difference between \eqref{wiK} and \eqref{wikappa}  yields
\begin{equation}
  \chi_S^\kappa (T)-\chi_P^K (T)=\frac{2}{m_s^2-\hat m^2}\left[m_s\condl (T)-2\hat m \conds (T)\right],
  \label{wichikappakaondif}
\end{equation}

The~interest of the above equation is the following: first, it proves that in the chiral  limit and at exact chiral restoration, i.e., where both $\hat m$ and $\condl$ vanish, the~kaon and the $\kappa$ states degenerate and then become chiral partners. Interestingly, they can be showed also to degenerate under $U(1)_A$, by choosing again a suitable transformation  \cite{GomezNicola:2017bhm}. On the other hand, the~light and strange combination showing up in \eqref{wichikappakaondif} is precisely the so-called subtracted quark condensate $\Delta_{l,s}$ thoroughly analyzed in lattice collaborations as the order parameter of the chiral transition, defined in that way to avoid finite-size divergences  of the form $\mean{\bar q_i q_i}\sim m_i/a^2$, with $a$ the lattice spacing \cite{Bazavov:2011nk,Buchoff:2013nra,Ratti:2018ksb}.  Therefore, \eqref{wichikappakaondif} allows us to describe through a measurable lattice quantity the relative strength of $O(4)\times U(1)_A$ breaking near $T_c$. All the above results for the $I=1/2$ channel provide very useful predictions to be confronted by lattice data, which so far are not available  in this channel for susceptibilities, only for screening masses as we are about to discuss below.

We~will finish this section by discussing another useful application of the previous WI when trying to make contact with lattice analyses. Among the properties that can be measured in the lattice for individual channels, i.e., physical states, is the variation of the screening masses with the medium parameters (essentially temperature). Such screening masses are nothing but the measure of the spatial falloff of euclidean propagators $K\sim \exp(-M^{sc}\vert z\vert)$, corresponding to taking the limit  $K(\omega=0,\vec{p}\rightarrow \vec{0})$ in Fourier space. On the other hand, the~susceptibilities in the same channel correspond to the inverse of squared pole masses, typically $\chi_P(T)\sim \left[M^{pole}\right]^{-2}$  as we have seen for instance in Section \ref{sec:f0500} regarding the saturated behaviour of the scalar susceptibility, where $K$ can be generically parametrized as $K^{-1}(\omega,\vec{p})\sim -\omega^2+A^2(T)\vert\vec{p}\vert^2+M^{pole}(T)^2$  around $p=0$ with with $A(T)=M^{pole}(T)/M^{sc}(T)$~\cite{Karsch:2003jg}. The~difference between screening and pole masses parametrized by $A(T)$ stems from the different spatial and temporal dependence of  self-energies  in the thermal bath and is  expected to change smoothly, at least below the transition   \cite{Schenk:1993ru,Ishii:2016dln}.  Assuming such smooth dependence also for possible residue contributions of the corresponding correlators leads to the following proposal of dominant scaling with temperature for screening masses,
\begin{equation}
\frac{M^{sc}_i(T)}{M^{sc}_i(0)} \sim \left[ \frac{\chi_i(0)}{\chi_i (T)} \right]^{-1/2} 
\label{susgenscaling}
\end{equation} 
for the different meson channels $i=\pi,K, \dots$.

Now, combining \eqref{susgenscaling} with the WIs \eqref{wipi}--\eqref{wikappa}, one gets a prediction for the temperature scaling of screening masses in terms of that of different combinations of light $\condl$ and strange $\conds$ quark condensates. In~the following, for simplicity, we neglect the second contribution in the right-hand side of \eqref{wiss} which is suppressed numerically by a $\hat m/m_s$ factor consistently with the lattice verification of that WI \cite{Buchoff:2013nra,Nicola:2016jlj}. 

\begin{figure}[h]
\centering
\includegraphics[width=10 cm]{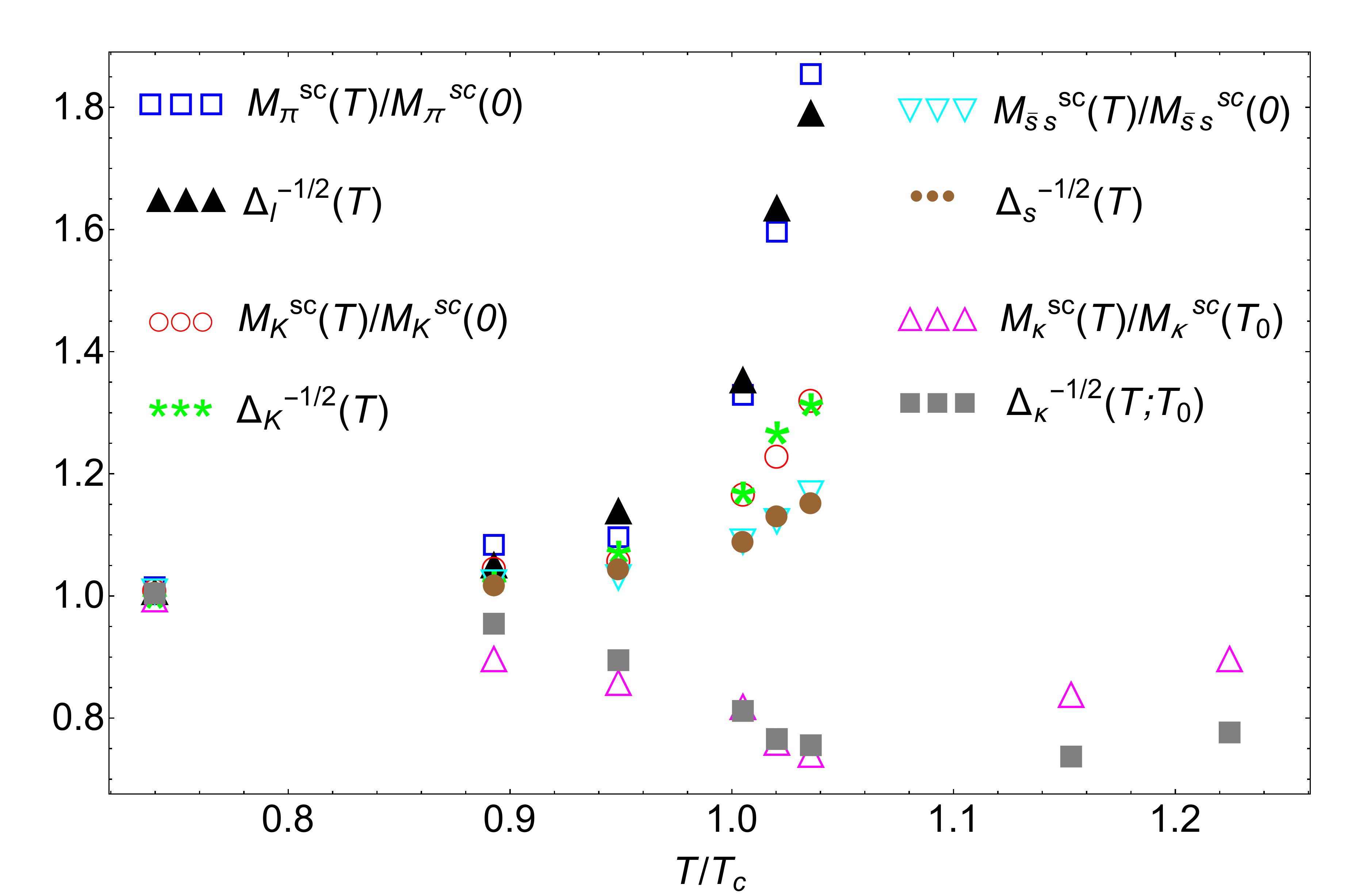}
\caption{Comparison of lattice data for screening masses with those for the combinations of quark condensates predicted by the scaling law \eqref{susgenscaling} and the WIs \eqref{wipi}--\eqref{wikappa}, from \cite{Nicola:2018vug}. The~lattice data are taken from 
\cite{Cheng:2010fe} (masses) and \cite{Cheng:2007jq} (condensates) with the same lattice action and resolution and $T_0=145$ MeV. the definition of the subtracted condensates $\Delta_i$ can be found in \cite{Nicola:2016jlj,Nicola:2018vug}.}
\label{fig:sm} 
\end{figure}

In~order to establish a measurable comparison with lattice data, one must  translate that prediction into properly subtracted lattice quark condensates $\Delta_i$  \cite{Nicola:2016jlj,Nicola:2018vug}. This allows for two fitting parameters related to those subtraction. In~Figure \ref{fig:sm} we show the result of such comparison, using the lattice data for screening masses in \cite{Cheng:2010fe}, which is the most complete screening mass analysis so far where all the channels of interest here are calculated within the same lattice setup. Consequently, we have used the quark condensate data in \cite{Cheng:2007jq} using that very same setup. Nevertheless, using more recent screening masses data, like those in \cite{Maezawa:2013nxa}, the~conclusions are similar, as discussed in \cite{Nicola:2016jlj}.  The~fitted parameters corresponding to the results in Figure \ref{fig:sm} are obtained by minimizing the sum of squared differences for the different channels, below the transition, and are given in  \cite{Nicola:2016jlj,Nicola:2018vug}. 

\vspace{-9pt}

Thus, the~main conclusion in view of the results in Figure \ref{fig:sm} is that the combination of the assumed scaling law \eqref{susgenscaling} and the WIs \eqref{wipi}--\eqref{wikappa} is indeed a very good approximation to describe the temperature evolution of screening masses below and up to the transition. Apart from the quantitative agreement showed in that figure, there is also a very important qualitative consequence regarding the comparative behaviour of the different channels from their dependence on the different quark condensate combinations from the WI-based scaling. Namely, the~stronger increase of $M_\pi^{sc}(T)$ can be understood from its dependence on the light quark condensate in \eqref{wipi}, since $\condl$ is meant to decrease strongly at the transition so its (subtracted) inverse would grow strongly as well. Similarly, the~softer behaviour of $M_K^{sc}(T)$ would be explained by its  $\conds$ contribution, which is supposed to present a much soft temperature decreasing behaviour around $T_c$ \cite{Bazavov:2011nk,Jankowski:2012ms}. Therefore, the~inverse  of the combination in \eqref{wiK} would still increase but at a softer rate than the pion channel. Likewise, the~$\bar s s$ channel, according to  \eqref{wiss}, is only dependent on $\conds$ once the second term is neglected, which again would justify the even softer growing behavious of $M_{\bar ss}^{sc}(T)$. The~case of the $\kappa$ channel is even more interesting.  The~combination of light and strange condensates in \eqref{wikappa} involves its difference rather than its sum. Thus, below $T_c$, the~decreasing behaviour of $\condl(T)$ while $\conds(T)$ remains almost constant is the dominant effect which makes their difference larger and then the inverse of such difference decreases with $T$. However, after the inflection point of $\condl(T)$ around $T_c$, it remains softly changing while the decreasing behaviour of $\conds(T)$ starts taking over, which explains the minimum around $T_c$ of $M_\kappa^{sc}(T)$. 

\subsection{$U(3)$ ChPT Analysis of the Scalar and Pseudoscalar Nonet} 
\label{sec:u3chpt}

In~the previous sections we have discussed the utility of the WIs regarding different aspects of the chiral transition. In~particular, the~connection with $U(1)_A$ restoration  and the qualitative understanding of the screening mass behaviour. Those are general results independent of the particular physical realization of those WIs in terms of the particle spectrum. However, in many instances one needs to know in addition the particular temperature evolution of, say, the~different susceptibilities involved in those identities, which allows us to extract further useful conclusions and asks for a particular theoretical framework describing the relevant degrees of freedom. 

A  recent  calculation in this direction has been performed within the $U(3)$ ChPT formalism, up to NLO in the combined $1/N_c$ and chiral expansion, for the quark condensates and all the susceptibilities of the scalar/pseudoscalar nonets \cite{Nicola:2016jlj,Nicola:2018vug}, as well as the topological susceptibility and the fourth-order cumulant of the topological charge distribution \cite{Nicola:2019ohb}. The~latter will be discussed in more detail in Section \ref{sec:topsus}. The~role of the $\eta'$ regarding $O(4)$ vs. $U(1)_A$ restoration has become clear from the previous sections and therefore a proper $U(3)$ ChPT  analysis is mandatory to explore those issues properly and to understand the meson realization of the different correlators involved.

\begin{figure}[h]
\centering
\includegraphics[width=7.5 cm]{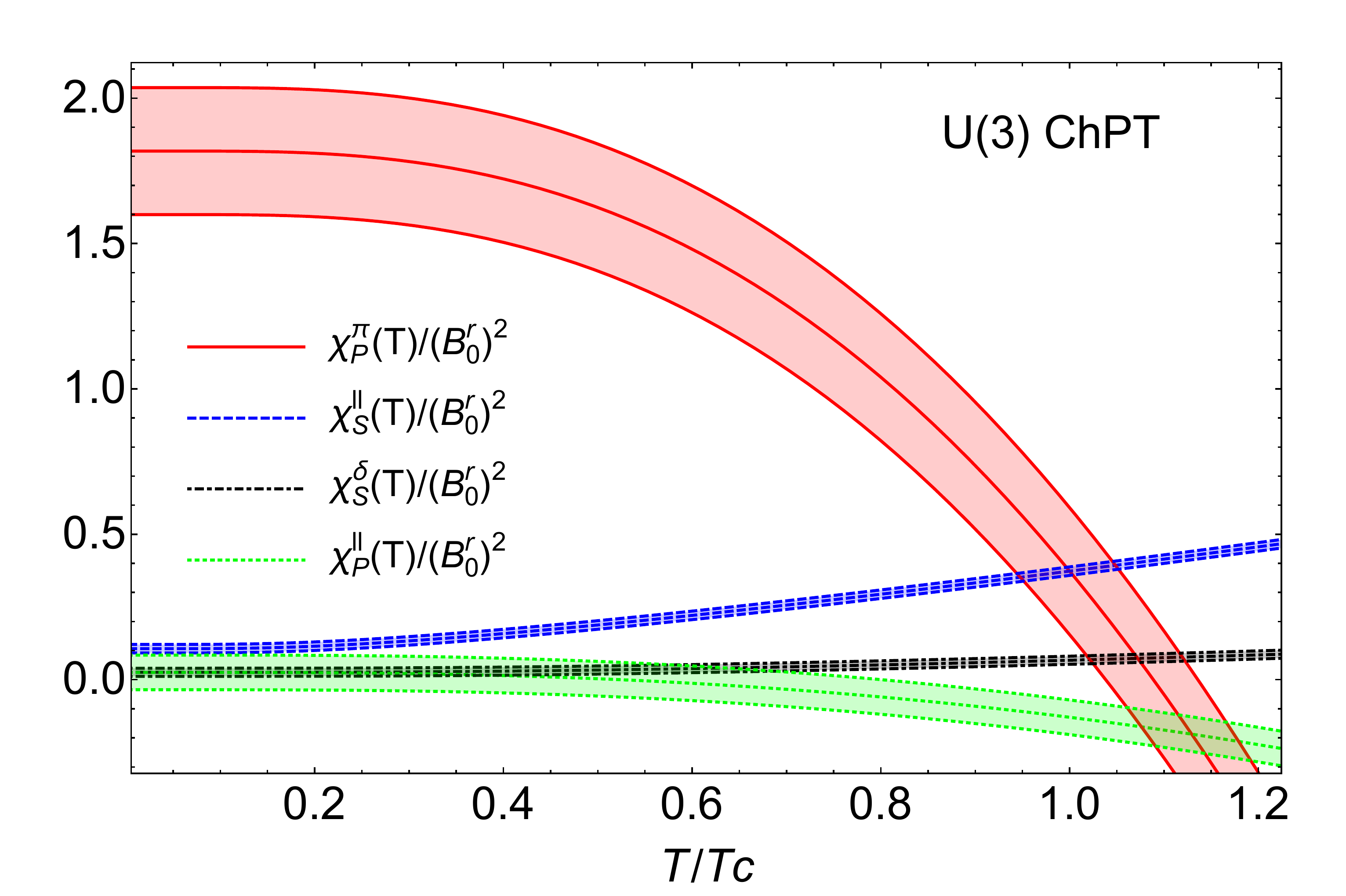}
\includegraphics[width=7.5 cm]{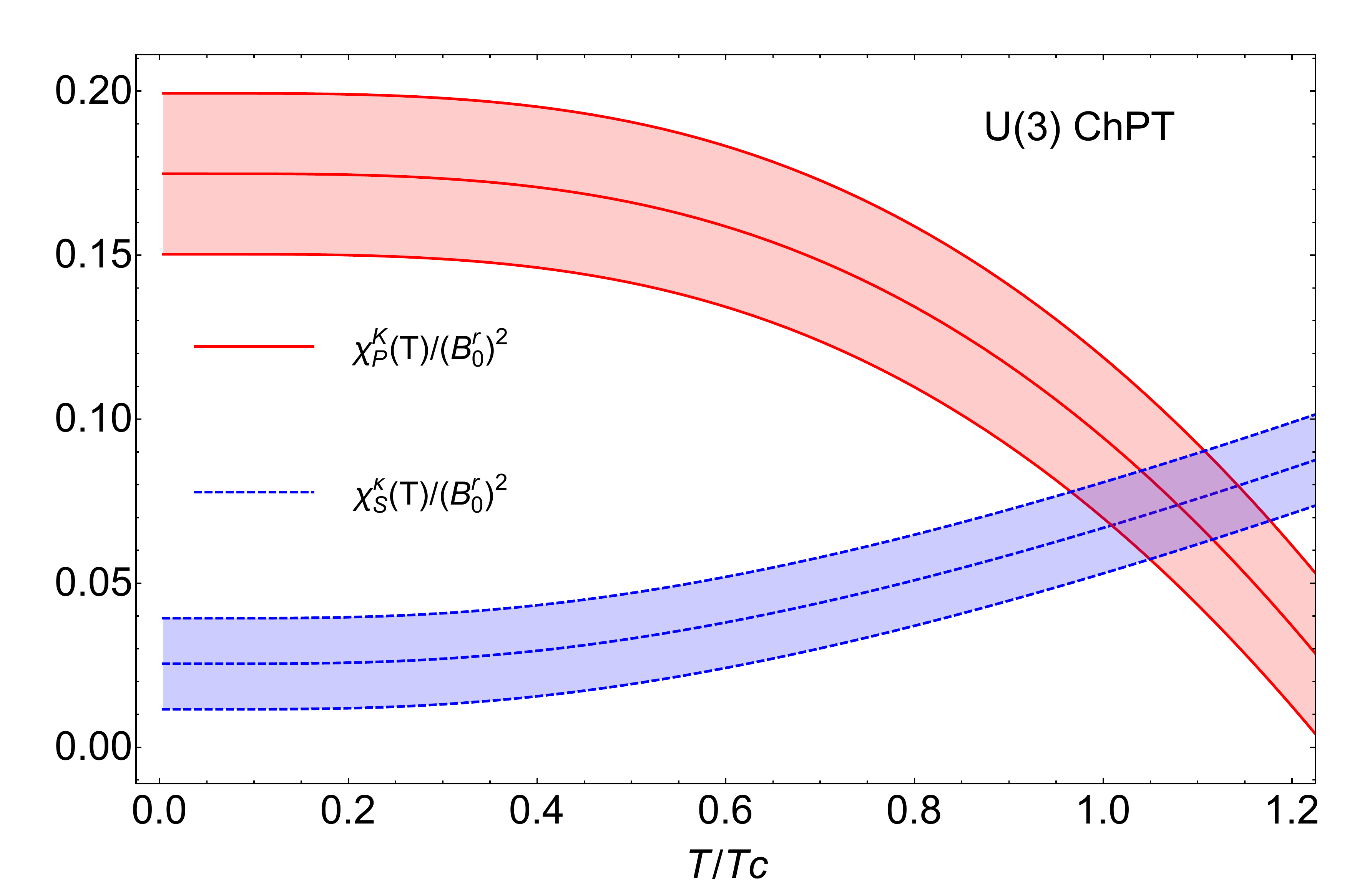}
\caption{$U(3)$ ChPT results from \cite{Nicola:2018vug} where the bands correspond to the LEC uncertainties given in~\cite{Guo:2015xva}. (\textbf{Left}): $I=0,1$ susceptibilities for  the $\pi^a$, $\delta^a$, $\sigma_l$, $\eta_l$ states in \eqref{Pop}, \eqref{Sop}, where $B_0^r$ is the renormalized chiral $B_0$ parameter. (\textbf{Right}): $I=1/2$ susceptibilities for $K$ and $\kappa$   states.}
\label{fig:u3sus} 
\end{figure}

\begin{figure}[h]
\centering
\includegraphics[width=10cm]{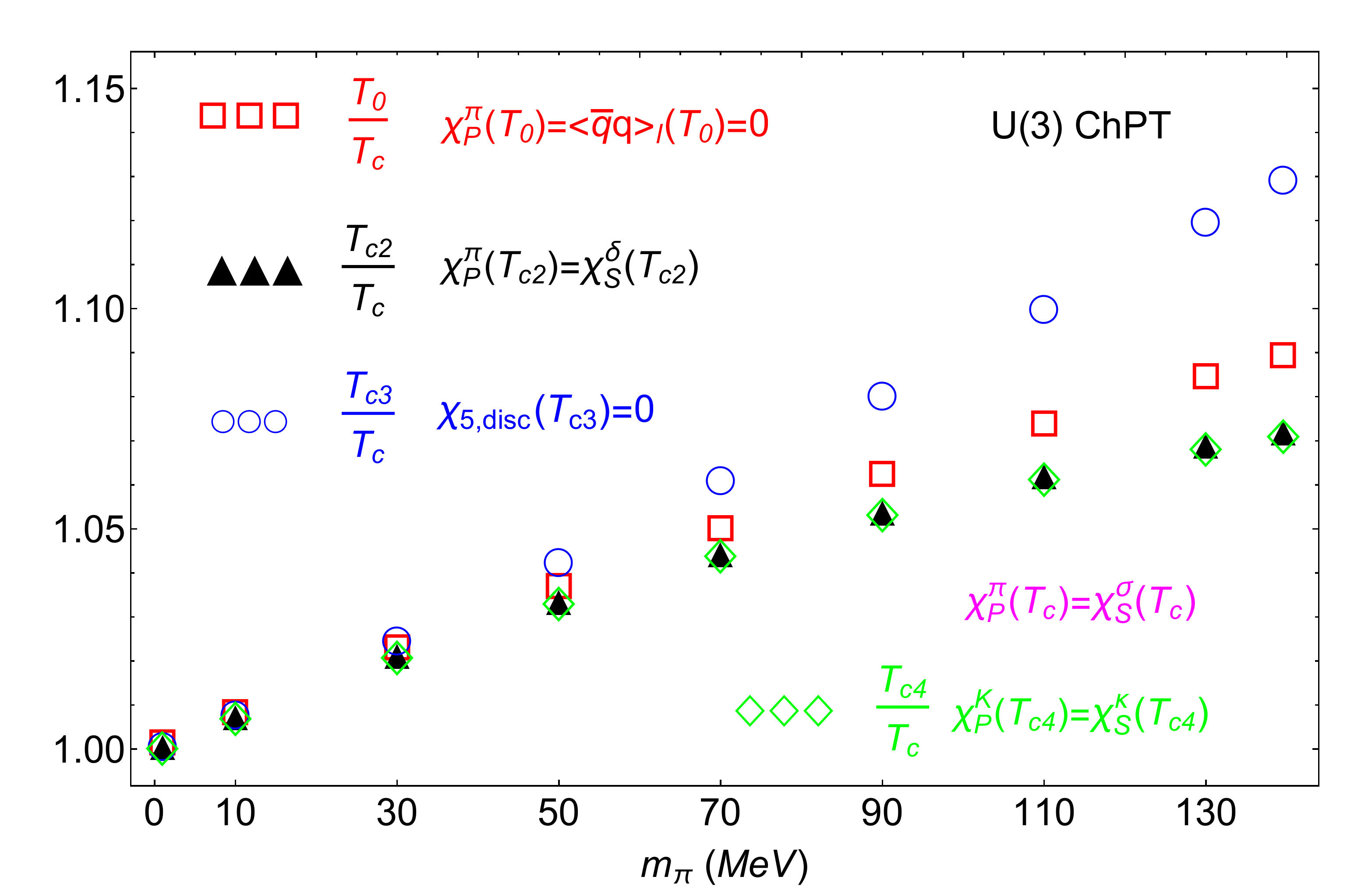}
\caption{$U(3)$ ChPT result for the chiral limit behaviour of different chiral and $U(1)_A$ restoration temperatures,  from \cite{Nicola:2018vug}.}
\label{fig:u3chiral} 
\end{figure}

In~fact, the~$U(3)$ ChPT analysis carried out in \cite{Nicola:2016jlj,Nicola:2018vug} has allowed us to verify explicitly in that formalism the WIs discussed in Section \ref{sec:ward}. As for the nonet susceptibilites, the~main results of that calculation  is shown in Figures \ref{fig:u3sus} and \ref{fig:u3chiral}.

\vspace{-9pt}

The~susceptibilities shown in Figure \ref{fig:u3sus} show the restoration pattern expected for massive quarks. Thus, setting $T_c$ as the chiral partner matching $\chi_P^\pi (T_c)=\chi_S^{ll} (T_c)$ in ChPT (although it does not reproduce a true degeneration above that temperature) the matching temperature $\chi_P^\pi (T_{c2})=\chi_S^{\delta} (T_{c2})$, which would correspond to $U(1)_A$ restoration is  given by $T_{c2}\simeq 1.07\,T_c$ for the central values of the LEC used and given in   \cite{Guo:2015xva}. Therefore, such $U(1)_A$ restoration effect takes place only slightly above $T_c$ and within the LEC uncertainty range.  Besides, the~difference between the $\pi-\eta_l$ indicating  $O(4)\times U(1)_A$  restoration, as commented above, through $\chi_{5,disc}(T_{c3})=0$ gives $T_{c3}\simeq 1.13 T_c$, again within the same range. Finally, in the  $I=1/2$ sectors the partner susceptibilities match at   $\chi_P^K(T_{c4})=\chi_S^\kappa(T_{c4})$ with $T_{c4}\simeq T_{c2}$. Recall that in the massive case, $K$-$\kappa$ degeneration would require $U(1)_A$ restoration as well, as discussed in Section \ref{sec:ward}. The~conclusion is then that ChPT would be consistent with $U(1)_A$ restoration, in terms of partner degeneration, not much above from chiral restoration, typically around 10\%. However, one must keep in mind the usual caveats of the ChPT framework regarding the critical region, since its applicability range is limited to low and moderate temperatures. Recall  also that in the physical mass case it is consistent that   different partner degeneration temperatures take different values, since we do not expect a sharp chiral transition. 

\vspace{-9pt}

More insight can be gained by taking the chiral limit, where ChPT is meant to provide more robust predictions. Thus, in Figure \ref{fig:u3chiral} we show the evolution of the different partner matching temperatures, as well as the vanishing condensate one $T_0$. The~results show clearly that all those temperatures converge to the same value, which is consistent with the $O(4)\times U(1)_A$ pattern  in the chiral limit predicted also by the WI analysis in Section \ref{sec:ward}. Moreover, the~fact that all pseudocritical $T_{ci}$ coincide is also consistent with the transition properties expected in the chiral limit and discussed in Section \ref{sec:intr}. 

The~analytic behaviour of those $T_{ci}$ in the chiral limit is also revealing and can be found in \cite{Nicola:2018vug}. The~gap between the $U(1)_A$ pseudocritical temperatures $T_{c3}$ and $T_{c2}$ is $\Od(m_\pi^2)$, which is also the gap between them and $T_0$. On the other hand, the~gap between $T_0$, $T_{c2}$ or $T_{c3}$ and the $O(4)$ $T_c$ is $\Od(m_\pi)$, i.e., larger in the chiral limit expansion.  This is consistent with our previous claims about the separation of chiral and $U(1)_A$ restoration, both gaps vanishing in the chiral limit. In~addition, the~hierarchy  $T_{c3}>T_0>T_{c2}>T_c$ is maintained for all masses, as seen in Figure \ref{fig:u3chiral}. Note that $T_{c3}>T_{c2}$,  $T_{c2}>T_c$ and $T_0>T_c$ are expected from our previous arguments. For $I=1/2$ , we see from Figure  \ref{fig:u3chiral} that $T_{c4}$ remains almost identical to $T_{c2}$, consistently with our discussion in Section \ref{sec:ward} on kaon-kappa~degeneration.

\subsection{The~Topological Susceptibility}
\label{sec:topsus}

As discussed in Section \ref{sec:ward}, the~topological susceptibility plays a direct role in the discussion about $O(4)$ and $U(1)_A$ restoration. In~particular, from the WI \eqref{wils5} and the discussion in that section, the~vanishing of $\chi_{top}$ signals also $O(4)\times U(1)_A$ restoration, as noted also in \cite{Azcoiti:2016zbi}. Apart from that, $\chi_{top}$ has itself a considerable relevance in different instances and, more importantly for the present work, its analysis through the ChPT formalism, both at $T=0$ and at finite temperature, provides a robust description when compared with lattice results, as we will see below.  

An important context where $\chi_{top}$ plays an important role  is the cosmological one, from the direct relation of $\chi_{top}$ with the axion mass, while the axion self-coupling is directly related to the fourth-order cumulant of the topological charge distribution, whose second-order cumulant is the topological susceptibility \eqref{chitopdef}. Updated theoretical results in that context, including estimates based on ChPT, can be found for instance in \cite{diCortona:2015ldu}.  

 Another important topic, relevant for the present work, where the topological susceptibility has played a crucial role, arises precisely from the WIs discussed in Section \ref{sec:ward}. In~particular, combining~\mbox{\eqref{wipi} and \eqref{wils5}} yields
 \begin{align}
\chi_{top}&=-\frac{1}{4}\left[m_q\condl+m_q^2 \chi_P^{ll}\right] \label{wi1}
\end{align}
 which establishes that near the chiral limit the topological susceptibility is proportional to the light quark condensate. That feature has been actually used extensively to extract the quark condensate in the lattice \cite{Bernard:2012fw,Aoki:2016frl,Burger:2018fvb} and is actually shared by the lowest order chiral Lagrangian calculation, which in $U(3)$ incorporates the $\eta'$ and is given by \cite{Nicola:2019ohb}
  \begin{equation}
\chi_{top}^{U(3),LO}= -\frac{1}{2}\condl^0 \frac{M_0^2 {\bar m}}{M_0^2+6B_0{\bar m}} 
\label{chitoploib}
\end{equation}
with  $\condl^0=-2B_0F^2$  the quark condensate in the chiral limit, $M_0$  the anomalous part of the $\eta'$ mass and $\displaystyle\bar m^{-1}=\sum_{i=u,d,s} m_i^{-1}$. Taking the limit $M_0\rightarrow\infty$ in \eqref{chitoploib} reproduces the $SU(2,3)$ results in \cite{Leutwyler:1992yt}.  The~fact that the topological susceptibility remains proportional to the quark masses through the $\bar m$ combination, vanishing in particular the chiral limit, supports the idea that a description based on ChPT is appropriate for this quantity.  The~situation is similar in some respect to the case of the scalar susceptibility analyzed in Section \ref{sec:f0500} in terms of the lightest $f_0(500)$ state, in the sense that the dominant dependence comes from inverse masses of light states.   Moreover, the~meson loop contributions giving rise to the quark mass dependence in the above expression for $\chi_{top}$ are crucial corrections to the quenched gluodynamics limit, obtained as  $m_i\rightarrow\infty$, where $\chi_{top}\sim M_0^2 F^2/6$ \cite{Witten:1979vv,Veneziano:1979ec}.

The~NLO corrections in $SU(N_f)$ ChPT can be found in \cite{Mao:2009sy,Bernard:2012fw}, while the $U(3)$ calculation up to NLO is provided in \cite{Nicola:2019ohb} including the temperature dependence of $\chi_{top}$.  
  The~dominant contribution comes from the $SU(2)$ analysis, thus confirming the previous claim about the accuracy of the low-energy description of $\chi_{top}$, while the contribution of $\eta'$ loops and $\eta-\eta'$ mixing corrections provided by the $U(3)$ formalism are of the same order as the $K,\eta$ $SU(3)$ ones. The~numerical predictions at $T=0$ are consistent with recent lattice analysis, like that in \cite{Bonati:2015vqz}. The~$U(3)$ calculation provides as an additional advantage a consistent way to analyze the dominant and subdominant contributions of the large-$N_c$ expansion of both the topological susceptibility and the fourth-order cumulant \cite{Nicola:2019ohb,Vonk:2019kwv} which are consistent with  the lattice results provided at different large-$N_c$ values in \cite{Vicari:2008jw}.

\begin{figure}[h]
\centering
\includegraphics[width=7.5cm]{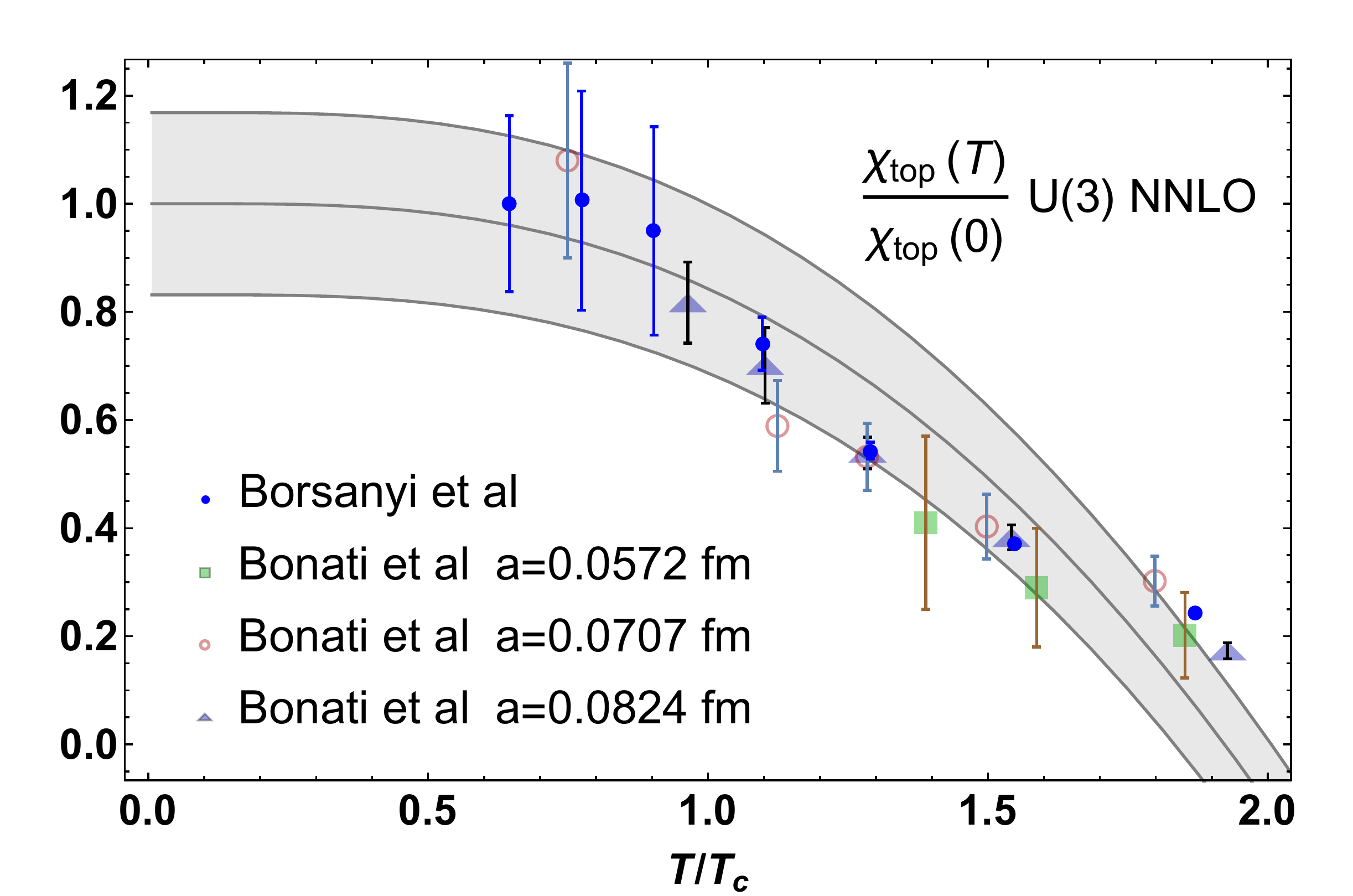}
\includegraphics[width=7.5cm]{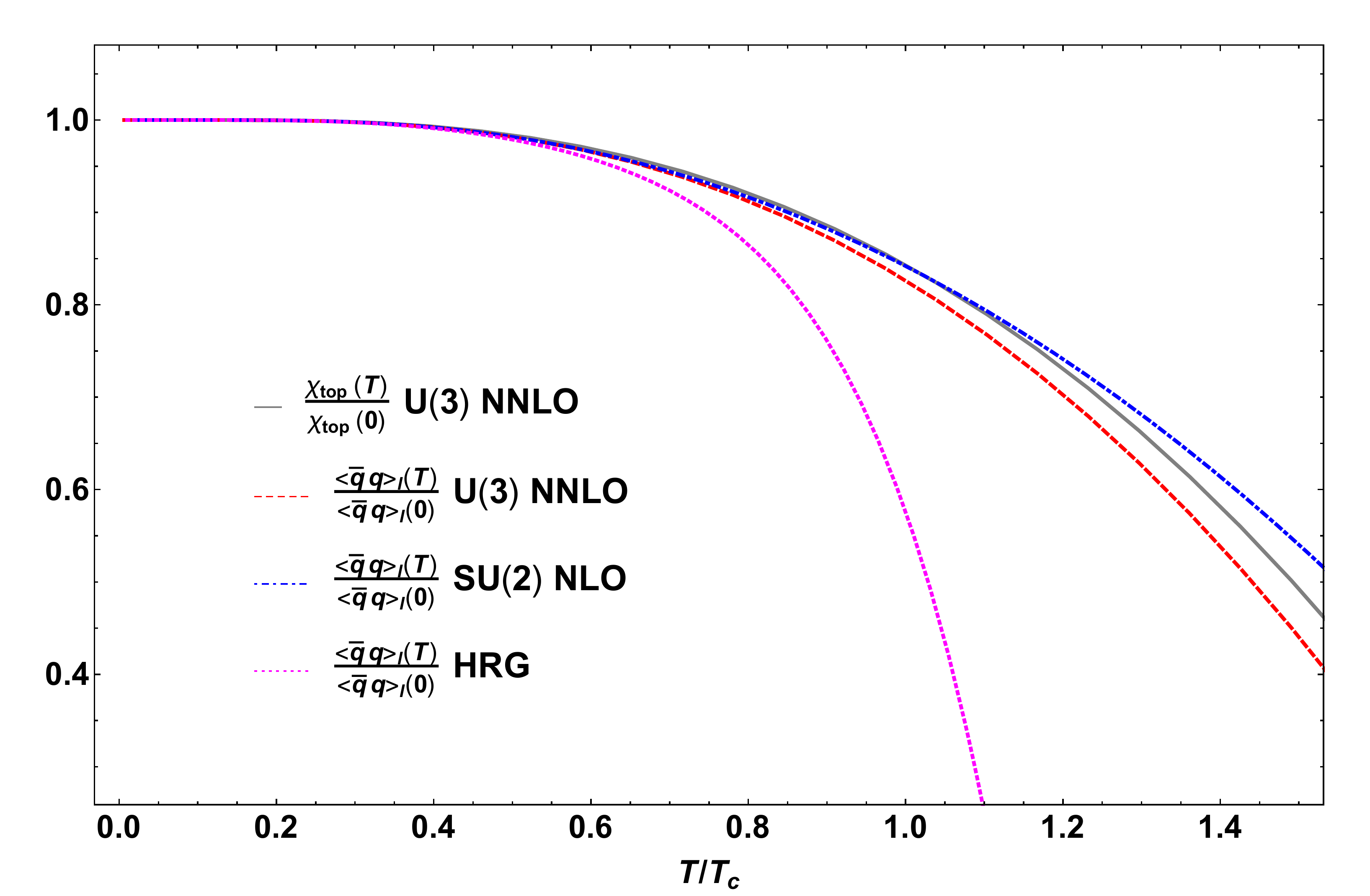}
\caption{$U(3)$ ChPT results for the $T$-dependence of the topological susceptibility \cite{Nicola:2019ohb}. (\textbf{Left}): result of the $U(3)$ calculation compared to the lattice results in \cite{Bonati:2015vqz,Borsanyi:2016ksw}, where the band corresponds to the LEC uncertainty. (\textbf{Right}): comparison of the $U(3)$ $\chi_{top}$ temperature scaling with the quark condensate in different approximations.}
\label{fig:u3topsus} 
\end{figure}

 As for the temperature evolution, the~result of the $U(3)$ analysis including the uncertainty bands coming from the LEC used, is shown in Figure \ref{fig:u3topsus}.
 
 \vspace{-6pt}
 
The~CHPT prediction is remarkably compatible with lattice results, even for temperatures close to the transition, which confirms once again the special characteristics of this observable which makes it suitable for an EFT  description. As mentioned above, the~dominant part comes from the $SU(2)$ contribution, even at finite temperature, and in fact the scaling with $T$ of $\chi_{top}(T)$ remains close to that of $\condl(T)$ calculated to the same order in ChPT. However, an important comment is in order here. While we expect that $\chi_{top}(T)$ does not deviates much from the ChPT curve, that is clearly not the case for the quark condensate, where, as discussed  in Section \ref{sec:effthe} we do expect sizable differences when all hadron states are taken into account. To highlight this fact, we have represented  in Figure \ref{fig:u3topsus} the quark condensate calculated in the HRG description as described also in Section  \ref{sec:effthe}. This  question on whether the topological susceptibility follows the quark condensate evolution for all temperatures  is intimately connected with the importance of the second term in the WI \eqref{wi1} and actually with our discussion about chiral and $U(1)_A$ restoration in terms of WI performed in Section \ref{sec:ward}. Thus, as we have seen in that section, the~vanishing of $\chi_{top}$ signals $O(4)\times U(1)_A$ restoration since it is proportional to $\chi_{5,disc}$ according to \eqref{wils5}. Therefore, the~second term in \eqref{wi1}, or in other words the deviations of $\chi_{top}$ from the quark condensate scaling, measure the strength of $U(1)_A$ breaking around the transition in the physical limit of massive quarks. The~topological susceptibility and its thermal dependence is then a suitable observable to connect the various ideas around chiral restoration and its nature that we have been discussing in this work.

\section{Conclusions and Discussion}
\label{sec:conc}

Over recent years our knowledge on the QCD phase diagram has considerably improved. The~physical phenomena of deconfinement and chiral symmetry restoration, predicted theoretically long time ago are nowadays realized both in experimental programs such as those carried out in Heavy-Ion Collisions  and in lattice simulations. The~precision achieved in those two fields has allowed us to test many theoretical developments and explore the main features of the transition, including the analysis of the  critical curve in the $(T,\mu_B)$ plane, the~possible existence of a critical point and  the nature of the transition in terms of its universality class. Those are however still open problems  and they concentrate  the main efforts of current theoretical, experimental and lattice research in this field. 

Effective theories have played recently a very important role regarding chiral symmetry restoration. In~this review, we have offered a current perspective of some of the main aspects for which this technique can be useful.  After discussing general aspects of this framework, we have concentrated on some particularly relevant problems as far as chiral symmetry restoration is concerned. Namely, the~importance of in-medium effects for resonances and the nature of the transition from the point of view of partner degeneration.

The~basic principle behind effective theories is to incorporate the appropriate degrees of freedom required for particular in-medium observables of interest, relying on the QCD chiral symmetry breaking pattern as a guiding principle to construct their properties and interactions. Thus, for many thermodynamical observables signaling the transition, such as the quark condensate or the pressure, accounting for all the possible hadron states  that can be excited, within the Hadron Resonance Gas framework, is crucial. That line of approach has proved very successful for many observables and corrections due to interactions and lattice masses have even improved the predictions. 

However, there are important issues for which that description is not the more adequate one. In~certain cases of interest, it is preferable to provide an accurate treatment of the lightest degrees of freedom and their interactions rather than to incorporate heavier states. In~that sense, it is sometimes crucial to account correctly for the spectral modifications of hadron states, including resonances, within the thermal  environment. That is the case for instance of the relevant states and interactions needed to describe correctly the hadron contributions to the direct photon and dilepton spectrum observed in heavy-ion collisions. Other prominent examples are those described in detail in this work and pertaining the scalar or the topological susceptibilities, for which an analysis based on Chiral Perturbation Theory complemented with additional physical requirements such as unitarity, allow us to achieve an accurate description of lattice results.  Thus, the~main expected properties of the scalar susceptibility, like its peak around the chiral transition, can be reproduced through a saturation approach by the thermal $f_0(500)$ state,  which we have described trough the pole generated by Unitarized Chiral Perturbation Theory at finite temperature in the $I=J=0$ channel. Likewise, the~topological susceptibility, which carries important information about chiral and $U(1)_A$ restoration, can be accurately described through a ChPT approach, both at zero and finite temperature, which eventually can be understood from its dependence with inverse quark masses. 

The~problem of the transition nature is also one of the main goals in the current research on this field and is intimately connected to the strength of $U(1)_A$ restoration around the QCD transition. Our~understanding of this problem is clearly improving over recent years, although lattice and theoretical analysis are not still fully conclusive. We~have reviewed here in detail  recent progress made on this particular issue through the use of Ward Identities combined with the $U(3)$ ChPT framework. Within that approach, it can be proved that in the limit of exact chiral restoration, for instance for two massless flavors, the~$U(1)_A$ symmetry is also restored in terms of chiral partner degeneration for the scalar/pseudoscalar nonets. The~use of these techniques is actually very powerful and give rise to other physically relevant predictions. Thus, in the $I=1/2$ sector where the lightest chiral partners are the kaon and kappa states, Ward Identities establish a quantitative way to measure the strength of $U(1)_A$ breaking in terms of a subtracted light quark condensate, measured in the lattice. Actually, this analysis motivates future lattice calculations for $I=1/2$. Another important application that we have discussed here is the temperature dependence of lattice screening masses, for which a WI-based scaling  provides a good quantitative and qualitative description. A detailed $U(3)$ ChPT calculation has allowed us to verify the above predictions of the WI framework, thus providing further support to this approach. 

Summarizing, the~framework of effective theories constitutes a modern, consistent and powerful theoretical tool to investigate several important problems regarding the QCD transition, especially concerning chiral symmetry restoration. The~various techniques and results reviewed here in that context are meant to shed light on the present understanding of the QCD phase diagram, thus joining efforts with the experimental and lattice current programs. 

 \vspace{6pt}


\acknowledgments{Useful comments, collaboration  and discussion with D.Cabrera, S.Cort\'es, A.Dobado, C.S. Fischer, F-K Guo, Z-H. Guo, F.Karsch, M.Laine, F.J.LLanes-Estrada, M.P. Lombardo, U.G.Meissner, J.Morales, S. Mukherjee, J.R. Pel\'aez, O. Phillipsen, J.Ruiz de Elvira, E. Ruiz Arriola, J.J. Sanz-Cillero, R.Torres Andr\'es, A. Vioque-Rodr\'iguez over recent years are acknowledged in connection with the different topics reviewed here. Work partially supported by  research contract FPA2016-75654-C2-2-P  (spanish ``Ministerio de Econom\'{\i}a y Competitividad"). This work has also received funding from the European Union Horizon 2020 research and innovation programme under grant agreement No 824093.}


\begin{thebibliography}{999}



\bibitem{Adams:2005dq} 
  Adams, J.; {et al}. [STAR Collaboration]. Experimental and theoretical challenges in the search for the quark gluon plasma: The~STAR Collaboration's critical assessment of the evidence from RHIC collisions. \emph{Nucl.~Phys.~A}\textbf{ 2005}, \emph{757}, 102--183.

  
  
  
\bibitem{Adcox:2004mh} 
  Adcox, K.; {et al}. [PHENIX Collaboration]. Formation of dense partonic matter in relativistic nucleus-nucleus collisions at RHIC: Experimental evaluation by the PHENIX collaboration. \emph{Nucl.\ Phys.\ A} \textbf{2005}, \emph{757}, 184--283.
  
  
  
  \bibitem{Ratti:2018ksb} 
  Ratti, C. Lattice QCD and heavy ion collisions: A review of recent progress.
  \emph{Rept. Prog. Phys.} \textbf{2018}, \emph{81},~084301.

\bibitem{Bazavov:2019lgz} 
  Bazavov, A.; Karsch, F.; Mukherjee, S.; Petreczky, P. Hot-dense Lattice QCD: USQCD whitepaper 2018.
  \emph{Eur.~Phys. J. A} \textbf{2019}, \emph{55}, 194.
  
  
\bibitem{Alford:2007xm} 
  Alford, M.G.; Schmitt, A.; Rajagopal, K.; Schafer, T. Color superconductivity in dense quark matter.  \emph{Rev.~Mod.~Phys.} \textbf{2008}, \emph{80}, 1455--1515.


  
  
\bibitem{Bazavov:2016uvm} 
  Bazavov, A.; Brambilla, N.; Ding, H.-T.; Petreczky, P.; Schadler, H.-P.; Vairo, A.; Weber, J.H. Polyakov loop in 2+1 flavor QCD from low to high temperatures. \emph{Phys. Rev. D}  \textbf{2016}, \emph{93}, 114502.
  
  
\bibitem{Smilga:1995qf} 
  Smilga, A.V.; Verbaarschot, J.J.M. Scalar susceptibility in QCD and the multiflavor Schwinger model. \mbox{\emph{Phys. Rev. D}} \textbf{1996}, \emph{54}, 1087.

  
\bibitem{Pisarski:1983ms} 
  Pisarski, R.D.; Wilczek, F. Remarks on the Chiral Phase Transition in Chromodynamics.
  \mbox{\emph{Phys. Rev. D}} \textbf{1984}, \emph{29}, 338--341.
  
 
 
 
 \bibitem{Aoki:2009sc}
  Aoki, Y.; Borsanyi, S.; Durr, S.; Fodor, Z.; Katz, S.D.; Krieg, S.; Szabo, K.K. The~QCD transition temperature: Results with physical masses in the continuum limit II.
  \emph{JHEP}  \textbf{2009}, \emph{0906}, 088.
  
  \bibitem{Borsanyi:2010bp} 
 Borsanyi, S.; Fodor, Z.; Hoelbling, C.; Katzc, S.D.; Krieg, S.; Ratti, C.; Szabo, K.K. Is there still any $T_c$ mystery in lattice QCD? Results with physical masses in the continuum limit III. \emph{JHEP} \textbf{2010}, \emph{1009}, 073.
  
  \bibitem{Bazavov:2011nk}
 Bazavov, A.;  Bhattacharya, T.;  Cheng, M.;  DeTar, C.; Ding, H.-T.; Gottlieb, S.;  Gupta, R.;  Hegde, P.; Heller,~U.M.; Karsch, F.; et al. The~chiral and deconfinement aspects of the QCD transition.
  \emph{Phys.~Rev.~D} \textbf{2012}, \emph{85}, 054503.


\bibitem{Bazavov:2018mes} 
  Bazavov, A.; Ding, H.-T.;  Hegde,  P.;  Kaczmarek. O.; Karsch,  F.;  Karthik, N.;  Laermann, E.; Lahiri, A.; Larsen,~R.; Li, S.-T.; et al. Chiral crossover in QCD at zero and non-zero chemical potentials.
  \emph{Phys. Lett. B} \textbf{2019}, \emph{795}, 15--21. 
  
  
  
\bibitem{Ding:2019prx} 
  Ding, H.T.;  Hegde, P.; Kaczmarek, O.;  Karsch, F.; Lahiri, A.;  Li, S.-T.;  Mukherjee, S.; Ohno, H.; Petreczky, P.;  Schmidt, C.; et al. Chiral Phase Transition Temperature in (2+1)-Flavor QCD.
  \emph{Phys. Rev. Lett. } \textbf{2019}, \emph{123},~062002.


\bibitem{DElia:2002tig} 
  D'Elia, M.; Lombardo, M.P. Finite density QCD via imaginary chemical potential.
  \emph{Phys. Rev. D} \textbf{2003}, {\emph{67}},~014505.
  
  
  
\bibitem{deForcrand:2006pv} 
  de Forcrand, P.; Philipsen, O. The~Chiral critical line of N(f) = 2+1 QCD at zero and non-zero baryon density. \emph{JHEP} \textbf{2007}, \emph{0701}, 077.
  
  
\bibitem{Fodor:2004nz} 
  Fodor, Z.; Katz, S.D. Critical point of QCD at finite T and mu, lattice results for physical quark masses. \emph{JHEP} \textbf{2004}, {\emph{0404}}, 050.
  
  
\bibitem{Aarts:2008wh} 
 Aarts, G. Can stochastic quantization evade the sign problem? The~relativistic Bose gas at finite chemical potential. \emph{ Phys. Rev. Lett.} \textbf{2009}, \emph{102}, 131601.
  
  
\bibitem{Adamczyk:2017iwn} 
  Adamczyk, L.; et al. [STAR Collaboration].  Bulk Properties of the Medium Produced in Relativistic Heavy-Ion Collisions from the Beam Energy Scan Program. \emph{Phys. Rev. C} \textbf{2017}, \emph{96}, 044904.




\bibitem{Andronic:2017pug} 
  Andronic, A.; Braun-Munzinger, P.; Redlich, K.; Stachel, J. Decoding the phase structure of QCD via particle production at high energy.  \emph{Nature} \textbf{2018}, \emph{561,} 321--330.



\bibitem{Bazavov:2014xya} 
  Bazavov, A.; Ding, H.-T.; Hegde, P.; Kaczmarek, O.; Karsch,  F.; Laermann,  E.; Maezawa, Y.;   Mukherjee,~S.; Ohno, H.; Petreczky, P.; et al. Additional Strange Hadrons from QCD Thermodynamics and Strangeness Freezeout in Heavy Ion Collisions.
  \emph{Phys. Rev. Lett.}  \textbf{2014}, \emph{113}, 072001.
  
  
\bibitem{Luo:2017faz} 
 Luo, X.; Xu, N. Search for the QCD Critical Point with Fluctuations of Conserved Quantities in Relativistic Heavy-Ion Collisions at RHIC: An Overview.
  \emph{Nucl. Sci. Tech.} \textbf{2017}, \emph{28}, 112.


\bibitem{Tanabashi:2018oca} 
  Tanabashi, M.;  et al. [Particle Data Group]. Review of Particle Physics. \emph{Phys. Rev. D} \textbf{2018}, \emph{98}, 030001.
  
  \bibitem{Hagedorn:1968zz} 
 Hagedorn, R. Hadronic matter near the boiling point. \emph{Nuovo Cim. A} \textbf{1968}, \emph{56}, 1027--1057.

\bibitem{Karsch:2003zq} 
  Karsch, F.; Redlich, K.; Tawfik, A. Thermodynamics at nonzero baryon number density: A Comparison of lattice and hadron resonance gas model calculations.
 \emph{ Phys. Lett. B} \textbf{2003},  \emph{571}, 67--74.

\bibitem{Karsch:2003vd} 
  Karsch, F.; Redlich, K.; Tawfik, A. Hadron resonance mass spectrum and lattice QCD thermodynamics. \emph{Eur.~Phys. J. C} \textbf{2003},  \emph{29}, 549--556.
  
\bibitem{Tawfik:2005qh} 
 Tawfik, A.; Toublan, D. Quark-antiquark condensates in the hadronic phase.
  \emph{Phys. Lett. B} \textbf{2005}, \emph{623}, 48--54.


\bibitem{Leupold:2006ih} 
  Leupold, S. Four-quark condensates and chiral symmetry restoration in a resonance gas model. \emph{J. Phys. G} \textbf{2006}, \emph{32}, 2199--2218.

\bibitem{Huovinen:2009yb} 
  Huovinen, P.; Petreczky, P. QCD Equation of State and Hadron Resonance Gas.
  \emph{Nucl. Phys. A} \textbf{ 2010}, \emph{837}, 26--53.
  
  \bibitem{Megias:2012kb} 
  Megias, E.; Arriola, E.R.; Salcedo, L.L. The~Polyakov loop and the hadron resonance gas model. \emph{ Phys. Rev. Lett.} \textbf{2012}, \emph{109}, 151601.

\bibitem{Jankowski:2012ms} 
  Jankowski, J.; Blaschke, D.; Spalinski, M. Chiral condensate in hadronic matter.
  \emph{Phys. Rev. D} \textbf{2013}, {\em 87}, 105018.
  
\bibitem{Andronic:2005yp} 
  Andronic, A.; Braun-Munzinger, P.; Stachel, J. Hadron production in central nucleus-nucleus collisions at chemical freeze-out.
  \emph{Nucl. Phys. A} \textbf{2006}, \emph{772}, 167--199.
  
\bibitem{Andronic:2012ut} 
  Andronic, A.; Braun-Munzinger, P.; Stachel, J.; Winn, M. Interacting hadron resonance gas meets lattice QCD.
  \emph{Phys. Lett. B} \textbf{2012}, \emph{718}, 80--85.
  
\bibitem{Huovinen:2017ogf} 
  Huovinen, P.; Petreczky, P. Hadron Resonance Gas with Repulsive Interactions and Fluctuations of Conserved Charges.
  \emph{Phys. Lett. B} \textbf{2018}, \emph{777}, 125--130.
  
\bibitem{Weinberg:1978kz} 
  Weinberg, S. Phenomenological Lagrangians.
  \emph{Physica A} \textbf{1979}, \emph{96}, 327--340.
  
    \bibitem{galekapustabook}  Kapusta, J.I.; Gale, C. {\em Finite Temperature Field Theory. Principles and Applications}; Cambridge University Press: Cambridge, UK, 2006.


  
\bibitem{Gasser:1983yg} 
  Gasser, J.; Leutwyler, H. Chiral Perturbation Theory to One Loop.
  \emph{Ann. Phys.} \textbf{1984}, \emph{158}, 142--210.


\bibitem{Gasser:1984gg} 
  Gasser, J.; Leutwyler, H. Chiral Perturbation Theory: Expansions in the Mass of the Strange Quark. \emph{Nucl.~Phys.~B} \textbf{1985},\emph{ 250}, 465--516.
  
\bibitem{Ecker:1988te} 
  Ecker, G.; Gasser, J.; Pich, A.; de Rafael, E. The~Role of Resonances in Chiral Perturbation Theory. \emph{Nucl.~Phys.~B} \textbf{1989}, \emph{321}, 311--342.
  
\bibitem{Ecker:1989yg} 
  Ecker, G.; Gasser, J.; Leutwyler, H.; Pich, A.; de Rafael, E. Chiral Lagrangians for Massive Spin 1 Fields. \emph{\mbox{Phys. Lett. B}} \textbf{1989}, \emph{223}, 425--432.
  
 
  
\bibitem{Meissner:1993ah} 
  Meissner, U.G. Recent developments in chiral perturbation theory.
  \emph{Rept. Prog. Phys.} \textbf{1993}, \emph{56}, 903--996.
  
  
  
\bibitem{Bernard:1995dp} 
 Bernard, V.; Kaiser, N.; Meissner, U.G. Chiral dynamics in nucleons and nuclei. \emph{Int. J. Mod. Phys. E}
 \textbf{1995}, \emph{4}, 193--346.
  
\bibitem{Ecker:1994gg} 
  Ecker, G. Chiral perturbation theory.
  \emph{Prog. Part. Nucl. Phys.} \textbf{1995}, \emph{35,} 1--80.



\bibitem{Pich:1995bw} 
  Pich, A. Chiral perturbation theory.
  \emph{Rept. Prog. Phys.} \textbf{1995}, \emph{58}, 563--610.
  
\bibitem{Witten:1979vv} 
  Witten, E. Current Algebra Theorems for the U(1) Goldstone Boson.
  \emph{Nucl. Phys. B} \textbf{1979}, \emph{156}, 269--283.


\bibitem{DiVecchia:1980ve} 
  Vecchia, P.D.; Veneziano, G. Chiral Dynamics in the Large n Limit. 
  \emph{Nucl. Phys. B} \textbf{1980}, \emph{171}, 253--271.
  
\bibitem{HerreraSiklody:1996pm} 
  Herrera-Siklody, P.; Latorre, J.I.; Pascual, P.; Taron, J. Chiral effective Lagrangian in the large N(c) limit: The~Nonet case. \emph{Nucl. Phys. B} \textbf{1997}, \emph{497}, 345--386.

\bibitem{Kaiser:2000gs} 
 Kaiser, R.; Leutwyler, H. Large N(c) in chiral perturbation theory.
  \emph{Eur. Phys. J. C} \textbf{2000}, \emph{17}, 623--649.
%


\bibitem{Aoki:2019cca} 
  Aoki, S.; Aoki, Y.;  Becirevic, D.; Blum, T.; Colangelo, G.; Collins, S.; Morte, M.D.; Dimopoulos, P.; Durr, S.;  Fukaya, H.;   et al. FLAG Review 2019.
  \emph{Eur. Phys. J. C} \textbf{2020}, \emph{80}, 113.

  
   \bibitem{Gerber:1988tt} 
  Gerber, P.; Leutwyler, H.  Hadrons Below the Chiral Phase Transition.
  \emph{Nucl. Phys. B} \textbf{1989}, \emph{321}, 387--429.
 
 
\bibitem{Leutwyler:1987th} 
 Leutwyler, H. {QCD}: Low Temperature Expansion and Finite Size Effects.
  \emph{Nucl. Phys. Proc. Suppl.} 
  \textbf{1988}, \emph{4}, 248--258.
  
  
\bibitem{FernandezFraile:2008vu} 
  Fernandez-Fraile, D.; G\'omez Nicola, A. Bulk viscosity and the conformal anomaly in the pion gas.
  \emph{Phys. Rev. Lett.} \textbf{2009}, \emph{102}, 121601.


\bibitem{FernandezFraile:2009mi} 
  Fernandez-Fraile, D.; Nicola, A.G. Transport coefficients and resonances for a meson gas in Chiral Perturbation Theory.
  \emph{Eur. Phys. J. C} \textbf{2009}, \emph{62}, 37.


\bibitem{FernandezFraile:2005ka} 
  Fernandez-Fraile, D.; Nicola, A.G. The~Electrical conductivity of a pion gas.
  \emph{Phys. Rev. D} \textbf{2006}, \emph{73}, 045025.
  
\bibitem{Dobado:2008vt} 
  Dobado, A.; Llanes-Estrada, F.J.; Torres-Rincon, J.M. Eta/s and phase transitions.
  \emph{Phys. Rev. D} \textbf{2009}, \emph{79},~014002.
  
\bibitem{Dobado:2011qu} 
 Dobado, A.; Llanes-Estrada, F.J.; Torres-Rincon, J.M. Bulk viscosity of low-temperature strongly interacting matter.
  \emph{Phys. Lett. B} \textbf{2011}, \emph{702}, 43.
  
\bibitem{Dobado:2011wc} 
  Dobado, A.; Llanes-Estrada, F.J.; Torres-Rincon, J.M. Bulk viscosity and energy-momentum correlations in high energy hadron collisions. \emph{Eur. Phys. J. C} \textbf{2012}, \emph{72}, 1873.
  
\bibitem{Das:2011vba} 
 Das, S.K.; Ghosh, S.; Sarkar, S.; Alam, J.E. Drag and diffusion coefficients of $B$ mesons in hot hadronic matter.
  \emph{Phys. Rev. D} \textbf{2012}, \emph{85}, 074017.
  
  
\bibitem{Tolos:2016slr} 
  Tolos, L.; Torres-Rincon, J.M.; Das, S.K. Transport coefficients of heavy baryons. \emph{Phys. Rev. D} \textbf{2016}, \emph{94},~034018.

  
  
\bibitem{Abhishek:2017pkp} 
  Abhishek, A.; Mishra, H.; Ghosh, S. Transport coefficients in the Polyakov quark meson coupling model: A~relaxation time approximation.
  \emph{Phys. Rev. D} \textbf{2018}, \emph{97}, 014005.
  
  

\bibitem{Dashen:1969ep} 
  Dashen, R.; Ma, S.K.; Bernstein, H.J. S Matrix formulation of statistical mechanics.
 \emph{ Phys. Rev.} \textbf{1969}, \emph{187}, 345.
  
    
\bibitem{Venugopalan:1992hy} 
 Venugopalan, R.; Prakash, M. Thermal properties of interacting hadrons.
  \emph{Nucl. Phys. A} \textbf{1992}, \emph{546}, 718.

\bibitem{Dobado:1998tv} 
  Dobado, A.; Pelaez, J.R. Chiral symmetry and the pion gas virial expansion.
  \emph{Phys. Rev. D} \textbf{1999}, \emph{59}, 034004.
 
\bibitem{Pelaez:2002xf}
 Pelaez, J.R. The~SU(2) and SU(3) chiral phase transitions within chiral perturbation  theory.
  \emph{Phys. Rev.  D} \textbf{2002}, \emph{66}, 096007.
  
  
\bibitem{GarciaMartin:2006jj}
Martin, R.G.; Pelaez, J.R. Chiral condensate thermal evolution at finite baryon chemical potential within Chiral Perturbation Theory.
\emph{Phys. Rev. D} \textbf{2006}, \emph{74}, 096003.

\bibitem{GomezNicola:2012uc} 
G\'omez Nicola, A.; Pelaez, J.R.; de Elvira, J.R. Scalar susceptibilities and four-quark condensates in the meson gas within Chiral Perturbation Theory.
  \emph{Phys. Rev. D} \textbf{2013}, \emph{87}, 016001.
  
  
   \bibitem{Broniowski:2015oha} 
  Broniowski, W.; Giacosa, F.; Begun, V.  Cancellation of the $\sigma$ meson in thermal models.
  \emph{Phys. Rev. C} \textbf{2015}, \emph{92},~034905.
  
  
\bibitem{Ferreres-Sole:2018djq} 
Ferreres-Sol\'e, S.; G\'omez Nicola, A.;  Vioque-Rodr\'iguez, A.
Role of the thermal $f_0$(500) in chiral symmetry restoration.
  \emph{Phys. Rev. D} \textbf{2019}, \emph{99}, 036018.
  
\bibitem{Albaladejo:2012te}
Albaladejo, M.; Oller, J. On the size of the sigma meson and its nature.
\emph{Phys. Rev. D} \textbf{2012}, \emph{86}, 034003.
  
  
\bibitem{Pelaez:2015qba} 
  Pelaez, J.R. From controversy to precision on the sigma meson: A review on the status of the non-ordinary $f_0(500)$ resonance.
  \emph{Phys. Rept.} \textbf{2016}, \emph{658}, 1.
  
\bibitem{GellMann:1960np} 
  Gell-Mann, M.; Levy, M. The~axial vector current in beta decay.
  \emph{Nuovo Cim.} \textbf{1960}, \emph{16}, 705.
  
  
  \bibitem{Hatsuda:1985eb} 
  Hatsuda, T.; Kunihiro, T. Fluctuation Effects in Hot Quark Matter: Precursors of Chiral Transition at Finite Temperature.
  \emph{Phys. Rev. Lett.} \textbf{1985}, \emph{55}, 158.
  
  
\bibitem{Bernard:1987im}
Bernard, V.; Meissner, U.G.; Zahed, I. Properties of the Scalar $\sigma$ Meson at Finite Density.
\emph{Phys. Rev. Lett.} \textbf{1987}, \emph{59}, 966.
  
  
  \bibitem{Bochkarev:1995gi} 
  Bochkarev, A.; Kapusta, J.I. Chiral symmetry at finite temperature: Linear versus nonlinear sigma models.
  \emph{Phys. Rev. D} \textbf{1996}, \emph{54}, 4066.
  
  \bibitem{Ayala:2000px} 
  Ayala, A.; Sahu, S. Pion propagation in the linear sigma model at finite temperature.
  \emph{Phys. Rev. D} \textbf{2000}, \emph{62},~056007.



 \bibitem{Masjuan:2008cp}
  Masjuan, P.; Sanz-Cillero, J.J.; Virto, J. Some Remarks on the Pade Unitarization of Low-Energy Amplitudes.
  \emph{Phys. Lett. B} \textbf{2008}, \emph{668}, 14. 
  
  
\bibitem{Truong:1988zp} 
  Truong, T.N. Chiral Perturbation Theory and Final State Theorem.
  \emph{Phys. Rev. Lett.} \textbf{1988}, \emph{61, }2526.
  
  
  
\bibitem{Dobado:1989qm}
  Dobado, A.; Herrero, M.J.; Truong, T.N. Unitarized Chiral Perturbation Theory for Elastic Pion-Pion Scattering.
  \emph{Phys. Lett. B} \textbf{1990}, \emph{235}, 134.
  
  
  
  \bibitem{Dobado:1996ps}
Dobado, A.; Pelaez, J.R. The~inverse amplitude method in Chiral Perturbation Theory.
  \emph{Phys. Rev.  D} \textbf{1997}, \emph{56},~3057.


\bibitem{Oller:1997ti} 
  Oller, J.A.; Oset, E. Chiral symmetry amplitudes in the S wave isoscalar and isovector channels and the $\sigma$, f$_0$(980), a$_0$(980) scalar mesons.
  \emph{Nucl. Phys. A} \textbf{1997}, \emph{620,} 438.
  Erratum in  \textbf{1999}, \emph{652}, 407.

\bibitem{Oller:1998hw} 
 Oller, J.A.; Oset, E.; Pelaez, J.R. Meson meson interaction in a nonperturbative chiral approach.
  \emph{Phys. Rev. D} \textbf{1999}, \emph{59}, 074001;
  Erratum in  \textbf{1999}, \emph{60}, 099906;    Erratum in \textbf{2007}, \emph{75}, 099903.



\bibitem{GomezNicola:2001as} 
 G\'omez Nicola, A.; Pelaez, J.R. Meson meson scattering within one loop chiral perturbation theory and its unitarization.
  \emph{Phys. Rev. D} \textbf{2002}, \emph{65}, 054009.
  
\bibitem{Pelaez:2003xd}
Pelaez, J.; G\'omez Nicola, A. Light meson resonances from unitarized chiral perturbation theory.
\emph{AIP Conf. Proc.} \textbf{2003}, \emph{660}, 102--115.
  
\bibitem{GomezNicola:2007qj} 
  G\'omez Nicola, A.; Pelaez, J.R.; Rios, G. The~Inverse Amplitude Method and Adler Zeros.
  \emph{Phys. Rev. D} \textbf{2008}, \emph{77}, 056006.

\bibitem{Albaladejo:2010tj}
Albaladejo, M.; Oller, J.; Roca, L. Dynamical generation of pseudoscalar resonances.
\emph{Phys. Rev. D} \textbf{2010}, \emph{82},~094019.



\bibitem{Dobado:1992jg}
Dobado, A.; Pelaez, J. On the large N(f) limit of chiral perturbation theory.
\emph{Phys. Lett. B} \textbf{1992}, \emph{286}, 136--146.


\bibitem{Dobado:1994fd}
Dobado, A.; Morales, J. Pion mass effects in the large N limit of chi(PT).
\emph{Phys. Rev. D} \textbf{1995}, \emph{52}, 2878--2890.


\bibitem{Cortes:2015emo}
Cort\'es, S.; G\'omez Nicola, A.; Morales, J. Large-$N$ pion scattering at finite temperature: The $f_0(500)$ and chiral restoration.
\emph{Phys. Rev. D} \textbf{2016}, \emph{93}, 036001.



\bibitem{Cortes:2016ecy}
Cort\'es, S.; G\'omez Nicola, A.; Morales, J. Chiral Symmetry Restoration for the large-$N$ pion gas.
\emph{Phys. Rev. D} \textbf{2016}, \emph{94}, 116008.

\bibitem{Gasser:1986vb} 
  Gasser, J.; Leutwyler, H. Light Quarks at Low Temperatures.
  \emph{Phys. Lett. B} \textbf{1987}, \emph{184}, 83.
  
  
  
\bibitem{Schenk:1993ru} 
Schenk, A. Pion propagation at finite temperature.
  \emph{Phys. Rev. D} \textbf{1993}, \emph{47}, 5138.
  
  
\bibitem{Goity:1989gs} 
  Goity, J.L.; Leutwyler, H. On the Mean Free Path of Pions in Hot Matter.
  \emph{Phys. Lett. B} \textbf{1989}, \emph{228}, 517.
  
\bibitem{Pisarski:1996mt} 
  Pisarski, R.D.; Tytgat, M. Propagation of cool pions.
  \emph{Phys. Rev. D} \textbf{1996}, \emph{54}, R2989.
  
  
  
\bibitem{Adamczyk:2015lme} 
  Adamczyk, L.;   et al. [STAR Collaboration]. Measurements of Dielectron Production in Au$+$Au Collisions at $\sqrt{s_{\rm NN}}$ = 200 GeV from the STAR Experiment.
  \emph{Phys. Rev. C} \textbf{2015}, \emph{92}, 024912.
  
  
\bibitem{Adare:2015ila} 
  Adare, A.;  Aidala, C.; Ajitanand, N.N.; Akiba, Y.; Akimoto, R.;  Alexander, J.;  Alfred,  M.;  Al-Ta'ani, H.;  Angerami, A.;  Aoki, K.; et al. Dielectron production in Au$+$Au collisions at $\sqrt{s_{NN}}$=200 GeV.
  \emph{Phys. Rev. C} \textbf{2016}, \emph{93}, 014904.
  
\bibitem{Acharya:2018nxm} 
  Acharya, S.; et al. [ALICE Collaboration]. Measurement of dielectron production in central Pb-Pb collisions at $\sqrt{{\textit{s}}_{\mathrm{NN}}}$ = 2.76 TeV.
  \emph{Phys.~Rev.~C} \textbf{2019}, \emph{99}, 024002.
  
\bibitem{Adamczewski-Musch:2019byl}
Adamczewski-Musch, J.;  et al. [HADES  Collaboration].  Probing dense baryon-rich matter with virtual photons.
\emph{Nat. Phys.} \textbf{2019}, \emph{15}, 1040--1045.
  
  \bibitem{Rapp:1999ej}
  Rapp, R.; Wambach, J. Chiral symmetry restoration and dileptons in relativistic heavy ion collisions.
 \emph{ Adv.~Nucl. Phys.} \textbf{2000}, \emph{25}, 1. 
  
  
\bibitem{Jung:2016yxl} 
 Jung, C.; Rennecke, F.; Tripolt, R.A.; von Smekal, L.; Wambach, J. In-Medium Spectral Functions of Vector- and Axial-Vector Mesons from the Functional Renormalization Group.
  \emph{Phys. Rev. D} \textbf{2017}, \emph{95}, 036020.
  
\bibitem{Rapp:2014hha} 
Rapp, R.; van Hees, H. Thermal Dileptons as Fireball Thermometer and Chronometer. \emph{Phys. Lett. B} \textbf{2016}, \emph{753}, 586.

\bibitem{Turbide:2003si}
Turbide, S.; Rapp, R.; Gale, C. Hadronic production of thermal photons.  
  \emph{Phys. Rev. C} \textbf{2004}, \emph{69}, 014903.
  
  
\bibitem{Paquet:2015lta} 
Paquet, J.F.; Shen, C.; Denicol, G.S.; Luzum, M.; Schenke, B.; Jeon, S.; Gale, C. Production of photons in relativistic heavy-ion collisions.
  \emph{Phys. Rev. C} \textbf{2016}, \emph{93}, 044906.
  
  
  
  
\bibitem{GomezNicola:2002tn}
G\'omez Nicola, A.; Llanes-Estrada, F.J.; Pelaez, J. Finite temperature pion scattering to one loop in chiral perturbation theory.
\emph{Phys. Lett. B} \textbf{2002}, \emph{550}, 55--64.

\bibitem{Dobado:2002xf}
Dobado, A.; G\'omez Nicola, A.; Llanes-Estrada, F.J.; Pelaez, J. Thermal rho and sigma mesons from chiral symmetry and unitarity.
\emph{Phys. Rev. C} \textbf{2002}, \emph{66}, 055201.

\bibitem{FernandezFraile:2007fv}
Fernandez-Fraile, D.; G\'omez Nicola, A.; Herruzo, E. Pion scattering poles and chiral symmetry restoration.
\emph{Phys. Rev. D} \textbf{2007}, \emph{76}, 085020.
  
\bibitem{Cabrera:2008tja}
Cabrera, D.; Fernandez-Fraile, D.; G\'omez Nicola, A. Chiral Symmetry and light resonances in hot and dense matter.
\emph{Eur. Phys. J. C} \textbf{2009}, \emph{61}, 879--892.

\bibitem{Nicola:2013vma}
G\'omez Nicola, A.; de Elvira, J.R.; Andres, R.T. Chiral Symmetry Restoration and Scalar-Pseudoscalar partners in QCD.
\emph{Phys. Rev. D} \textbf{2013}, \emph{88}, 076007.

\bibitem{Weldon:1983jn}
Weldon, H. Simple Rules for Discontinuities in Finite Temperature Field Theory.
\emph{Phys. Rev. D} \textbf{1983}, \emph{28}, 2007.


\bibitem{Ghosh:2009bt}
Ghosh, S.; Sarkar, S.; Mallik, S. Analytic structure of rho meson propagator at finite temperature.
\emph{Eur. Phys. J.~C} \textbf{2010}, \emph{70}, 251--262.

\bibitem{GomezNicola:2004gg}
G\'omez Nicola, A.; Llanes-Estrada, F.J.; Pelaez, J. Finite temperature pion vector form-factors in chiral perturbation theory.
\emph{Phys. Lett. B} \textbf{2005}, \emph{606}, 351--360.

\bibitem{Song:1996dg}
Song, C.; Koch, V. Pion electromagnetic form-factor at finite temperature.
\emph{Phys. Rev. C} \textbf{1996}, \emph{54}, 3218--3231.

\bibitem{Gao:2019idb}
Gao, R.; Guo, Z.; Pang, J. Thermal behaviors of light scalar resonances at low temperatures.
\emph{Phys. Rev. D} \textbf{2019}, \emph{100}, 114028.


\bibitem{Hanhart:2008mx} 
Hanhart, C.; Pelaez, J.R.; Rios, G. Quark mass dependence of the rho and sigma from dispersion relations and Chiral Perturbation Theory.
  \emph{Phys. Rev. Lett.}  {\bf 2008}, {\em 100}, 152001.
  
  \bibitem{Pelissetto:2013hqa} 
Pelissetto, A.; Vicari, E. Relevance of the axial anomaly at the finite-temperature chiral transition in QCD.
\emph{Phys. Rev. D} \textbf{2013}, \emph{88}, 105018.

 \bibitem{Eser:2015pka} 
Eser, J.; Grahl, M.; Rischke, D.H. Functional Renormalization Group Study of the Chiral Phase Transition Including Vector and Axial-vector Mesons.
  \emph{Phys. Rev. D} \textbf{2015}, \emph{92}, 096008.

 \bibitem{Fejos:2015xca} 
Fejos, G. Functional dependence of axial anomaly via mesonic fluctuations in the three flavor linear sigma model.
  \emph{Phys. Rev. D} \textbf{2015}, \emph{92}, 036011.
 
 \bibitem{Gross:1980br} 
Gross, D.J.; Pisarski, R.D.; Yaffe, L.G. QCD and Instantons at Finite Temperature.
  \emph{Rev. Mod. Phys.} \textbf{1981}, \emph{53},~43.

\bibitem{Mitter:2013fxa} 
Mitter, M.; Schaefer, B.J. Fluctuations and the axial anomaly with three quark flavors.
  \emph{Phys. Rev. D}  \textbf{2014}, \emph{89},~054027.
  
  \bibitem{Kapusta:1995ww} 
Kapusta, J.I.; Kharzeev, D.; McLerran, L.D. The~Return of the prodigal Goldstone boson.
  \emph{Phys. Rev. D} \textbf{1996}, \emph{53}, 5028.

  
  \bibitem{Csorgo:2009pa} 
Csorgo, T.; Vertesi, R.; Sziklai, J. Indirect observation of an in-medium $\eta$' mass reduction in $\sqrt{s_{NN}}=200$ GeV Au+Au collisions.
  \emph{Phys. Rev. Lett.} \textbf{2010}, \emph{105}, 182301.
  
  \bibitem{Kotov:2019dby} 
Kotov, A.Y.; Lombardo, M.P.; Trunin, A.M. Fate of the $\eta?$ in the quark gluon plasma.
  \emph{Phys. Lett. B} \textbf{2019}, \emph{794},~83.



\bibitem{Shuryak:1993ee} 
Shuryak, E.V. Which chiral symmetry is restored in hot QCD?
  \emph{Comments Nucl. Part. Phys.} \textbf{1994}, \emph{21}, 235.

  
\bibitem{Cohen:1996ng} 
Cohen, T.D. The~High temperature phase of QCD and U(1)-A symmetry.
  \emph{Phys. Rev. D} \textbf{1996}, \emph{54}, R1867.

\bibitem{Lee:1996zy} 
Lee, S.H.; Hatsuda, T. U-a(1) symmetry restoration in QCD with N(f) flavors.
  \emph{Phys. Rev. D} \textbf{1996}, \emph{54}, R1871.
  
  \bibitem{Meggiolaro:2013swa} 
Meggiolaro, E.; Morda, A. Remarks on the $U(1)$ axial symmetry and the chiral transition in QCD at finite temperature.
\emph{Phys. Rev. D} \textbf{2013}, \emph{88}, 096010.


  
\bibitem{Heller:2015box} 
Heller, M.; Mitter, M. Pion and $\eta$-meson mass splitting at the two-flavor chiral crossover.
\emph{Phys. Rev. D} \textbf{2016}, \emph{94}, 074002.


\bibitem{Ishii:2015ira} 
Ishii, M.; Yonemura, K.; Takahashi, J.; Kouno, H.; Yahiro, M.
Determination of $U(1)_{\rm A}$ restoration from pion and $a_0$-meson screening masses: Toward the chiral regime.
\emph{Phys. Rev. D} \textbf{2016}, \emph{93}, 016002.

\bibitem{Glozman:2018jkb}
Glozman, L. Chiralspin Symmetry and Its Implications for QCD.
\emph{Universe} \textbf{2019}, \emph{5}, 38.

\bibitem{Nicola:2016jlj} 
  G\'omez Nicola, A.; de Elvira, J.R. Pseudoscalar susceptibilities and quark condensates: Chiral restoration and lattice screening masses.
  \emph{JHEP} \textbf{2016}, \emph{1603}, 186.


 \bibitem{Azcoiti:2016zbi} 
Azcoiti, V. Topology in the SU(Nf) chiral symmetry restored phase of unquenched QCD and axion cosmology.
  \emph{Phys. Rev. D} \textbf{2016}, \emph{94}, 094505.


 \bibitem{GomezNicola:2017bhm} 
  G\'omez Nicola, A.; de Elvira, J.R. Patterns and partners for chiral symmetry restoration.
  \emph{Phys. Rev. D} \textbf{2018}, \emph{97}, 074016.
  
  \bibitem{Nicola:2018vug} 
  G\'omez Nicola, A.; Elvira, J.R.D. Chiral and $U(1)_A$ restoration for the scalar and pseudoscalar meson nonets.
  \emph{Phys. Rev. D} \textbf{2018}, \emph{98}, 014020. 

\bibitem{Buchoff:2013nra} 
Buchoff, M.I.;  Cheng, M.; Christ, N.H.;  Ding, H.-T.; Jung, C.; Karsch, F.;   Lin, Z.; Mawhinney, R.D.; Mukherjee,~S.; Petreczky,  P.; et al. QCD chiral transition, U(1)A symmetry and the Dirac spectrum using domain wall fermions.
  \emph{Phys. Rev. D} \textbf{2014}, \emph{89}, 054514.


\bibitem{Aoki:2012yj}
Aoki, S.; Fukaya, H.; Taniguchi, Y. Chiral symmetry restoration, eigenvalue density of Dirac operator and axial U(1) anomaly at finite temperature.
  \emph{Phys. Rev. D} \textbf{2012}, \emph{86}, 114512.

\bibitem{Cossu:2013uua} 
Cossu, G.; Aoki, S.; Fukaya, H.; Hashimoto, S.; Kaneko, T.; Matsufuru, H.; Noaki, J.I. Finite temperature study of the axial U(1) symmetry on the lattice with overlap fermion formulation.
  \emph{Phys. Rev. D} \textbf{2013}, \emph{87}, 114514.
  Erratum in  \textbf{2013}, \emph{88}, 019901.


\bibitem{Tomiya:2016jwr} 
  Tomiya, A.; Cossu, G.; Aoki, S.; Fukaya, H.; Hashimoto, S.; Kaneko, T.; Noaki, J. Evidence of effective axial U(1) symmetry restoration at high temperature QCD.  \emph{Phys. Rev. D} \textbf{2017}, \emph{96}, 034509.
    
\bibitem{Brandt:2016daq} 
  Brandt, B.B.; Francis, A.; Meyer, H.B.; Philipsen, O.; Robaina, D.; Wittig, H. On the strength of the $U(1)_A$ anomaly at the chiral phase transition in $N_f=2$ QCD.
  \emph{JHEP} \textbf{2016}, \emph{1612}, 158.
   

\bibitem{Brandt:2019ksy}
Brandt, B.B.; C\`e, M.; Francis, A.; Harris, T.; Meyer, H.B.; Wittig, H.; Philipsen, O. Testing the strength of the $\text{U}_A(1)$ anomaly at the chiral phase transition in two-flavour QCD.
\emph{arXiv} \textbf{2019}, arXiv:1904.02384.

\bibitem{Lee:2017uvl} 
Lee, J.W.; Lucini, B.; Piai, M. Symmetry restoration at high-temperature in two-color and two-flavor lattice gauge theories. 
 \emph{ JHEP} \textbf{2017}, \emph{1704}, 036.
  
  \bibitem{Chandrasekharan:2010ik} 
Chandrasekharan, S.; Li, A. Anomaly and a QCD-like phase diagram with massive bosonic baryons.
  \emph{JHEP} \textbf{2010}, \emph{1012}, 021.
  
   \bibitem{Aarts:2015mma} 
Aarts, G.; Allton, C.; Hands, S.; J\"ager, B.; Praki, C.; Skullerud, J.I. 
`Nucleons and parity doubling across the deconfinement transition.
  \emph{Phys. Rev. D} \textbf{2015}, \emph{92}, 014503. 
  
  
  \bibitem{Aarts:2017rrl} 
 Aarts, G.; Allton, C.; Boni, D.D.; Hands, S.; J\"ager, B.; Praki, C.; Skullerud, J.I. Light baryons below and above the deconfinement transition: Medium effects and parity doubling.
  \emph{JHEP} \textbf{2017}, \emph{1706}, 034.



\bibitem{Bochicchio:1985xa}
Bochicchio, M.; Maiani, L.; Martinelli, G.; Rossi, G.C.; Testa, M. Chiral Symmetry on the Lattice with Wilson Fermions.
  \emph{Nucl. Phys. B} \textbf{1985}, \emph{262}, 331.


\bibitem{Boucaud:2009kv}
Boucaud, P.; Leroy, J.P.; Yaouanc, A.L.; Micheli, J.; Pene, O.; Rodriguez-Quintero, J. Quark pseudoscalar vertex and quark mass function with clover fermions: Spontaneous symmetry breaking, OPE, symmetry restoration at small volume. \emph{Phys. Rev. D} \textbf{2010}, \emph{81}, 094504.
    
    
    \bibitem{Karsch:2003jg}
Karsch, F.; Laermann, E. 
Thermodynamics and in medium hadron properties from lattice QCD.
  \emph{arXiv} \textbf{2003}, arXiv:hep-lat/0305025




\bibitem{Ishii:2016dln} 
Ishii, M.; Kouno, H.; Yahiro, M. Model prediction for temperature dependence of meson pole masses from lattice QCD results on meson screening masses.
  \emph{Phys. Rev. D} \textbf{2017}, \emph{95}, 114022.
  
  \bibitem{Cheng:2010fe}
  Cheng, M.;  Datta,  S.;  Francis, A.; van der Heide, J.;  Jung, C.;  Kaczmarek, O.;  Karsch, F.;   Laermann, E.;  Mawhinney, R.D.; Miao, C.; et al.  Meson screening masses from lattice QCD with two light and the strange quark.
  \emph{Eur. Phys. J. C} \textbf{2011}, \emph{71}, 1564.
  
   \bibitem{Cheng:2007jq} 
  Cheng, M.;  Datta, S.;  Francis, A.; van der Heide, J.;  Jung, C.; Kaczmarek, O.;  Karsch, F.;   Laermann, E.;  Mawhinney, R.D.;  Miao, C.; et al. The~QCD equation of state with almost physical quark masses. \emph{Phys.~Rev.~D} \textbf{2008}, \emph{77}, 014511.
  
  \bibitem{Maezawa:2013nxa} 
  Maezawa, Y.; Bazavov, A.; Karsch, F.; Petreczky, P.; Mukherjee, S. Meson screening masses at finite temperature with Highly Improved Staggered Quarks.
\emph{ arXiv} {\bf 2013}, arXiv:1312.4375
  
\bibitem{Nicola:2019ohb}
G\'omez Nicola, A.; Elvira, J.R.D.; Vioque-Rodr\'iguez, A. The~QCD topological charge and its thermal dependence: The role of the $\eta'$.
\emph{JHEP} \textbf{2019}, \emph{11}, 086.

 \bibitem{Guo:2015xva} 
Guo, X.K.; Guo, Z.H.; Oller, J.A.; Sanz-Cillero, J.J. Scrutinizing the $\eta$-$\eta'$ mixing, masses and pseudoscalar decay constants in the framework of U(3) chiral effective field theory.
  \emph{JHEP} \textbf{2015}, \emph{1506}, 175.
  
\bibitem{diCortona:2015ldu} 
di Cortona, G.G.; Hardy, E.; Vega, J.P.; Villadoro, G. The~QCD axion, precisely.
  \emph{JHEP} \textbf{2016}, \emph{1601}, 034.
  
\bibitem{Bernard:2012fw} 
Bernard, V.; Descotes-Genon, S.; Toucas, G. Topological susceptibility on the lattice and the three-flavour quark condensate.
  \emph{JHEP} \textbf{2012}, \emph{1206}, 051.
  
  \bibitem{Aoki:2016frl} 
Aoki, S.; Aoki, Y.; Bernard, C.; Blum, T.;  Colangelo, G.;  Della Morte, M.;  Durr, S.;   El-Khadra, A.X.; Fukaya,~H.;  Horsley, R.; et al. Review of lattice results concerning low-energy particle physics.
  \emph{Eur. Phys. J. C} \textbf{2017}, \emph{77},~112.
  
\bibitem{Burger:2018fvb} 
Burger, F.; Ilgenfritz, E.M.; Lombardo, M.P.; Trunin, A. Chiral observables and topology in hot QCD with two families of quarks.
  \emph{Phys. Rev. D} \textbf{2018}, \emph{98}, 094501.
  
\bibitem{Leutwyler:1992yt} 
Leutwyler, H.; Smilga, A.V. Spectrum of Dirac operator and role of winding number in QCD.
  \emph{Phys. Rev. D} \textbf{1992}, \emph{46}, 5607.
  
  \bibitem{Veneziano:1979ec} 
Veneziano, G. U(1) Without Instantons.
  \emph{Nucl. Phys. B} \textbf{1979}, \emph{159}, 213.
  
\bibitem{Mao:2009sy} 
Mao, Y-Y.; Chiu, T-W. Topological Susceptibility to the One-Loop Order in Chiral Perturbation Theory.
  \emph{Phys.~Rev. D} \textbf{2009}, \emph{80}, 034502.
  
   \bibitem{Bonati:2015vqz} 
Bonati, C.; D'Elia, M.; Mariti, M.; Martinelli, G.; Mesiti, M.; Negro, F.; Sanfilippo, F.; Villadoro, G. Axion phenomenology and $\theta$-dependence from $N_f = 2+1$ lattice QCD.
  \emph{JHEP} \textbf{2016}, \emph{1603}, 155.
  
  
\bibitem{Vonk:2019kwv} 
Vonk, T.; Guo, F.K.; Mei\ss ner, U.G. Aspects of the QCD $\theta$-vacuum.
  \emph{JHEP} \textbf{2019}, \emph{1906}, 106.
  
\bibitem{Vicari:2008jw} 
Vicari, E.; Panagopoulos, H. Theta dependence of SU(N) gauge theories in the presence of a topological term. \emph{Phys. Rept.} \textbf{2009}, \emph{470}, 93.

  
\bibitem{Borsanyi:2016ksw} 
  Borsanyi, S.;  Fodor, Z.;  Guenther, J.; Kampert, K.-H.;  Katz, S.D.; Kawanai, T.; Kovacs, T.G.; Mages, S.W.;  Pasztor, A.;  Pittler,   F.; et al. Calculation of the axion mass based on high-temperature lattice quantum chromodynamics. \emph{Nature} \textbf{2016},\emph{ 539}, 69.
  
\end{thebibliography}
\end{document}